\documentclass[openany]{book}
\usepackage{makeidx}
\usepackage{enumitem}
\usepackage{multicol}
\usepackage{color}
\usepackage{ifpdf}
\usepackage[pdftex]{graphicx}
\DeclareGraphicsExtensions{.eps, .pdf, .jpg, .tif}
\usepackage{ulem}
\usepackage{amssymb}
\usepackage{amscd}
\usepackage{amsfonts}
\usepackage{amsmath}
\usepackage{dsfont}
\usepackage{bm}
\usepackage{eufrak}
\usepackage{float}
\usepackage{mathptmx}
\usepackage[pdftex]{hyperref}
\usepackage{bookmath}
\usepackage{mathfrak}
\usepackage{mymatrices}
\usepackage{myslides}
\setlist{leftmargin=25pt,itemsep=2pt,topsep=0pt,parsep=-2pt,partopsep=-2pt}
\def\sec#1#2{{\section{\large\sl #1}\label{#2}}}

\renewcommand{\vec}[1]{{\boldsymbol{#1}}}
\newcommand{\vhat}[1]{\hat{\boldsymbol{#1}}}
\renewcommand{\onehalf}{{{1}\over {2}}}
\renewcommand{\onethird}{{{1}\over {3}}}
\renewcommand{\onefifth}{{{1}\over {5}}}
\renewcommand{\onesixth}{{{1}\over {6}}}
\renewcommand{\threehalves}{{{3}\over {2}}}
\newcommand{\sevensixths}{{{7}\over {6}}}
\newcommand{\sixsevenths}{{{6}\over {7}}}
\newcommand{\foursevenths}{{{4}\over {7}}}
\newcommand{\onetwelveth}{{{1}\over {12}}}
\newcommand{\twotwentysevenths}{{{2}\over{27}}}

\begin{document}
\frontmatter
\thispagestyle{empty}
\vspace*{-3cm}
{\footnotesize
\hspace*{-2.5cm} Published in {\it HELIUM THREE}										\newline
\hspace*{-2.0cm} Volume 26 of {\sl Modern Problems in Condensed Matter Physics}, Chapter 5, pp. 255--311.			\newline
\hspace*{-2.0cm} Edited by W. P. Halperin and L. P. Pitaevskii, North-Holland, 1990.	\newline
}
\vspace*{4cm}

\centerline{\LARGE\bf Collective Modes and Nonlinear Acoustics in}	
\smallskip
\centerline{\LARGE\bf Superfluid $^3$He-B}
\font\twelverm=cmr12
\bigskip
\bigskip
\centerline{\large Ross H. McKenzie$^{\dag}$                   and J. A. Sauls}
\bigskip
\centerline{Department of Physics and Astronomy}
\centerline{Northwestern University}
\centerline{2145 Sheridan Road}
\centerline{Evanston, Illinois 60208}
\smallskip
\centerline{April 1990}
\vspace*{8cm}
\noindent{\footnotesize$^{\dag}$Current address: Department of Physics, Ohio State University, Columbus, OH 43210.} 
\vfill
\eject

\hspace*{-1.5cm}
\begin{minipage}{1.5\textwidth}
\tableofcontents
\end{minipage}
\mainmatter
\pagenumbering{arabic}
\setcounter{chapter}{5}
\setcounter{page}{1}
\vspace*{1.3cm}
\sec{Introduction}{Introduction}

Ultrasound has been used extensively to study the collective excitations in superfluid $^3$He.
Early ultrasonic studies of order parameter collective modes provided important confirmation of
the identifications of the {\it A}- and {\it B}-phases based on NMR. More recent experimental
studies have discovered a remarkable spectrum of order parameter collective modes that clearly
reveal the underlying broken symmetries of the superfluid phases. Theoretical studies have
further elucidated the nature of the collective modes in terms of the condensate of Cooper
pairs and their couplings to external fields, quantified the energy spectrum and underlying
molecular fields of the Fermi liquid, and predicted a number of new physical effects, many of
which have striking similarities to the interaction of electromagnetic fields with the atoms
and molecules. The article by Halperin and Varoquaux \cite{hal90} in this volume gives a
comprehensive review of ultrasonic experiments in superfluid $^3$He as well as their
theoretical interpretation in terms of collective modes of the order parameter. Most studies of
collective modes in superfluid $^3$He have delt with the linear response of the condensate. In
this article we present recent theoretical work on the role of order parameter collective modes
in the nonlinear response of $^3$He-B to acoustic waves. Many of the nonlinear acoustic
phenomena we describe have analogs in nonlinear optical excitation of atoms and molecules.

The subject of order parameter collective modes originated with questions that were raised
about the gauge invariance of the original BCS theory \cite{and58a}. Anderson \cite{and58}
provided a gauge-invariant formulation of the pairing theory and elucidated the role of
collective modes in superconductors (see also \cite{ric59a,nam60,amb61}). These modes may be
broadly classified into (i) {\it Goldstone modes}, associated with a spontaneously broken
symmetry, and (ii) {\it exciton modes}, excitations of the condensate of Cooper pairs which
involve time-dependent deformation of the superconducting order parameter \cite{mar69}.
Goldstone modes reflect the degeneracy of the order parameter under time-independent and
spatially uniform gauge transformations and rotations, and are the low-energy excitations of
the condensate. Broken gauge symmetry in superconductors and superfluid $^3$He leads to a phase
mode, oscillations in the overall phase of the order parameter. This mode is closely related to
the collisionless sound mode in superfluid $^3$He and exhibits a phonon-like dispersion
relation, $\omega=c_1 q$, where the velocity $c_1$, in the limit $T\rightarrow 0$ and
$\omega\ll\Delta$, is identical to the hydrodynamic sound velocity of the normal phase of
liquid $^3$He. The phase mode was predicted by Anderson \cite{and58} and Bogoliubov, Tolmachev
and Shirkov \cite{bogoliubov58}, and is essential to spectroscopic studies of the additional
order parameter collective modes in $^3$He.\footnote{In most superconductors the phase mode is
dynamically uninteresting for studying the other possible excitations of the superconducting
state because it couples directly to long-wavelength charge fluctuations, and consequently
oscillates at the plasma frequency, $\omega_{pl}\gg\Delta$. In dirty superconductors the phase
mode is observable in a narrow temperature region near $T_c$ as a weakly damped collisionless
second-sound mode \cite{car75,sch75}.} Several authors also predicted the existence of exciton
modes in superconductors with excitation energies, $\omega < 2\Delta$, corresponding to excited
bound states of Cooper pairs \cite{and58,bogoliubov58,tsu60,vak61,bar61}. Most notable from the
viewpoint of collective modes in superfluid $^3$He is the work of Vdovin \cite{vdo63} who
predicted the order parameter collective modes for the Balian-Werthamer state, and calculated
the dispersion relations, long before the discovery of superfluid $^3$He. Although exciton
modes have never been definitively observed in superconductors to our knowledge, the analogous
modes in superfluid $^3$He have been studied extensively. The main reason that exciton modes
are readily observed in superfluid $^3$He, while not in most superconductors, is that $^3$He
is an {\it unconventional} superfluid with an order parameter that breaks the rotational
invariance in both spin- and orbital space; and therefore, belongs to a higher dimensional
representation of the full symmetry group of the normal phase. This implies that there is a
spectrum of pairing states - the ground pair state and excited pair states - belonging to the
same representation, and therefore bound by the same pairing interaction. It is well
established that the superfluid phases of $^3$He belong to the $S = 1,\ l = 1$ (spin-triplet,
p-wave) representation of the rotation group $SO(3)\times SO(3)$, giving rise to an order
parameter with nine complex amplitudes and a corresponding spectrum of eighteen collective
modes, many of which lie well below the pair-breaking edge of $2\Delta$.

Quite generally the order parameter can be identified as the pair amplitude,
\be
F_{\alpha\beta}(\vec{r},\vec{x}) 
=
\langle\psi_{\alpha}(\vec{x}+\vec{r}/2)\psi_{\beta}(\vec{x}-\vec{r}/2)\rangle 
\;.
\ee 
Since pairing occurs in a narrow band in momentum space near the Fermi momentum, $p_f$, it is
convenient to use a {\it mixed} Fourier-space representation,

\be
F_{\alpha\beta}(\vec{p},\vec{x}) 
= 
\int\limits d^{3}r\,e^{-i\vec{p}\cdot\vec{r}}\,F_{\alpha\beta}(\vec{r},\vec{x}) 
\;,
\ee 
with $\vert\vec{p}\vert\simeq p_{f}$. Thus, $F_{\alpha\beta}(\vec{p},\vec{x})\simeq
F_{\alpha\beta}(p_{f}\vhat{p},\vec{x})$ is a function of the center-of-mass coordinate,
$\vec{x}$, of the Cooper pair for non-uniform states of the superfluid, and the direction
$\vhat{p}$ describes the orbital motion of the pair. In addition, the order parameter depends on
the spin configuration $(\alpha,\beta)$ of the Cooper pairs, and can be decomposed into
spin-singlet $(S=0)$ and spin-triplet $(S = 1)$ amplitudes,
\be
F_{\alpha\beta}(\vhat{p}) = f_{0}(\vhat{p})(i\sigma_{y})_{\alpha\beta}
                         + \vec{f}(\vhat{p})\cdot(i\vec{\sigma}\sigma_{y})_{\alpha\beta}
\;,
\ee 
where $\vec{f}(\vhat{p})$ represents the three odd-parity, spin-triplet amplitudes.\footnote{The
parities of $f_{0}(\vhat{p})$ and $\vec{f}(\vhat{p})$ are fixed by the exchange anti-symmetry of
$F_{\alpha\beta}(\vec{x},\vhat{p})$ and the odd [even] exchange symmetry of the spin matrices,
$(i\sigma_{y})_{\alpha\beta}\;[(i\vec{\sigma}\sigma_{y})_{\alpha\beta}]$.} In equilibrium the
superfluid phases of $^3$He are described by an order parameter in which only the spin-triplet
part is non-zero.\footnote{The identification of the order parameters for the A- and B-phases
from NMR, ultrasound and thermodynamic data is discussed by Leggett \cite{leg75}, which also
provides a more general review of the theory of superfluid $^3$He.} The rotational invariance
of the normal phase of $^3$He suggests that the condensate of Cooper pairs is also defined by a
single orbital angular momentum channel, known to be the p-wave channel.\footnote{This is
strictly true only at the superfluid transition; however, for $^3$He-B the equilibrium order
parameter in zero field is pure p-wave at all temperatures.} An alternative order parameter to
the pair field, $\vec{f}(\vec{x},\vhat{p})$, is the weak-coupling gap function,
\be
\vec{\Delta}(\vec{x},\vhat{p})
=
\int\frac{d\Omega^{\prime}}{4\pi}\;V^{t}(\vhat{p}\cdot\vhat{p}')\;
\vec{f}(\vec{x},\vhat{p}^{\prime})\,, 
\ee 
where $V^t$ is the pairing interaction in the spin-triplet channel; for equilibrium states of
$^3$He these two order parameters contain essentially the same information. A trivial extension
of the above to include time-dependent pair amplitudes describes non-equilibrium states of the
condensate described by a fluctuation of the order parameter,
\be
\vec{d}(\vhat{p};\vec{x},t) 
=
\int\frac{d\Omega'}{4\pi}\,V^{t}(\vhat{p}\cdot\vhat{p}')\, 
\left[\vec{f}(\vhat{p}';\vec{x},t)-\vec{f}_{eq}(\vhat{p}')\right] 
\;,
\ee 
where $\vec{f}(\vhat{p};\vec{x},t)$ is the time-dependent pair amplitude defined in Sec. 
(\ref{Review}).

The equilibrium B-phase is decribed by the an order parameter of the Balian-Werthamer (BW)
class \cite{bal63},
\be\label{BW}
\Delta_{i}(\vhat{p}) = \Delta(T)\,R_{i\alpha}[\vhat{n},\theta]\;e^{i\Phi}\;(\vhat{p})_{\alpha}
\;,
\ee
where $R_{i\alpha}[\vhat{n},\theta]$ is an orthogonal matrix defining a relative rotation of
the spin and orbital angular momentum quantization axes, $\Phi$ is the global phase, and
$\Delta(T)$ is the magnitude of the gap. If we neglect the weak dipolar interaction then the
free energy of $^3$He is separately invariant under spin rotations and orbital rotations. Thus,
the BW states described by different relative rotations of the spin- and orbital coordinates
are degenerate, as are BW states differing by a gauge transformation. The parameters
$\left[\vhat{n},\theta,\Phi\right]$ that define the equilibrium B-phase order parameter are
{\it soft} degrees of freedom; long-wavelength variations of these fields cost little energy on
the scale of $T_c$. The nuclear dipolar interaction, which is small on the scale of
$T_{c}\,(E_{dipolar}\sim 10^{-7}K\ll T_{c}\sim 10^{-3}K)$, partially lifts this degeneracy by
fixing the rotation angle $\theta=\cos^{-1}(-1/4)$ \cite{leg73a}; but the direction of the axis
of rotation $\vhat{n}$ remains a soft degree of freedom. Many of the remarkable properties of
superfluid $^3$He are related to these soft degrees freedom and are described in detail
elsewhere in this volume.

The B-phase, described by one of the the BW states, is ``pseudo-isotropic'' in the sense that
it is invariant under the transformations generated by the total ``twisted'' angular momentum
$\vec{J}=\vec{L}+\mathop{\ul{R}}(\vhat{n},\theta)^{-1}\cdot\vec{S}$, and so has total ``twisted
angular momentum'' $J=0$. For pure $l=1$ pairing the order parameter collective modes
correspond to the natural oscillations of the $3\times 3$ complex matrix $D_{ij}(\vec{x},t)$
defined in terms of the time-dependent order parameter for pairing fluctuations,
\be\label{ac}
d_{i}(\vhat{p};\vec{x},t) = D_{ij}(\vec{x},t)\vhat{p}_{j}
\,.
\ee 
The eighteen order parameter modes are classified \cite{mak76} by the quantum numbers
$J^{\zeta}$ and $M$, where $J=\lbrace 0,1,2\rbrace$ is the total angular momentum,
$M=\{-J,\ldots,0,\ldots +J\}$ is the magnetic quantum number, and $\zeta=\{+,-\}$ 
is the signature
of the pair amplitude under the particle-hole transformation. The $J=2^{+}$ and $J=2^{-}$ modes
are of particular interest because they have excitation energies which lie well below the
pair-breaking edge of $2\Delta(T)$ [$\omega_{2^{+}}\simeq 1.1\;\Delta(T)$ and
$\omega_{2^{-}}\simeq 1.5\;\Delta(T)$], couple to sound and are only weakly damped by
quasiparticles collisions. The linear interaction of zero sound with the order parameter
collective modes of $^3$He for frequencies in the range of $10-100$ MHz has a number of
similarities with the interaction of light with atoms, molecules and solids, including:
\begin{itemize}
\item[-] a spectroscopic classification of modes (excited states)
      in terms of quantum numbers for rotational and discrete symmetries,
\item[-] sharp resonance features in the phase velocity, group velocity and
      attenuation of sound (light) when its frequency and wave-vector matches that of
      the collective modes (excited states),
\item[-] a linear Zeeman splitting of the collective modes (excited
      states) in magnetic field, and
\item[-] a nonlinear splitting of modes (excited
      states) in large magnetic fields due in part to level repulsion.\footnote{For
      a detailed discussion of the collective mode
      spectrum in $^3$He-B see the review by Halperin
      and Varoquaux in this volume.}
\end{itemize}
Partly because of the similarities between the ultrasonic absorption spectrum of superfluid
$^3$He and optical systems, as well as the sharpness of the collective mode resonances, $^3$He
is an ideal media in which to look for acoustic analogues of nonlinear optical effects.

Nonlinear optical effects can be broadly divided into two classes. One class of effects occur
because intense radiation induces a macroscopic population of one or more of the excited states
of the medium. Consequently, the occupation of the ground state of the system must be included
as a dynamical variable in the constitutive equations. Examples include population inversion,
saturation effects and self-induced transparency \cite{allen75}. A second class of nonlinear
effects are parametric processes such as harmonic generation, stimulated Raman scattering and
two-photon absorption \cite{blo65,yariv75}. Usually the population of the ground state need not
be treated as a dynamical variable and can be assumed to have its equilibrium value. Except for
a theoretical paper by Serene \cite{ser84} on third-harmonic generation, previous
investigations, both experimental and theoretical, of nonlinear sound in superfluid $^3$He have
all involved effects belonging to the first class above. We briefly review these before turning
to the main subject of this article, parametric processes.

Polturak {\it et al.} \cite{pol81a} observed saturation of the attenuation of zero sound in
$^3$He-B for frequencies nearly resonant with the $J=2^{+}$ mode, as well as propagation
delay, pulse sharpening and break-up of narrow sound pulses. They pointed out that these
observations were qualitatively similar to self-induced transparency. Sauls \cite{sau81d}
constructed a phenomenological set of nonlinear equations, which are consistent with the
symmetries of superfluid $^3$He-B, reduce to the known linear equations, and are analogous to
the Maxwell-Bloch equations for the optical self-induced transparency. He found that the
relationship between the pulse width and the pulse velocity for the soliton solutions of these
equations was in good agreement with the results of Polturak {\it et al.} However, Rouff and
Varoquaux \cite{rou83} questioned the interpretation of the experiments of Polturak, {\it et
al.} in terms of self-induced transparency and pointed out that the energy density required for
the formation of solitons in Sauls' phenomenological theory, which is of the order of the
superfluid condensation energy density, is two orders of magnitude larger than the estimated
energy density in the sound pulses of Polturak {\it et al.} The discrepancy with the energy
density of the phenomenological theory, combined with the observation of soliton-like
propagation, underscores the need for a more fundamental theory of nonlinear sound propagation
in superfluid $^3$He as well as further experimental efforts to study the nonlinear dynamics of
the order parameter.\footnote{We note that Namaizawa \cite{nam87} recently obtained nonlinear
equations which are similar to the phenomenological equations of \cite{sau81d}, but with
significantly different coupling parameters.}

Nonlinear effects in the A-phase have also been observed. The attenuation of low-frequency zero
sound in the $^3$He-A is dominated by the pair-breaking process near the nodes of the
anisotropic energy gap. Avenel {\it et al.} \cite{ave85} observed a decrease in the attenuation
as the sound intensity was increased and showed that the reduction was due to a saturation
effect from the creation of a non-equilibrium distribution of quasiparticles over a small
region of the Fermi surface. Kopp and W\"olfle \cite{kop87a} have recently derived dynamical
equations for the quasiparticle distribution function, similar to the Maxwell-Bloch equations
of nonlinear optics, which they use to describe the observations reported by Avenel {\it et al}
\cite{ave85}.

Parametric processes, involving the absorption and emission of excitations with differing
frequencies, are common in nonlinear systems and occur in diverse fields such as optics
\cite{shen84}, plasma physics \cite{kau79}, electronics \cite{yar69}, and acoustics
\cite{bun86}. The simplest parametric process is a three-wave resonance in which nonlinear
interactions allow two modes with frequencies $\omega_{1}$ and $\omega_{2}$ and wavevectors
$\vec{q}_{1}$ and $\vec{q}_{2}$ to excite a third mode with frequency $\omega_{3}$ and
wavevector $\vec{q}_{3}$ given by
\vspace*{-2mm}
\be\label{ad}
\omega_{3}  = \omega_{1}  + \omega_{2}
\,,
\qquad
\vec{q}_{3} = \vec{q}_{1} + \vec{q}_{2}
\,.
\vspace*{-2mm}
\ee
Often the first wave, known as the {\it pump wave}, is of high intensity. The other waves are
known as the {\it idler} and {\it signal waves}. The rate at which the process occurs depends
on the intensity of the pump wave. The reverse process, the excitation of modes $1$ and $2$ by
mode $3$, is also allowed. Such a parametric process is interpreted as the decay of a quanta of
mode 3 into quanta of modes $1$ and $2$, and Eq. (\ref{ad}) expresses the conservation of
energy and momentum. In Sec. (\ref{Stimulated}) we discuss two particular parametric processes:
\begin{enumerate}
\item[(1)] the production of a real squashon\footnote{We shall often refer to a quantum of the 
$J=2^{+}$, or real squashing mode, as a {\it real squashon}.} by two zero-sound phonons
({\it two phonon absorption}),
\item[(2)] the decay of a zero-sound phonon into a real squashon and a second zero-sound phonon
({\it stimulated Raman scattering}).
\end{enumerate}
Other processes such as {\it third harmonic generation} are discussed as extensions of the
above two processes in Sec. (\ref{Generation}). 

Two important questions need to be answered by
a theory of parametric processes in superfluid $^3$He:
\begin{enumerate}
\item[(a)] When is a particular parametric process forbidden by the symmetry of the ground state 
of $^3$He, {\it i.e.} what are the selection rules?
\item[(b)] What energy density of the pump wave is required for experimental detection of
parametric excitation of the modes?
\end{enumerate}
Liquid $^3$He possesses a number of symmetries and approximate symmetries, which determine the
selection rules for parametric processes that occur. An example is the decay process for zero
sound in normal $^3$He proposed by Ketterson \cite{ket83} in which a zero-sound phonon decays
into a spin wave by absorbing a $q=0$ magnon from a macroscopic population of such magnons that
have been prepared by tipping the magnetization relative to the static field with an {\it
rf}-pulse. However, such a decay process is not allowed \cite{sau84c}, without dipolar
interactions, because of the nearly exact invariance of the density matrix under separate
rotations in spin and position space. Similarly, Serene \cite{ser83a} has shown that the
approximate particle-hole symmetry of the $^3$He Fermi liquid determines important selection
rules for the linear coupling of zero sound to the order parameter collective modes in
superfluid $^3$He. This symmetry also determines important selection rules for parametric
processes involving zero sound as discussed in Secs. (\ref{Conservation}) and
(\ref{Nonlinear}).
In fact because of particle-hole symmetry processes (1) and (2) above are allowed only for real
squashons ($J=2^+$), not imaginary squashons ($J=2^-$). In Sec. (\ref{Stimulated}) we discuss
the coupling strengths of real squashons to sound via parametric processes and argue that the
three-wave processes should be observable when the pump-wave energy density is of the order of
the superfluid condensation energy density. Finally, in Sec. (\ref{Generation}) we show that
although the generation of third harmonics and anti-stokes waves are limited by dispersion,
they may still be observable, at least for sufficiently short sound path lengths.

Section (\ref{Review}) is a review of the quasiclassical theory of superfluid $^3$He which
provides the basis for our treatment of sound propagation and collective modes in the B-phase
starting with the linear response theory in Sec. (\ref{Linear}). These results are important
for the development of parametric nonlinear effects that follow. We begin with a discussion of
the conservation laws and general properties of the constitutive equations for collisionless
sound propagation in $^3$He-B.

\vspace*{-5mm}
\sec{Conservation laws and constitutive equations}{Conservation}

The density fluctuation $\delta n(\vec{x},t)$ and current density $\vec{J}(\vec{x},t)$ satisfy
the continuity equation for particle conservation
\be\label{ae} 
\pder{\delta n}{t} + \pder{J_{k}}{x_{k}} = 0 
\,,
\ee 
and the equation for momentum conservation
\be\label{af}
m\,\pder{J_{k}}{t} + \pder{}{x_{l}}\Pi_{kl} = 0
\,,
\ee 
where $\Pi_{kl}$ is the energy-momentum stress tensor for the fluid. The Fourier transforms of
(\ref{ae}) and (\ref{af}) combine to give,
\be\label{ag}
\omega^{2}\delta n(\vec{q},\omega) =\frac{1}{m}\,\vec{q}_{k}\vec{q}_{l}\Pi_{kl}(\vec{q},\omega) 
\,.
\ee 
It is straight-forward to show that in the collisionless limit the trace of the energy-momentum
tensor is related to the density fluctuation by
\be\label{ah}
\Pi_{kk}=3\,m\,c_{1}^{2}\,\delta n
\,,
\ee 
where $c_{1}$ is the velocity of hydrodynamic first sound, 
\be\label{ai}
c_{1}^{2}=\frac{1}{3}\,v_f^2\,(1 + F_{0}^{s})\,(1+F_{1}^{s}/3)
\,,
\ee 
$v_{f}$ is the Fermi velocity, and $(F_{0}^{s}\,,\,F_{1}^{s})$ are Landau's Fermi-liquid
parameters (see {\it e.g.} \cite{bay78}). The wave equation (\ref{ag}) can then be written
\be\label{aj}
(\omega^{2} - c_{1}^{2}\,q^{2})\,\delta n = 2\,c_{1}^{2}\,q^{2}\;\delta\Pi
\,,
\ee 
where $\delta\Pi$ is proportional to the traceless part of the energy-momentum stress tensor
\be\label{ak}
\delta\Pi = \frac{1}{2mc_{1}^{2}}
\,
(\vhat{q}_{k}\vhat{q}_{l} - \frac{1}{3}\,\delta_{kl})\,\Pi_{kl} 
\;.
\ee 
It is important to note that although the wave equation is linear in $\delta n$ and $\delta\Pi$
it still describes nonlinear sound propagation because the longitudinal stress $\delta\Pi$ is a
nonlinear functional of the density fluctuation and, in general, the amplitudes of the other
collective modes of the system which couple to zero sound.\footnote{Note that for low
amplitude, low frequency sound and $T\rightarrow 0$, there are no quasiparticle excitations,
nor collective modes, in which case $\delta\Pi\rightarrow 0$, and we recover the
Anderson-Bogoliubov result for the density wave.}
The relationship between the fluctuating stress $\delta\Pi$ and the density fluctuation
$\delta n$ must be obtained from a more microscopic theory than hydrodynamics. Under certain
conditions (see below) the constitutive relation is schematically of the form
\be\label{al}
\delta\Pi
\simeq \chi^{(1)}\,\delta n 
+ \chi^{(2)}\,(\delta n)^{2} 
+ \chi^{(3)}\,(\delta n)^{3} + \ldots 
\,.
\ee 
In the linear response limit the frequency dependent phase velocity $v(\omega)$ and attenuation
of sound $\alpha(\omega)$ are related to the real and imaginary parts of $\chi^{(1)}$,
respectively; while the higher-order susceptibilities govern the nonlinear acoustic response.
The detailed form of the constitutive equation connecting $\delta\Pi$ and $\delta n$ reflects
the symmetry of the ground state and the properties of liquid $^3$He. A similar situation
occurs in nonlinear optics where the electromagnetic field satisfies
$\left(\partial_{t}^{2} - c^2\nabla^2\right)\,\vec{E} = -4\pi\,\partial_{t}^{2}\vec{P}$
where $\vec{P}$ is the electric polarization. The constitutive relation connecting $\vec{P}$
and $\vec{E}$ can be nonlinear and reflects the symmetry and material properties of the
optically active medium (see {\it e.g.} \cite{shen84}).

Serene \cite{ser83a} has discussed the constraints which the symmetries of $^3$He place on the
linear response functions for zero sound and the couplings to different order parameter
collective modes. Liquid $^3$He is a system with a high degree of symmetry. The normal phase is
rotationally invariant in both spin and orbital spaces, translationally invariant,
time-reversal invariant and reflection symmetric. In addition, the normal Fermi liquid
possesses an approximate discrete symmetry called particle-hole symmetry. The density of
quasiparticle states $N(\xi_{\vec{p}})$ varies on the scale of the Fermi energy, $E_{f}$.
Consequently, for the low-energy properties of the liquid it is generally an excellent
approximation to take the density of states to be constant. This approximation is closely
related to the symmetry of the low-energy effective Hamiltonian under interchange of
quasiparticles and quasiholes. Particle-hole symmetry is, however, only an approximate symmetry
since the density of states is not precisely an even function of the excitation energy,
$\xi_{\vec p}$. In fact, several striking manifestations of the small particle-hole {\it
asymmetry} of liquid $^3$He have been observed in the superfluid phases, all of which occur
when particle-hole asymmetry is connected with another broken symmetry\footnote{The $A_{1}$
phase of superfluid $^3$He is stable in a small temperature range proportional to the magnetic
field only because of particle-hole asymmetry. Similarly, the tiny gyromagnetic effect in the
NMR of rotating $^3$He-B, interpreted in terms of a ferromagnetic moment of the vortex lines,
is also due to particle-hole asymmetry.} or the resonant response of a collective mode.
Particle-hole symmetry leads to a selection rule forbidding the excitation of the $J=2^{+}$
mode by a weak ({\it i.e.} in the linear response limit) zero-sound field. The experimental
observation of resonance peaks in the ultrasonic absorption due to the $J=2^{+}$ modes in
spatially uniform $^3$He-B \cite{gia80,mas80,ave80} are a direct consequence of particle-hole
asymmetry \cite{koc81}. We discuss this case in more detail below.

A quasiparticle with energy $\xi_{\vec{p}}=p^{2}/2m^{*}-E_{f}>0$ above the Fermi surface is
mapped into a quasiparticle just below the Fermi surface with energy $-\xi_{\vec{p}}$ and
spin rotated by $\pi$ by a unitary transformation,
\be\label{an}
a_{\vec{p}\alpha} \rightarrow \cC\,a_{\vec{p}\alpha}\cC^{\dagger}
= 
(i\sigma_2)_{\alpha\beta}\;a_{\ul{\vec{p}}\beta}^{\dagger}
\,,
\ee 
where $\ul{\vec{p}}$ is the vector with $\ul{\vhat{p}}=\vhat{p}$ and
$\xi_{\ul{\vec{p}}}=-\xi_{\vec{p}}$.\footnote{An explicit construction of the operator $\cC$ is
given in Ref. \cite{fis85}.} 
This transformation leaves the low-energy effective Hamiltonian invariant to leading order in
$\varpi_{c}/E_{F}$, where $\varpi_{c}$ is the quasiparticle ``bandwidth'' \cite{ser83a}.
Assuming exact particle-hole symmetry of the normal Fermi liquid, the order parameter and
density operators transform according to
\be
\hspace*{-3mm}
\cC\vec{d}_{op}(\vhat{p};\vec{x},t)\cC^{\dagger} 
=
+\vec{d}_{op}(\vhat{p};\vec{x},t)^{\dagger}
\,,\,\,
\cC\delta n_{op}(\vec{x},t)\cC^{\dagger}
=
-\delta n_{op}(\vec{x},t) 
\,,
\ee
where
\ber
\delta n_{op}(\vec{x},t) 
&=&
\psi^{\dagger}_{\alpha}(\vec{x},t)\psi_{\alpha}(\vec{x},t)
-
\langle\psi^{\dagger}_{\alpha}\psi_{\alpha}\rangle_{eq}
\\
&=&
\int\frac{d^{3}q}{(2\pi)^{3}}\,e^{i\vec{q}\cdot\vec{x}}\,N(0)
\int^{\varpi_{c}}_{-\varpi_{c}}d\xi_{\vec{p}} 
\int\frac{d\Omega_{p}}{4\pi}
	\,a^{\dagger}_{\vec{p}+\vec{q},\alpha}(t)\,a_{\vec{p},\alpha}(t) 
\nonumber\\
\nonumber\\
\vec{d}_{op}(\vhat{p};\vec{x},t) 
&=& 
\int\frac{d\Omega_{p}}{4\pi}\,V^{t}(\vhat{p}\cdot\vhat{p}') 
\int^{\varpi_{c}}_{-\varpi_{c}}\,d\xi_{\vec{p}'}
\int\,d^{3}r\,e^{-i\vec{p}'\cdot\vec{r}}
\\
&\times&
\frac{-i}{2}
\left[\sigma_{2}\vec{\sigma}\right]_{\beta\alpha}
\, 
\psi_{\alpha}(\vec{x}+\vec{r}/2,t) 
\psi_{\beta}(\vec{x}-\vec{r}/2,t) 
\nonumber\\
&=&
\int\frac{d^{3}q}{(2\pi)^{3}}\,e^{i\vec{q}\cdot\vec{x}}\,
\int\frac{d\Omega_{p}}{4\pi}\,V^{t}(\vhat{p}\cdot\vhat{p}') 
\int^{\varpi_{c}}_{-\varpi_{c}}\,d\xi_{\vec{p}'}
\nonumber\\
&\times&
\frac{-i}{2}
\left[\sigma_{2}\vec{\sigma}\right]_{\beta\alpha}
\,a_{\vec{p}'+\vec{q}/2,\alpha}(t)
\,a_{\vec{p}'-\vec{q}/2,\beta}(t) 
\;.
\nonumber
\eer
It then follows that the operators corresponding to the real and imaginary parts of the order
parameter fluctuation have signatures $+$ and $-$, respectively, under $\cC$, {\it i.e.}
\be\label{aq}
\cC\vec{d}^{\pm}_{op}(\vhat{p};\vec{x},t)\cC^{\dagger}
=
\pm\vec{d}^{\pm}_{op}(\vhat{p};\vec{x},t) 
\;,
\ee 
where 
$\vec{d}^{\pm}_{op}=\frac{1}{2}\left(\vec{d}_{op}\pm\vec{d}_{op}^{\dagger}\right)$.

The consequences of these symmetry relations and exact particle-hole symmetry of the equilibrium
density matrix, $e^{-\beta\cH}$, for the linear response of the density and order parameter are
easily deduced \cite{ser83a}. Consider the linear dispersion relations between $\vec{d}^{\pm}$
and the density fluctuation $\delta n$
\be\label{ar}
K_{ij}^{\pm}(\vec{q},\omega,\vec{\Delta})\;d_{j}^{\pm}(\vec{q},\omega) 
=
L_{ij}^{\pm}(\vec{q},\omega,\vec{\Delta})\;\delta n(\vec{q},\omega) 
\;.
\ee 
For a real equilibrium order parameter, like the homogenous BW state,\footnote{Any overall phase
factor is removed by a uniform gauge transformation.}
$\vec{\Delta}\xrightarrow[]{\cC}\vec{\Delta}$; thus, if (\ref{ar}) is to be invariant under
$\cC$, then $L_{ij}^{+}$ must vanish. Therefore, if particle-hole symmetry were an exact
symmetry the density would not couple to oscillations of the real part of the order parameter,
$\vec{d}^{-}$. However, the coupling of sound to the imaginary part of the order parameter,
$\vec{d}^{-}$, is not forbidden by particle-hole symmetry. Excitation of the $J=2^-$ mode can be
interpreted as a phonon of energy $\hbar\omega$, momentum $\hbar\vec{q}$, and particle-hole
symmetry $\zeta=-$, exciting the superfluid from the ground state, with $\zeta =+$, to an
excited state (e.g. an imaginary squashon) with energy $\hbar\omega$, momentum $\hbar\vec{q}$,
and $\zeta =-$. However, particle-hole symmetry is weakly broken in $^3$He. Consequently, there
is a weak coupling between the modes with $\zeta = +$ and zero sound. In the linear response
limit the dynamical equations for the $J=2^{\pm}$ modes are of the form
\be\label{as}
\hspace*{-3mm}
\lambda(\omega)\negthickspace
\left[\omega^{2}+i\omega\Gamma_{\pm}-\omega_{M\pm}^{2}-c^{2}_{M\pm}\,q^{2}\right]\negthickspace
D^{\pm}_{M}(\vec{q},\omega)
\negthickspace=\negthickspace
\frac{6\Delta^{2}}{1+F^{s}_{0}}
\beta_{M}^{\pm}\,\delta n(\vec{q},\omega) 
\;,
\ee
where $D^{\pm}_{M}(\vec{q},\omega)$ is the amplitude of the mode with magnetic quantum number
$M,\zeta=\pm$, $1/\Gamma_{\pm}$ is the lifetime of the corresponding mode due to quasiparticle
collisions, and $\lambda(\omega)$ is the frequency- and temperature-dependent response function
given in (\ref{ay}). The coupling ``constant'' $\beta^{-}_{M}$ for the $\zeta = -1$ modes is the
of order one, while $\beta^{+}_{M}$ is small, of order $N'(E_{f})$ $\Delta/N(E_{f})$ $\sim$
$\Delta/E_{f}\ll 1$, where $N(E_{f})$ and $N'(E_{f})$ are the density of states and its slope at
the Fermi surface \cite{koc81}. The fluctuations in the stress tensor and the order parameter
are related by
\be\label{at}
\delta\Pi(\vec{q},\omega)
=
\frac{N(E_f)}{1+F_0^s}\,
\sum_{M=-2}^{+2}\left[\beta_{M}^{+}\; D^{+}_{M}(\vec{q},\omega)
+
                      \beta_{M}^{-}\; D^{-}_{M}(\vec{q},\omega)\right]
\;.
\ee 
The precise definitions of the collective mode amplitudes, frequencies and coupling functions
are given in Sec. (\ref{Linear}). Equations (\ref{as}) and (\ref{at}) describe the interaction
of zero sound with the $J=2^{+}$ and $J=2^{-}$ modes in the linear response limit. In
particular, these constitutive equations describe the resonant absorption and anomalous
dispersion of zero sound resulting from the excitation of the $J=2^{\pm}$ modes. In addition, a
nonlinear term which is quadratic in $\delta n$ is allowed by particle-hole symmetry for the
real modes, but not for the imaginary modes. At higher sound amplitudes a driving term on the
right side of (\ref{as}), which is second order in the density, becomes important. Similarly,
the stress tensor has a term which is bilinear in $\delta n$ and the $J = 2^+$ amplitude
$D^{+}_{M}$. In Sec. (\ref{Nonlinear}) these nonlinear couplings are shown to have the form
\be\label{au}
\delta\Pi(\omega) 
= 
\frac{1}{(1+F^{s}_{0})\Delta^2}\sum_{M}\int\,d\nu\;A_{M}(\omega,\nu,\omega -\nu)\, 
\delta n(\nu)\;D^{+}_{M}(\omega-\nu) 
\;,
\ee 
\ber\label{av}
\lambda(\omega)\negthickspace
\left[\omega^{2}+2i\omega\Gamma_{+}-\omega_{M+}^{2}-c_{M+}^{2}\,q^{2}\right]\negthickspace
D^{+}_{M}(\vec{q},\omega) 
\qquad\qquad\qquad\qquad\quad
\nonumber
&&
\\
=
\frac{6}{N(E_{f})^{2}}
\int\,d\nu\,\vert\frac{c_{1}(\vec{q}-\vec{s})}{\omega-\nu}\vert^{2}\, 
A_{M}(\nu-\omega,\nu,-\omega)^{*}\,\delta n(\nu)\;\delta n(\omega-\nu) 
\,,
&&
\eer 
where $A_{M}$ is a dimensionless function of temperature of order one except near $T_c$, and the
wave-vector dependence of the right-hand side of these equations has been suppressed for
clarity. Note that the same function $A_{M}$ appears in both equations. These equations together
with Eq. (\ref{aj}) describe the interaction of the $J=2^{+}$ mode with two zero-sound modes and
are central equations of this paper; their consequences are discussed in Sec. (\ref{Stimulated})
and (\ref{Generation}).
\vspace*{3mm}
\sec{Review of the quasiclassical theory}{Review}

The starting point for our derivation of the nonlinear constitutive equations in (\ref{au}) and
(\ref{av}) is the quasiclassical theory of superfluid $^3$He; we follow the notation in the
review article by Serene and Rainer \cite{ser83} as much as possible. However, for technical
reasons we do not start directly from Eilenberger's transport equations, but rather from an
expansion of the low-energy Dyson equation. Thus, it is useful to briefly review the
quasiclassical equations on superfluid $^3$He and their relation to the Dyson equation.

The dynamical theory of superfluid $^3$He has as its main components: (i) a transport equation
describing the evolution of the quasiparticle distribution function, (ii) a time-dependent gap
equation for the order parameter, and (iii) inputs describing the initial state of the
quasiparticles and the condensate. The quasiclassical theory is derived from the full many-body
perturbation theory and Landau's observation that the low-lying excitations in liquid $^3$He are
quasiparticles obeying Fermi statistics. This latter hypothesis corresponds to assuming that the
normal-state self-energy is a slowly varying function of momentum near the Fermi surface, while
the Green's function is dominated by the quasiparticle pole at $|\vec p | = p_f$.

The dynamical equations governing superfluid $^3$He are formulated in terms of $4\times 4 $
matrix Green's functions which are defined in terms of products of Nambu spinors that combine
the spin and particle-hole degrees of freedom,
\be\label{bh}
\Psi =
\left(\psi_{\uparrow}\;\psi_{\downarrow}\;
      \psi^{\dag}_{\uparrow}\;\psi^{\dag}_{\downarrow}\right) 
\;.
\ee 
For example, the {\it retarded} Green's function is 
\be\label{bj}
\hat{G}^{R}(\vec{x},t;\vec{x}',t')_{ab}
=
-i\theta(t-t')\left\langle\{\Psi_{a}(\vec{x},t)\,,\,\Psi_{b}^{\dag}(\vec{x}',t')\}\right\rangle 
\,;\quad a,b=\left\{1,2,3,4\right\} 
\;,
\ee 
where $\{A\,,\,B\} = AB + BA$ and $\langle ...\rangle $ denotes the average over the statistical
ensemble. The structure of the matrix Green's function in particle-hole space is
\be\label{bk}
\hat {G}=\begin{pmatrix}G	&	F	\cr	\bar{F}	&	\bar{G}	\end{pmatrix} 
\;,
\ee 
where $G_{\alpha\beta}\sim \left\langle\{\psi_{\alpha}\,,\,\psi_{\beta}^{\dag}\}\right\rangle$
is the conventional one-particle retarded Green's function and $F_{\alpha\beta}$ is the
corresponding {\it anomalous} Green's function, and is related to the Cooper pair amplitude
$\left\langle\psi_{\alpha}\psi_{\beta}\right\rangle$. The barred quantities obey the symmetry
relations (see Ref. \cite{ser83} for a summary of these and other symmetry relations),
\ber
\bar {G}^{R}_{\alpha\beta}\left( {\vec {x},t;\vec {x}',t'}\right) =
     -G^{A}_{\beta\alpha}\left( {\vec {x}',t';\vec {x},t}\right) 
\; ,
\nonumber\\
\bar {F}^{R}_{\alpha\beta}\left( {\vec {x},t;\vec {x}',t'}\right) =
     -F^{A}_{\beta\alpha}\left( {\vec {x}',t';\vec {x},t}\right)^{*}
\; .
\eer 
It is convenient to introduce the mixed Fourier-space representation for any of the Green's
functions,
\be\label{bl}
\hat{G}(\vec{p},\vec{x};\epsilon,t)\negthickspace
\negthickspace = \negthickspace
\int\negthickspace d^{3}r\negthickspace\negthickspace
\int^{+\infty}_{-\infty}
\negthickspace\negthickspace\negthickspace\negthickspace 
d\tau e^{i(\epsilon\tau-\vec{p}\cdot\vec{r})}
\;
\hat{G}(\vec{x}+\onehalf\vec{r},t+\onehalf\tau;\vec{x}-\onehalf\vec{r},t-\onehalf\tau) 
\;,
\ee 
and often to Fourier transform the center of mass variables as well,
\be
\hat{G}(\vec{p},\vec{q};\epsilon,\omega) 
= 
\int d^{3}x
\int^{+\infty}_{-\infty}dt\,e^{i(\omega t-\vec{q}\cdot\vec{x})}
\; 
\hat{G}(\vec{p},\vec{x};\epsilon,t)
\;.
\ee

The central object of the quasiclassical theory is not the full Green's function, but the
quasiclassical propagator,
\be
\hat{g}(\hat{p},\vec{x};\epsilon,t) 
=
\frac{1}{a}\int_{-\varpi_{c}}^{+\varpi_{c}}
d\xi_{\vec {p}}\,\hat{\tau}_{3}\,\hat{G}(\vec{p},\vec{x};\epsilon,t)
\;,
\ee 
in which the sharp structure in $\hat{G}(\vec{p},\vec{x};\epsilon,t)$ due to the quasiparticles
with $p\simeq p_f$ [{\it i.e.} $\xi_{\vec{p}} = v_f(p-p_f)\simeq 0$] is integrated out. The
appearance of the matrix
\ber
\hat{\tau}_{3}=\begin{pmatrix}	I	&	0	\cr		0	&	-I	\end{pmatrix}
\,,
\nonumber
\eer
where $I$ the $2\times 2$ identity matrix in spin space, in the definition of $\hat{g}$ is
conventional. The cutoff, $\varpi_c$, separates the low-energy quasiparticle states from the
high-energy incoherent part of the spectral function,\footnote{Obviously $\varpi_c$ is not well
defined; however, the ambiguity turns out to be of no significance through first-order in the
small energy scale $\delta\sim T/E_f$. The cutoff appears explicitly only in the logarithmically
divergent weak-coupling gap equation, and can always be eliminated in favor of the physically
relevant superfluid transition temperature, $T_c$. For a detailed discussion of this procedure
see Ref. \cite{ser83}.} and the factor $a$ is the weight of the quasiparticle pole in the
spectral function. The central equation of the quasiclassical theory follows from the {\it
low-energy} Dyson equation,\footnote{$\hat{G}$ satisfies both a left- and right-handed
low-energy Dyson equation with $\hat{\tau}_{3}\hat{G}$ and the differential operator
interchanged.} which holds for energies $\vert\epsilon\vert <\varpi_{c}$ and
momenta such that $\vert\xi_{\vec{p}}\vert <\varpi_{c}$ [$\xi_{\vec{p}}\equiv v_{f}(p-p_{f})$],
\be\label{bp}
\left(\epsilon\hat{\tau}_{3} - \xi_{\vec{p}}\hat{1} - \hat{\sigma}\right)
\otimes\hat{\tau}_{3}\hat{G}
= a\hat{1}
\;,
\ee 
where $\hat{\sigma}(\hat{p},\vec{q};\epsilon,\omega)$ is the quasiclassical self energy and the
convolution operator $\otimes$ is defined by
\ber\label{bm}
\hat{A}\otimes\hat{B}(\vec{p},\vec{q};\epsilon,\omega) 
\negthickspace=\negthickspace
\int\frac{d^{3}s}{(2\pi)^{3}}\negthickspace\negthickspace
\int\frac{d\nu}{2\pi}\negthickspace\negthickspace
&\,&
\negthickspace\negthickspace\negthickspace\negthickspace
\negthickspace\negthickspace\negthickspace\negthickspace
\negthickspace\negthickspace
\hat{A}(\vec{p}+\onehalf\vec{s},\vec{q}-\vec{s};\epsilon+\onehalf\nu,\omega-\nu) 
\\
&\times&
\negthickspace\negthickspace\negthickspace\negthickspace
\hat{B}(\vec{p}-\onehalf(\vec{q}-\vec{s}),\vec{s};
        \epsilon-\onehalf(\omega-\nu),\nu) 
\,.
\nonumber
\eer

The self energy is a functional of $\hat{g}$, and a formal perturbation expansion for
$\hat{\sigma}$ in terms of $\hat{g} $ has been derived by Rainer and Serene \cite{rai76}. The
contributions to $\hat{\sigma}$ are classified by their order in the small parameters,
$\delta/E_{f}$, where $\delta$ represents any of the low-energy scales
$\delta=\{k_{B}T_{c},\hbar\omega,\hbar qv_{f},v_{ext}\}$. The first-order terms are the {\it
mean fields}, while the second-order terms describe the leading order effects due to
quasiparticle collisions and strong-coupling corrections to the mean fields. In what follows we
work in the high-frequency limit, $\omega\gg 1/\tau$, where $\tau$ is the quasiparticle
lifetime, and we will typically neglect collision effects as well as strong-coupling
corrections.

The quasiclassical theory describes the nonequilibrium dynamics of superfluid $^3$He by making
use of the Keldysh method \cite{kel65} (see also the recent review \cite{ram86}) which requires
three propagators: $\hat{g}^{R},\hat{g}^{A},\hat{g}^{K}$ representing the low-energy parts of
the full Green's functions $\hat{G}^{R},\hat{G}^{A}$, and $\hat{G}^{K}$. The retarded and
advanced Green's functions, $\hat{G}^{R}$ and $\hat{G}^{A}$, describe the quantum states of the
superfluid, while the Keldysh Green's function, $\hat{G}^{K}$, describes the occupation of these
states\footnote{This interpretation is strictly true only in the low-frequency limit,
$\hbar\omega\ll\Delta$, but the information contained in $\hat{g}^{R}$, $\hat{g}^{A}$,
$\hat{g}^{K}$, and the utility of these functions extends well beyond this regime to frequencies
$\Delta < \hbar\omega\ll E_f $.} and is defined by
\be\label{bt}
G_{ab}^{K}(\vec{x},t;\vec{x}'t') 
=
-i\left\langle\left[\Psi_{a}(\vec{x},t)\,,\,\Psi_{b}^{\dag}(\vec{x}',t')\right]\right\rangle 
\;.
\ee 
These three Green's functions can be written as elements of a $2\times 2$ 
matrix in ``Keldysh space''
\be\label{bu}
\ul{\hat{G}}
=
\begin{pmatrix}	\hat{G}^{R}	&	\hat{G}^{K}	\cr
				0			&	\hat{G}^{A}	\end{pmatrix}
\;,
\ee 
and the corresponding equation of motion for the low-energy part of the nonequilibrium matrix
Green's function is the familiar Dyson equation, {\it lifted} to include the Keldysh ``degree of
freedom'',
\be\label{bv}
\left(\epsilon\ul{\hat{\tau}}_{3} - \xi_{\vec{p}}\ul{\hat{1}} - \ul{\hat{\sigma}}\right) 
\otimes
\ul{\hat{\tau}}_{3} \ul{\hat{G}} = a\,\ul{\hat{1}}
\;,
\ee 
where the self energy matrix $\ul{\hat{\sigma}}$ (in Keldysh space) is of the form
\be\label{bx} 
\hat{\ul{\sigma}}
=
\begin{pmatrix}	\hat{\sigma}^{R}	&	\hat{\sigma}^{K}	\cr
					0				&	\hat{\sigma}^{A}	\end{pmatrix}
\;.
\ee 
The off-diagonal term, $\hat{\sigma}^K$, describes collision effects as
do the imaginary parts of $\hat{\sigma}^{R,A}$. However, in the mean-field
approximation ({\it i.e.} the leading order expansion
of $\ul{\hat{\sigma}}$ in $\delta$) these terms are zero and the components
of $\ul{\hat{\sigma}}$ collapse to one,
\ber\label{by}
\hat{\sigma}^{K}(\vhat{p},\vec{x};\epsilon,t) 
&=& 
0
\,,
\nonumber\\
\hat{\sigma}^{R}(\vhat{p},\vec{x};\epsilon,t) 
&=&
\hat{\sigma}^{A}(\vhat{p},\vec{x};\epsilon,t) 
=
\hat{\sigma}(\vhat{p},\vec{x};t) 
\;.
\eer
Furthermore, for $^3$He-B in homogenous thermal equilibrium the only nonzero mean field is the
``gap function'', or off-diagonal mean field $\hat {\Delta} $, representing the formation of
Cooper pairs. In $4\times 4$ Nambu space the gap function takes the form,
\be\label{bs}
\hat{\sigma}_{0}(\vhat{p})\equiv\hat{\Delta}(\vhat{p}) 
=
\begin{pmatrix}	
		0			&	\vec{\Delta}(\vhat{p})\cdot i\vec{\sigma}\sigma_{2}		\cr

\vec{\Delta}(\vhat{p})^{*}\cdot i\sigma_{2}\vec{\sigma}		&			0
\end{pmatrix}
\,,
\ee 
where $\{\sigma_{j}\;\vert\; j=1,2,3\}$ are the Pauli matrices and $\vec{\Delta}(\vhat{p})$ is
given by (\ref{BW}) for $^3$He-B.

In the low-frequency limit $(\hbar\omega\ll\Delta)$ the Keldysh propagator, $\hat{g}^K$, is
simply related to the quasiparticle distribution function which satisfies Landau's kinetic
equation (see {\it e.g.} \cite{ser83} and \cite{ram86}). This interpretation breaks down at
higher frequencies, $(\hbar\omega\sim\Delta)$; nevertheless, $\hat{g}^K$ still determines the
observable properties ({\it e.g.} sound and spin waves), and is calculated by solving the
quasiclassical transport equation, or the low-energy Dyson equation, as discussed below. For
many purposes knowledge of the full quasiclassical propagator
$\hat{g}^{K}(\vhat{p},\vec{x};\epsilon,t)$, is not necessary, but rather the equal-time (or
energy-integrated) propagator,
\be\label{bz}
\delta\check{g}(\vhat{p},\vec{x};t) 
=
\int\frac{d\epsilon}{2\pi i}\;\delta\hat{g}^{K}(\vhat{p},\vec{x};\epsilon,t) 
=
\delta\hat{g}^{K}(\vhat{p},\vec{x};t+0^{+},t) 
\;,
\ee 
where $\delta\hat{g}^{K}=\hat{g}^{K}-\hat{g}^{K}_{0}$ and
$\delta\hat{\sigma}=\hat{\sigma}-\hat{\sigma}_{0}$ represent the nonequilibrium deviations of
the propagator and self energy from their equilibrium values,
$\hat{g}_{0}^{K}\,,\,\hat{\sigma}_{0}$ (given below).

In order to represent the $ 4\times 4$ matrices $\delta\check{g}$ and
$\delta\hat{\sigma}$ in Nambu space it is convenient, both in terms of calculations
and physical interpretation, to introduce a particular basis set of
$4\times 4$ matrices,
\be\label{ca}
\left\{\hat{\gamma}_{a}\;\vert\; a=1,\ldots,16\right\} 
=
\left\{\hat{\Sigma}_{\mu},\;\hat{\tau}_{3}\hat{\Sigma}_{\mu},\;
       \hat{\delta}_{\mu},\;\hat {\tau}_{3}\hat {\delta}_{\mu}\;\vert\;\mu=0,2,3,4\right\}
\;,
\ee 
where
\ber\label{cb}
\hspace*{-3mm}
\hat{\Sigma}_{0} 
&=& 
\begin{pmatrix}	I	& 	0 \cr 0	 &	I	\end{pmatrix}
\,,\quad\quad\,\,
\hat{\Sigma}_{j} = \begin{pmatrix}	\sigma_{j}	&			0	\cr
										0		& -\sigma_{2}\sigma_{j}\sigma_{2} \end{pmatrix}
\,,
\\
\hat{\delta}_{0} 
&=& 
\begin{pmatrix}		0		&	i\sigma_{2}	\cr
									i\sigma_{2} &	0 			\end{pmatrix}
\,,\quad
\hat{\delta}_{j} = \begin{pmatrix}			0				&	i\sigma_{j}\sigma_{2}	\cr
									i\sigma_{2}\sigma_{j}	&		0 	\end{pmatrix}
\,,
\,j= 1,2,3
\,.
\nonumber
\eer
These matrices satisfy the orthonormality conditions,
\be\label{cc}
\frac{1}{4}\,\mbox{Tr}\left[\hat{\gamma}_{a}\hat{\gamma}_{b}\right] 
=
e_{a}\,\delta_{ab}
\,,
\ee
where 
$e_{a}=-1$ for the anti-hermitian matrices $\hat{\delta}_{\mu}$ ({\it i.e.} $ a=9,10,11,12 $)
and $e_{a}=+1$ for the remaining hermitian matrices.

The general expansions of $\delta\check{g}$ and $\delta\hat{\sigma}$ in terms of the
$\{\hat{\gamma}_a\}$ basis are
\be\label{cd}
\delta\check{g}
=
\sum^{16}_{a=1}\,\hat{\gamma}_{a}\delta\,g_{a},\qquad\delta\hat{\sigma}
=
\sum^{16}_{a=1}\hat{\gamma}_{a}\delta\sigma_{a}
\,.
\ee
It is also convenient to introduce the following notation for the coefficients
\ber\label{cea,ceb}
\delta\hat{\sigma}
&=&
\varepsilon^{+}\hat{1} 
+ \varepsilon^{-}\hat{\tau}_{3} 
+ \vec{d}^{+}\cdot\vhat{\delta}
+ \vec{d}^{-}\cdot\hat{\tau}_{3}\hat{\vec{\delta}}
+ \ldots
\,,
\nonumber\\
\delta\check {g}
&=&
\delta\,g^{+}\hat{1}
+ \delta g^{-}\hat{\tau}_{3}
+ \delta\vec{f}^{+}\cdot\vhat{\delta}
+ \delta\vec {f}^{-}\cdot\hat{\tau}_{3}\vhat{\delta}
+ \ldots
\,,
\eer
since the functions $\delta g^{\pm}$, etc. obey simple symmetry relations, $\delta
g^{\pm}(\vhat{p},\vec{x};t)=\pm\delta g^{\pm}(\pm\vhat{p},\vec{x};t)$, etc. These are the main
quantities needed for the description of the interaction of sound with the order parameter. The
mass density and current fluctuations are related to $\delta g^{+}$ and $\delta g^{-}$,
respectively, while the real and imaginary parts of the spin-triplet order parameter are
related to $\delta\vec{f}^{+}$ and $\delta\vec{f}^{-}$, respectively. We will often refer to
$\delta g^{\pm}$ as the quasiclassical ``distribution functions'', and $\delta\vec{f}^{\pm}$ as
the ``pair amplitudes''. Furthermore, the mean fields and equal-time propagators in
(\ref{cea,ceb}) are related by the mean-field constitutive equations,
\ber
\varepsilon^{\pm}(\vhat{p},\vec{q};\omega) 
&=&
\onehalf
\int\frac{d\Omega_{\vhat{p}'}}{4\pi}\,A^{s}(\vhat{p}\cdot\vhat{p}^{'})\,
\delta\,g^{\pm}(\vhat{p}^{'},\vec{q};\omega)
\,,
\label{cf}
\\
\vec{d}^{\pm}(\vhat{p},\vec{q};\omega) 
&=&
\onehalf
\int\frac{d\Omega_{\vhat{p}'}}{4\pi}\,V^{t}(\vhat{p}\cdot\vhat{p}^{'})\,
\delta\vec{f}^{\pm}(\vhat{p}^{'},\vec{q};\omega)
\,,
\label{cfa}
\eer
where $A^{s}(\vhat{p}\cdot\vhat{p}')$ is the spin-symmetric scattering amplitude for
quasiparticles in the forward direction and $V^{t}(\vhat{p}\cdot\vhat{p}')$ is the pairing
interaction for spin-triplet Cooper pairs with zero total momentum. Both of these functions may
be expanded in Legendr\'e polynomials,
\be\label{cg}
A^{s}(x) = \sum^{\infty}_{l=0}\,A^{s}_{l}\,P_{l}(x)
\,,\quad
V^{t}(x) = \sum_{l\geq 1}^{odd}\,(2l+1)\,V_{l}\,P_{l}(x) 
\,.
\ee
The $A^{s}_{l}$ are related to the conventional Landau Fermi-liquid parameters $F^{s}_{l}$,
\be\label{ch}
A^{s}_{l}=\frac{F_{l}^{s}}{1+F^{s}_{l}/(2l+1)}
\,,
\ee 
while $V_l$ is the pairing interaction in the angular momentum channel $l$. The dominant
contribution to $V^t$ is the $l=1$ term, and is responsible for the formation of p-wave Cooper
pairs. The Landau parameter $F^s_0$ is large in $^3$He ($F^s_0\simeq 10 - 100 $), reflecting
the stiffness of $^3$He, as well is $F^s_1 (\simeq 6 - 12) $, which is related to the large
effective mass. For the rest of the article we assume that the only nonzero parameters are
$F^s_0 , F^s_1 ,$ and $V^t_1$. This simplifies the calculations and is a good approximation for
the type of semiquantitative results given here. We have not discussed the exchange field which
plays an important role in all magnetic phenomena of $^3$He. We assume zero magnetic field
throughout most of this article, but occasionally refer to the effects a field has on the
collective mode spectrum, or coupling strength of these modes to zero sound.

The density fluctuation $\delta n$ and the stress fluctuation $\delta\Pi$ discussed in Sec.
(\ref{Conservation}) are related to the quasiclassical distribution function, $\delta g^{+}$,
by
\be\label{cj}
\delta n=\frac{N(E_{f})}{1+F_{0}^{s}}\,\delta g_{0}
\,,
\ee 
\be\label{stress}
\delta\Pi=\frac{N(E_{f})}{1+F_{0}^{s}}\,\delta g_{2}
\; ,
\ee
where 
\be\label{ck}
\delta g_{l}(q,\omega) 
=
\int\frac{d\Omega_{\vhat{p}}}{4\pi}\,P_{l}(\vhat{p}\cdot\vhat{q})\,
\delta g^{+}(\vhat{p},\vec{q};\omega)\,;\,\,l=0,2 
\,,
\ee 
and $N(E_{f})$ is the density of states at the Fermi surface for one spin population.
Similarly, the longitudinal current density is proportional
to $\delta g_1$,
\be\label{currq}
J_{q}=N(E_{f})\,v_{f}\,\delta g_{1}
\,.
\ee 

If the left and right-handed Dyson's equations are subtracted from one another and integrated
over $\xi_{\vec {p}} $ the result is a transport-like equation \cite{eil68,lar69,eli71} for the
quasiclassical propagators $\hat{g}^{a}(\vhat{p},\vec{q};\epsilon,\omega)$, where
$a=\{R,A,K\}$. In the mean-field approximation all three propagators obey the same equation,
\be\label{cl}
\hat{Q}\left[\hat{g}_{a},\hat{\sigma}\right]
\equiv 
\epsilon\left[\hat{\tau}_{3}\,,\,\hat{g}^{a}\right] 
+
\onehalf\omega\left\{\hat{\tau}_{3}\,,\,\hat{g}^{a}\right\}
-
\eta\,\hat{g}^{a}
-\hat{\sigma}\circ\hat{g}^{a}
+\hat{g}^{a}\circ\hat{\sigma}
= 0
\,,
\ee 
where $\eta=v_{f}\vhat{p}\cdot\vec{q}$ and the convolution is defined by 
\be\label{cm}
\hat{A}\circ\hat{B}(\vhat{p},\vec{x};\epsilon,\omega) 
=
\int\frac{d\nu}{2\pi}\hat{A}(\vhat{p},\vec{x};\epsilon+\onehalf\nu,\omega-\nu)
                     \hat{B}(\vhat{p},\vec{x};\epsilon-\onehalf(\omega-\nu),\nu) 
\,.
\ee 

In addition to these homogeneous transport equations the quasiclassical propagators satisfy the
normalization conditions first derived by Eilenberger \cite{eil68},
\ber
\hat{g}^{a}\circ\hat{g}^{a} = -2\pi^{3}\,\delta(\omega)\,\hat{1}\,,\,a=R,A
\,,\,
\hat{g}^{K}\circ\hat{g}^{A} + \hat{g}^{R}\circ\hat{g}^{K} = 0
\,.
\eer
These conditions together with the full quasiclassical equations for the self energy are
discussed in detail in Ref. \cite{ser83}.

Here we summarize the elementary, but important equilibrium
solutions of these equations. In homogeneous equilibrium the transport
equation simplifies to a matrix equation,
\be
\left[\epsilon\hat{\tau}_{3} - \hat{\sigma}_{0}(\vhat{p})\,,\,\hat{g}^{a}_{0}\right] = 0
\,,
\ee 
with the normalizations,
\ber\label{equlibrium-normalizations}
\hat{g}^{a}_{0}(\vhat{p},\epsilon)^{2} = -\pi^{2}\,\hat{1}\,,\,a=R,A
\,,\quad
\hat{g}^{K}_{0}\,\hat{g}^{A}_{0} + \hat{g}^{R}_{0}\,\hat{g}^{K}_{0} = 0
\,.
\eer
In addition, the equilibrium Keldysh propagator is proportional to the distribution function,
\be
\hat{g}_{0}^{K}(\vhat{p},\epsilon) 
=
\tanh\left(\frac{\epsilon}{2T}\right)
\left[\hat{g}_{0}^{R} - \hat{g}_{0}^{A}\right]
\,.
\ee 
Furthermore, the only non-zero mean field is the off-diagonal gap function,
$\hat{\sigma}_{0}(\vhat{p})=\hat{\Delta}(\vhat{p})$, given in (\ref{bs}) for triplet pairing.
We also assume that the order parameter is ``unitary'', in which case
$\hat{\Delta}(\vhat{p})\hat{\Delta}(\vhat{p}) = -
\vert\vec{\Delta}(\vhat{p})\vert^{2}\,\hat{1}$. The corresponding equilibrium solutions to the
transport equation and normalization conditions are,
\ber
\hat{g}_{0}^{({{R}\atop{A}})}
&=&
\pi\frac{\hat{\Delta}(\vhat{p}) - \epsilon\hat{\tau}_{3}}
        {\sqrt{(\epsilon + i\varsigma)^{2} - \vert\vec{\Delta}(\vhat {p})\vert^{2}}}
\,,
\\
\hat{g}_{0}^{K}
&=&
2\pi i\,\,\sgn(\epsilon)\,\tanh\left(\frac{\epsilon}{2T}\right)
\frac{\hat{\Delta}(\vhat{p}) - \epsilon\hat{\tau}_{3}}
     {\sqrt{\epsilon^{2} - \vert\vec{\Delta}(\vhat{p})\vert^{2}}}\,
\Theta\left(\epsilon^{2} - \vert\vec{\Delta}(\vhat{p})\vert^{2}\right) 
\,.
\eer
Using the equilibrium solution for $\hat{g}_{0}^{K}$ in the mean-field equation for 
$\vec{\Delta}(\vhat{p})$ we obtain the self-consistency equation,
\be\label{gap}
\vec{\Delta}(\vhat{p}) 
=
\int\frac{d\Omega_{\vhat{p}'}}{4\pi}\,V^{t}(\vhat{p}\cdot\vhat{p}')
\int^{\varpi_{c}}_{\vert\vec{\Delta}(\vhat{p}')\vert} 
d\epsilon\,\tanh\left(\frac{\epsilon}{2T}\right) 
\frac{\vec{\Delta}(\vhat{p}')}{\sqrt{\epsilon^{2}-\vert\vec{\Delta}(\vhat{p}'})\vert^{2}}
\,,
\ee 
where $\varpi_{c} $ is the cutoff defining the low-energy states. For p-wave pairing, ($V_1 >
V_{l\ne 1}$) the BW state (\ref{BW}), with
$\vert\vec{\Delta}\left({\hat{p}}\right)\vert^{2}=\Delta\left(T\right)^{2}$, is the
lowest-energy solution of Eq. \ref{gap} \cite{bal63}. The magnitude of the gap is then the
solution of the BCS gap equation,
\be\label{BCSgap}
\frac{1}{V_{1}}
=
\int^{\varpi_{c}}_{\Delta(T)} d\epsilon\,
\frac{\tanh\left(\frac{\epsilon}{2T}\right)}
     {\sqrt{\epsilon^{2}-\Delta(T)^{2}}}
\,.
\ee 
The superfluid transition temperature, $T_c$, is fixed by the gap equation for
$\Delta\rightarrow 0$, $1/V_{1} = \cK(T_{c})$, where
\be
\cK(T) = \int^{\varpi_{c}}_{0}\,\frac{d\epsilon}{\epsilon}\,
         \tanh\left(\frac{\epsilon}{2T}\right)
         \simeq\ln\left(1.13\,\frac{\varpi_{c}}{T}\right) 
\,.
\ee 
The ill-defined cutoff and pairing interaction can be eliminated in favor of the transition
temperature. The gap $\Delta(T)$ then becomes a function only of the reduced temperature
through the equation,
\be
\ln\left(T/T_{c}\right) 
=
\int^{\infty}_{\Delta(T)}d\epsilon\,
\tanh\left(\frac{\epsilon}{2T}\right)
\left[\frac{1}{\sqrt{\epsilon^{2} - \Delta(T)^{2}}} - \frac{1}{\epsilon}\right] 
\,,
\ee 
where $\varpi_{c}/\Delta\rightarrow\infty $ in all convergent integrals. These equilibrium
functions are important inputs to the linear and weakly nonlinear response functions of
$^3$He-B.

Applications of the quasiclassical transport equations to nonequilibrium problems in superfluid
$^3$He have been considered by several authors. Kopnin \cite{kop78} and Eckern \cite{eck81}
investigated the orbital and spin dynamics of the superfluid phases, while Kieselmann and
Rainer \cite{kie83}, and Zhang, {\it et al.} \cite{zha86}, studied Andreev scattering of
quasiparticle wavepackets from an inhomogenous order parameter field characteristic of
superfluid $^3$He near a surface. The theory of Andreev scattering in $^3$He is reviewed in
this volume by Kurkij\"arvi and Rainer \cite{kur90}. In the present context of high-frequency
collective modes in $^3$He-B the quasiclassical theory has been used to investigate the linear
response to zero sound \cite{sau81b,sau82,fis86,fis88b}, as well as parametric nonlinear
effects in $^3$He-B by Serene \cite{ser84} and the authors \cite{mck89a}.

For the nonlinear response there are distinct advantages, at
least in weak-coupling theory, to calculating $\hat{g}$ 
by directly integrating the low-energy Dyson equation
(\ref{bp}), which can be rewritten as
\be\label{cn}
\frac{1}{a}\ul{\hat{\tau}}_{3}\ul{\hat{G}}
=
\ul{\hat{G}}_{0} 
+ 
\ul{\hat{G}}_{0}\otimes\delta\ul{\hat{\sigma}}\otimes\frac{1}{a}
\ul{\hat\tau}_{3}\ul{\hat{G}}
\,,
\ee 
where $\ul{\hat{G}}_{0}$ is the low-energy equilibrium propagator for the superfluid phase
(given in Eqs. \ref{GK0}-\ref{cv} below), and $\delta\hat{\sigma}$ is the nonequilibrium mean
field. For weak nonlinearities we can formally expand the nonequilibrium quasiclassical
propagator $\delta\hat{g}$ in powers of the nonequilibrium self energy $\delta\hat{\sigma}$,
\be\label{cq}
\delta\hat{\ul{g}}
=
\int d\xi_{\vec{p}}\;
\left\{\hat{\ul{G}}_{0}\otimes\delta\hat{\ul{\sigma}}\otimes\hat{\ul{G}}_{0}
\,+\,
\hat{\ul{G}}_{0}\otimes\delta\hat{\ul{\sigma}}
                \otimes\hat{\ul{G}}_{0}
                \otimes\delta\hat{\ul{\sigma}}
                \otimes\hat{\ul{G}}_{0}\,+\,\ldots\right\} 
\,.
\ee 
The well-known weak-coupling linear response functions and collective mode spectrum of $^3$He is
obtained directly from the leading order term in Eq. (\ref{cq}) and the self-consistency
equations for $\delta\hat{\sigma}$, Eqs. (\ref{cf}) and (\ref{cfa}). The weak nonlinear response
functions are obtained from the terms quadratic in $\delta\hat{\sigma}$ in Eq. (\ref{cq}).
\vspace*{3mm}
\sec{Linear response}{Linear}

The quasiclassical distribution function and nonequilibrium pair amplitude are obtained from
the Keldysh component of (\ref{cq}), which to linear order in $\delta\hat{\sigma}$ is
\ber\label{cr}
&&\hspace*{-13mm}\delta\hat{g}^{K}(\vhat{p},\vec{q};\epsilon,\omega)
\\
&&\hspace*{-9mm}=
\int d\xi_{\vec{p}} 
\bigg[\hat{G}_{0}^{R}(\xi_{\vec{p}}+\onehalf\eta,\epsilon+\onehalf\omega)\,
      \delta\hat{\sigma}(\vhat{p},\vec{q};\omega)\,
      \hat{G}_{0}^{K}(\xi_{\vec{p}}-\onehalf\eta,\epsilon-\onehalf\omega) 
\nonumber\\
&&+\hspace*{3mm}\hat{G}_{0}^{K}(\xi_{\vec{p}}+\onehalf\eta,\epsilon+\onehalf\omega)\,
      \delta\hat{\sigma}(\vhat{p},\vec{q};\omega)\,
      \hat{G}_{0}^{A}(\xi_{\vec{p}}+\onehalf\eta,\epsilon+\onehalf\omega)
\bigg]
\,.
\nonumber
\eer
The propagators $\hat{G}_{0}^{R,A,K}$ are solutions of the low-energy equilibrium Dyson
equation,
\be
\left[
\left(\epsilon\pm i\varsigma\right)\hat{\tau}_{3}
+
\xi_{\vec{p}}\hat{1} 
- 
\hat{\Delta}(\vhat{p})
\right]
\hat{G}_{0}^{({{R}\atop{A}})} = \hat{1}
\,,
\ee 
with
\be\label{GK0}
\hat{G}_{0}^{K} = \tanh\left(\frac{\epsilon}{2T}\right) 
                  \left[\hat{G}_{0}^{R} - \hat{G}_{0}^{A}\right] 
\,,
\ee 
and have the simple form,
\be\label{cs}
\hat{G}_{0}^{R} = h^{R}\hat{H}\,,\quad
\hat{G}_{0}^{A} = h^{A}\hat{H}\,,\quad
\hat{G}_{0}^{K} = h^{K}\hat{H}\,,\quad 
\ee 
where 
$\hat{H}=\xi_{\vec{p}}\hat{1}+\epsilon\,\hat{\tau}_{3}-\hat{\Delta}$, and 
\be\label{cu}
h^{R}=\left(h^{A}\right)^{*} = \frac{1}{(\epsilon+i\varsigma)^{2}-E_{\vec{p}}^{2}}
\,, 
\ee 
with $\varsigma\rightarrow 0^{+}$, $E_{\vec{p}}^{2}=\xi_{\vec{p}}^{2}+\Delta^{2}$, and 
\be\label{cv}
h^{K}=\frac{i\pi}{E_{\vec{p}}}\,\tanh\left(\frac{\epsilon}{2T}\right)\,
      \left[\delta(\epsilon-E_{\vec{p}}) - \delta(\epsilon+E_{\vec{p}})\right] 
\,.
\ee 
We have assumed the long-wavelength limit, $q\ll k_{f}$, in order to write
$\xi_{\vec{p}\pm\vec{q}/2}=\xi_{\vec{p}}\pm\onehalf\eta$, where
$\eta=v_{f}\,\vhat{p}\cdot\vec{q}$. Note that the quasiclassical propagators
$\hat{g}_{0}^{R,A,K}$ are obtained by direct integration of 
$\hat{G}_{0}^{R,A,K}$ over $\xi_{\vec{p}}$.

By integrating Eq. (\ref{cr}) over excitation energies, $\epsilon$, we obtain the linear
response functions for the equal-time quasiclassical propagator,
\be\label{cw}
\delta g_{a}(\vhat{p},\vec{q};\omega) 
=
\sum^{16}_{b=1}\,\epsilon_{a}\,Y_{ab}(\vhat{p},\vec{q};\omega)\,
                 \delta\sigma_{b}(\vhat{p},\vec{q};\omega) 
\,,
\ee 
where the coefficients are 
\ber
Y_{ab}(\vhat{p},\vec{q};\omega) 
&=&
\frac{1}{4}\int\frac{d\epsilon}{2\pi i}
\int d\xi_{\vec{p}}\,
\hspace*{5cm}
\label{cx}
\\
&\times&
\mbox{Tr}
	\bigg[
	\hat{\gamma}_{a}\,
	\hat{G}_{0}^{R}(\xi_{\vec{p}}+\onehalf\eta,\epsilon+\onehalf\omega)\,
	\hat{\gamma}_{b}\,
	\hat{G}_{0}^{K}(\xi_{\vec{p}}-\onehalf\eta,\epsilon-\onehalf\omega) 
\nonumber
\\
&+&\quad\,
	\hat{\gamma}_{a}\,
	\hat{G}_{0}^{K}(\xi_{\vec{p}}+\onehalf\eta,\epsilon+\onehalf\omega)\,
	\hat{\gamma}_{b}\,
	\hat{G}_{0}^{A}(\xi_{\vec{p}}-\onehalf\eta,\epsilon-\onehalf\omega) 
	\bigg] 
\,.
\nonumber
\eer 

There is considerable redundancy among the various response functions $Y_{ab}$. First of all the
identities
$\mbox{Tr}[\hat{A}\hat{B}]=\mbox{Tr}[\hat{B}\hat{A}]$ and
$h^{R}=h^{A*}$ imply the relations
\be\label{cy}
\mbox{Re}\,Y_{ab}(\vhat{p},\vec{q},\omega)=\mbox{Re}\,Y_{ba}(\vhat{p},-\vec{q},-\omega)
\,.
\ee 
A similar identity for the imaginary part,
\be\label{cz}
\mbox{Im}\,Y_{ab}(\vhat{p},\vec{q},\omega)=-\mbox{Im}\,Y_{ba}(\vhat{p},-\vec{q},-\omega)
\,.
\ee
also holds if $\mbox{Tr}[\hat{\gamma}_{a}\hat{H}\hat{\gamma}_{b}\hat{H}]$ is real, as is the
case for nonmagnetic ground states. These identities require time-reversal symmetry of the
ground state and are analogous to the Onsager relations of irreversible thermodynamics (see {\it
e.g.} \cite{kreuzer81}). Together with the identities that follow from particle-hole symmetry,
gauge and Galilean invariance, they greatly reduce the number of independent coefficients
$Y_{ab}$. The symmetry relations that follow from gauge and Galilean invariance are obtained by
first considering how the conservation laws for particle number and momentum are related to the
quasiclassical transport equation, Eq. (\ref{cl}). Although this equation is nonlinear in 
$\hat{g}$, it contains two important linear relations connecting the low-order moments of the
quasiclassical distribution functions $\delta g^{\pm}(\vhat{p},\vec{q};\omega)$. By projecting
out the diagonal terms in particle-hole space from (\ref{cl}) and integrating over $\epsilon$ we
obtain,
\ber\label{proto1}
\omega\delta g^{+}
\negthickspace\negthickspace - \negthickspace
\eta\delta g^{-}
\negthickspace\negthickspace - \negthickspace
2\omega\varepsilon^{+}
\negthickspace = \negthickspace
-2\gamma\vec{\Delta}\cdot\vec{d}^{-}
\negthickspace\negthickspace + \negthickspace
2\vec{\Delta}\cdot\delta\vec{f}^{-}
\negthickspace\negthickspace - \negthickspace
\vec{d}^{-}\circ\delta\vec{f}^{+}
\negthickspace + \negthickspace
\vec{d}^{+}\circ\delta\vec{f}^{-}
\,, 
\\
\label{proto2}
\omega\delta g^{-}
\negthickspace\negthickspace - \negthickspace
\eta\delta g^{+}
\negthickspace\negthickspace - \negthickspace
2\omega\varepsilon^{-} 
 =  0
\,,
\hspace*{6cm}
\eer
where $\onehalf\gamma$ is the same integral appearing in the weak-coupling gap equation, Eq.
(\ref{BCSgap}). These equations are closely related to the conservation laws for particle number
and momentum. Projecting out the $l=0$ term of (\ref{proto1}) gives
\be
\omega\delta g_{0}-qv_{f}\,\delta g_{1}-2\omega\varepsilon_{0} = 0
\,,
\ee
which upon using the identities (\ref{ch}), (\ref{cj}), (\ref{currq}) and the self-consistency
equation, Eq. (\ref{cf}) for $\varepsilon_0$, reduces to the continuity equation, Eq.
(\ref{ae}). Note that the right side of Eq. (\ref{proto1}) vanishes upon integration over the
Fermi surface when the self-consistency equations, Eqs. (\ref{cfa}) and (\ref{gap}), for
$\vec{d}^-$ and $\vec{\Delta}$ are imposed. Similarly, the $l=1$ projection of Eq.
(\ref{proto2}) gives
\be
\omega\delta g_{1}-\onethird qv_{f}\,\delta g_{0}
                  -\twothirds qv_{f}\,\delta g_{2}
                  -\twothirds\omega\varepsilon_{1} = 0
\,,
\ee
from which the second conservation law, Eq. (\ref{af}), for the longitudinal current follows.
Alternatively, Eqs. (\ref{proto1}) and (\ref{proto2}) can be used to obtain linear relations
between the response functions. The functional derivative of ( \ref{proto2}) - {\it i.e. } the
proto-momentum conservation law - with respect to any of the mean fields $\sigma_{a}$, except
$\varepsilon^{-}$ ({\it i.e.} $a\ne 2$), gives
\be
\omega\,Y_{2a}\, - \eta\,Y_{1a}=0\,;\quad a\ne 2
\,.
\ee 
Then using the full Onsager-like relations, Eqs. (\ref{cy})
and (\ref{cz}), we also obtain,
\be\label{dh}
\omega\,Y_{a2}\, - \eta\,Y_{a1}=0\,;\quad a\ne 2
\,.
\ee 
Similarly, the functional derivative of Eq. (\ref{proto1}) - the proto-contiunity equation -
with respect to $\sigma_{a}$ for $\sigma_{a}\ne\varepsilon^{+}$ ({\it i.e.} $a\ne 1$), combined
with the Onsager-like relations yields,
\be\label{dl}
\omega\,Y_{a1}\, - \eta\,Y_{a2} - 2\sum^{3}_{b=1}\,\Delta_{b}\,Y_{a,b+13} = 0
\,;\quad a\ne 1
\,.
\ee 
If we now consider the expansion of the distribution function, $\delta g^+$, and the pair
amplitudes, $\delta\vec{f}^{\pm}$, in the mean fields, then it is clear from Eq. (\ref{dh}) that
$\varepsilon^{+}$ and $\varepsilon^{-}$ always appear in the linear combination,
\be\label{invariant1}
\delta g^{+},\delta\vec{f}^{\pm}
\sim
\left(\omega\varepsilon^{+} + \eta\varepsilon^{-}\right) +\,\ldots
\,.
\ee
By combining (\ref{dh}) and (\ref{dl}) we obtain,
\be\label{dm}
\left(\omega^{2}-\eta^{2}\right)\,Y_{a1} 
- 
2\omega\sum^{3}_{b=1}\,\Delta_{b}\,Y_{a,b+13} = 0\,;\qquad a\ne 1,2
\,.
\ee 
Thus, the mean fields, $\varepsilon_{\pm}$ and $\vec{\Delta}\cdot\vec{d}^{-}$, 
if they enter the expansions of $\delta\vec{f}^{\pm}$ do so in the linear combination,
\be\label{invariant2}
\delta\vec{f}^{\pm}
\sim
\Delta^{2}(\omega\varepsilon^{+} + \eta\varepsilon^{-})
-
\onehalf(\omega^{2}-\eta^{2})\vec{\Delta}\cdot\vec{d}^{-} + \,\ldots
\,.
\ee 
The reasons for these invariant combinations of Landau molecular fields, $\varepsilon_{\pm}$,
and the imaginary part of the order parameter fluctuation, $\vec{d}^-$, are hinted at by their
originating from the conservation laws for particle number and momentum; Eqs. (\ref{invariant1})
and (\ref{invariant2}) reflect the covariance of the transport equations for $^3$He under gauge
and Galilean transformations.

Consider the following transformation of the quasiclassical propagator,
\be\label{da}
\hat{g}(\vhat{p},\vec{x};t,t')
\rightarrow\hat{g}'
=
\hat{U}\left[\hat{\Lambda}(\vhat{p},\vec{x};t)\right]\,
	\hat{g}(\vhat{p},\vec{x};t,t')\,
\hat{U}\left[\hat{\Lambda}(\vhat{p},\vec{x};t')\right]^{\dag}
\,,
\ee 
where 
\be\label{db}
\hat{U}\left[\hat{\Lambda}(\vhat{p},\vec{x};t)\right] = e^{-i\hat{\Lambda}(\vhat{p},\vec{x};t)}
\,. 
\ee 
The self energy, $\hat{\sigma}$, transforms similarly. The particular choices,
$\hat{\Lambda}=\phi(\vec{x},t)\,\hat{\tau}_{3}$ and 
$\hat{\Lambda}=p_{f}\,\vhat{p}\cdot\vec{X}(t)\,\hat{1}$, 
correspond to the gauge and Galilean transformations, respectively. 
More generally, for any $\hat{\Lambda}$ that commutes with $\hat{\tau}_{3}$ ({\it i.e.} any
$\hat{\Lambda}$ that is diagonal in Nambu space, which includes local spin rotations generated
by $\hat{\Lambda}=\vec{\theta}\cdot\hat{\vec{\Sigma}}$), then if the pair ($\hat{g}$,
$\hat{\sigma}$) satisfy the quasiclassical transport equation,
$\hat{Q}\left[\hat{g}\,,\,\hat{\sigma}\right] = 0$, then
\be\label{QC-transformed}
\hat{U}\left[\hat{\Lambda}\right]\,
	\hat{Q}\left[\hat{g}\,,\,\hat{\sigma}\right]\,
\hat{U}\left[\hat{\Lambda}\right]^{\dag}
=
\hat{Q}\left[\hat{g}'\,,\,\hat{\sigma}' + \hat{\sigma}_{\nabla}\right] = 0
\,.
\ee
The form of the transport equation is invariant, but the local gauge, spin, or Galilean
transformation generates an additional mean field associated with the space or time variations
of the generator $\hat{\Lambda}(\vhat{p},\vec{x};t)$,
\be\label{de}
\hat{\sigma}_{\nabla}(\vhat{p},\vec{x};t,t')
=
-i\delta(t-t')
\left[
	\hat{U}\,v_{f}\vhat{p}\cdot\vec{\nabla}\hat{U}^{\dag} +
	\hat{U}\,\hat{\tau}_{3}\pder{}{t}\,\hat{U}^{\dag}
\right] 
\; .
\ee 
For the generalized Galilean transformation,
$\hat{\Lambda}(\vhat{p},\vec{x};t)=\Lambda(\vhat{p},\vec{x};t)\,\hat{1}$,
\be\label{dg}
\hat{\sigma}_{\nabla}
=
\delta(t-t')\left[
	v_{f}\,\vhat{p}\cdot\vec{\nabla}\,\Lambda\,\hat{1}
+	\pder{\Lambda}{t}\,\hat{\tau}_{3}
\right]
\,.
\ee 
It is then clear that the linear combination 
$(\omega\varepsilon^{+}+\eta\varepsilon^{-})$ 
contributing to $\delta g^{+},\delta\vec{f}^{\pm}$ remains invariant under 
a Galilean transformation as required by Eq. (\ref{da}) with 
$\hat{\Lambda}=\Lambda(\vec{x};t)\,\hat{1}$;
\ber
\varepsilon^{+}\xrightarrow[]{\hat{\Lambda}}
\varepsilon^{+}+\varepsilon^{+}_{\nabla}=\varepsilon^{+}+2i\eta\Lambda\,,
\nonumber
\\
\varepsilon^{-}\xrightarrow[]{\hat{\Lambda}}
\varepsilon^{-}+\varepsilon^{-}_{\nabla}=\varepsilon^{+}-2i\omega\Lambda\,,
\nonumber
\\
\left(\omega\varepsilon^{+}+\eta\varepsilon^{-}\right)\xrightarrow[]{\hat{\Lambda}}
\left(\omega\varepsilon^{+}+\eta\varepsilon^{-}\right) 
\,.
\eer 
Similarly, for gauge transformations, 
$\hat{\Lambda}(\vhat{p},\vec{x};t) =\Lambda(\vhat{p},\vec{x};t)\,\hat{\tau}_{3}$, 
the shift in $\hat{\sigma}$ is given by 
\be\label{dk}
\hat{\sigma}_{\nabla}
=
\delta(t-t')\left[
	\pder{\Lambda}{t}\,\hat{1}\, 
+
	v_{f}\,\vhat{p}\cdot\vec{\nabla}\,\Lambda\,\hat{\tau}_{3}
\right]
\,.
\ee 
The linear combination, 
$\Delta^{2}(\omega\varepsilon^{+}+\eta\varepsilon^{-})
-\onehalf(\omega^{2}-\eta^{2})\vec{\Delta}\cdot\vec{d}^{-}$,
is then gauge invariant, implying that any contribution to 
$\delta\vec{f}^{\pm}$ from the Landau molecular fields must be 
accompanied by the imaginary part (``phase'') of the order parameter
fluctuation projected along the direction $\vec{\Delta}$.

The above formalism can be used to reduce the linear response functions, $Y_{ab}$, that
determine $\delta\check{g}(\vhat{p},\vec{q};\omega)$ to two independent functions. The
quasiclassical distribution function and spin-triplet order parameter fluctuations to linear
order in the nonequilibrium mean fields become,
\ber
\delta g^{+}
&=&
(1-\lambda)\frac{2\omega}{\omega^{2}-\eta^{2}}
	\left(\omega\varepsilon^{+}+\eta\varepsilon^{-}\right) 
+
\frac{\omega\lambda}{\Delta^{2}}\,\vec{\Delta}\cdot\vec{d}^{-}
\,,
\label{dna}
\\
\delta\vec{f}^{-}
&=&
\gamma\vec{d}^{-}
+
\frac{\lambda}{2\Delta^{2}}
	\bigg[\left(\omega^{2}-\eta^{2}-4\Delta^{2}\right)\vec{d}^{-}
\nonumber\\
		&&\qquad\qquad
		+ 4\vec{\Delta}\left(\vec{\Delta}\cdot\vec{d}^{-}\right) 
		  - 2\vec{\Delta}\left(\omega\varepsilon^{+}+\eta\varepsilon^{-}\right)
		\bigg]
\,,
\label{dnb}
\\
\delta\vec{f}^{+}
&=&
\gamma\vec{d}^{+}
+
\frac{\lambda}{2\Delta^{2}}
	\bigg[\left(\omega^{2}-\eta^{2}\right)\vec{d}^{+}
		- 4\vec{\Delta}\left(\vec{\Delta}\cdot\vec{d}^{+}\right)
	\bigg] 
\,,
\label{dnc}
\eer
where $\gamma$ is given by 
\be\label{dp}
\gamma = 2\int^{\varpi_{c}}_{\Delta}d\epsilon\,
         \frac{\tanh\left(\epsilon/2T\right)}{\sqrt{\epsilon^{2}-\Delta^{2}}}
\,,
\ee 
the same integral that appears in the equilibrium gap equation, and
\ber
\lambda(\omega,\eta) 
\negthickspace\negthickspace\negthickspace&=&\negthickspace\negthickspace\negthickspace
\Delta^{2}\int^{+\infty}_{-\infty}\,\frac{d\epsilon}{2\pi i}\, 
\frac{(2\epsilon\omega+\eta^{2})\beta(\epsilon+\onehalf\omega) 
     -(2\epsilon\omega-\eta^{2})\beta(\epsilon-\onehalf\omega)}
	 {(4\epsilon^{2}-\eta^{2})(\omega^{2}-\eta^{2})+4\eta^{2}\Delta^{2}}
\,,
\nonumber\\
\beta(\epsilon) 
\negthickspace\negthickspace\negthickspace&=&\negthickspace\negthickspace\negthickspace
2\pi i\,\frac{\tanh\left(\epsilon/2T\right)}{\sqrt{\epsilon^{2}-\Delta^{2}}}\,
        \Theta\left(\epsilon^{2}-\Delta^{2}\right)
\,,
\eer
is the response function introduced by Tsuneto \cite{tsu60}. This function plays an important
role in determining the coupling strength of zero sound to the collective modes, the
Fermi-liquid renormalizations of the collective mode frequencies, the g-factors for the Zeeman
splittings of the $J=2^{\pm}$ modes, as well as the response of the quasiparticle excitations
for frequencies above the pair-breaking edge. So far as we are interested in the coupling of
sound to the $J=2^{\pm}$ modes we can typically assume the high-frequency ($\omega\sim\Delta$),
long-wavelength ($qv_f\ll\omega$) limits, in which case the $q\rightarrow 0$ response function
is most important,
\be\label{ay}
\lambda(\omega,T)
\equiv
\Delta^{2}\bar{\lambda}(\omega,T) 
=
\Delta(T)^{2}\int_{\Delta(T)}^{\infty}\,
\frac{d\epsilon}{\sqrt{\epsilon^{2}-\Delta(T)^{2}}}\,
\frac{\tanh\left(\epsilon/2T\right)}{\left(\epsilon^{2}-\omega^{2}/4\right)}
\,.
\ee 
%
\begin{figure}[t]
\includegraphics[width=0.95\textwidth]{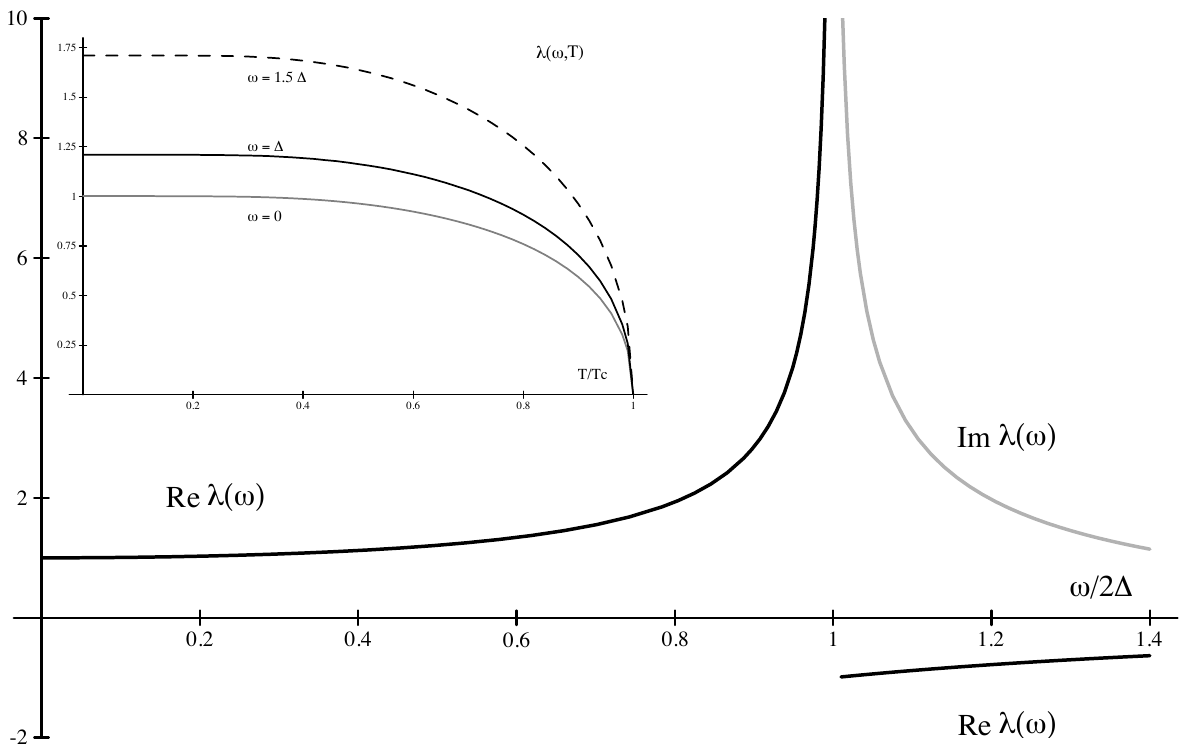}
\caption{
The response function $\lambda(\omega,T)$ as a function
of frequency for $T=0$. The inset shows $\lambda(\omega,T)$ 
as a function of temperature at $\omega = \Delta(T)$.
}
\label{Figure:Tsuneto_Function}
\end{figure}
At frequencies below the pair-breaking edge, $\omega < 2\Delta$, $\lambda(\omega)$ is real and
positive; the frequency and temperature dependence is shown in Fig.
\ref{Figure:Tsuneto_Function}. Note that $\lambda(\omega)$ is nominally of order one, except
near $T_c $ where $\lambda\sim (T_c -T)$, and for $\omega\rightarrow 2\Delta$ where $\lambda$
has a weak singularity, $\lambda\sim(2\Delta-\omega)^{-1/2}$. Above the pair-breaking edge
$\lambda(\omega)$ has an imaginary part,
\be
\mbox{Im}\,\lambda(\omega,T) 
	= 
		\frac{\pi}{2}\left(\frac{2\Delta}{\omega}\right)\,               
        \frac{\tanh(\omega/4T)}{\sqrt{(\omega/2\Delta)^2-1}}\,,\quad
		\omega >2\Delta
\,,
\ee 
which determines (in part) the absorption of zero sound by pair breaking processes.

For a purely p-wave pairing interaction the equilibrium gap equation implies that 
$\gamma=2/V_{1}$. This relation is used to eliminate the pairing interaction
$V_1$ and the cutoff-dependent integral $\gamma$ from the time-dependent gap equation
that determines the order parameter collective modes. The solutions, Eqs. 
(\ref{dna},\ref{dnb},\ref{dnc}), are identical with those obtained by solving the
linearized quasiclassical transport equations \cite{sau81b,sau87}.

The dispersion relations for the order parameter collective modes are found by substituting the
solutions (\ref{dnb}) and (\ref{dnc}) into the nonequilibrium gap equation, Eq. 
(\ref{cfa}). For a
purely p-wave pairing interaction, $\vec{d}^{+}$ and $\vec{d}^{-}$ have the form,
\be\label{dq}
d^{\pm}_{i}(\vhat{p},\vec{q};\omega) = D^{\pm}_{ij}(\vec{q},\omega)\,\vhat{p}_{j}
\,,
\ee 
and satisfy the equations,
\ber\label{dr}
\int\frac{d\Omega_{\vhat{p}}}{4\pi}\,\lambda(\omega,\eta)\,
\big[ 
	  (\omega^{2}-\eta^{2})\vec{d}^{+}
	- 4\vec{\Delta}(\vec{\Delta}\cdot\vec{d}^{+})
\big]_{i}\,\,\vhat{p}_{j} &=& 0
\;,
\\
\int\frac{d\Omega_{\vhat{p}}}{4\pi}\,\lambda(\omega,\eta)\,
\big[
	  (\omega^{2}-\eta^{2}-4\Delta^{2})\vec{d}^{-}
	+ 4\vec{\Delta}(\vec{\Delta}\cdot\vec{d}^{-}) 
&&
\nonumber\\
	- 2\vec{\Delta}\left(\omega\varepsilon^{+}+\eta\varepsilon^{-}\right)
\big]_{i}\,\,\vhat{p}_{j} &=& 0
\,.
\hspace*{1cm}
\label{ds}
\eer 
Since the equilibrium state is the pseudo-isotropic state give in Eq. (\ref{BW}), with $J=0$ and
an isotropic gap, the correct basis functions describing the excitations of the condensate are
the spherical tensors, $t^{J,M}_{ij}$,
\be\label{dt}
D_{ij}^{\pm}=\sum^{2}_{J=0}\,\sum^{+J}_{M=-J}\,D_{J,M}^{\pm}\,t^{J,M}_{ij}
\,,
\ee 
where 
\ber\label{du}
\hspace*{-5mm}
t^{0,0}_{ij}
&=&
\frac{1}{\sqrt{3}}\,\delta_{ij}\; ,
\nonumber\\
\hspace*{-5mm}
t^{1,M}_{ij}
&=&
\frac{1}{\sqrt{2}}\,\epsilon_{ijk}\,
\vec{u}^{M}_{k}\,;\quad 
\vec{u}^{0}_{k}=\vhat{z}_{k}\,,\quad 
\vec{u}^{\pm 1}_{k}=-\frac{1}{\sqrt{2}}\,\left(\vhat{x}\pm i\vhat{y}\right)_{k}
\,,
\eer
and
\ber\label{ba}
t_{ij}^{2,0}
&=&
\sqrt{\frac{3}{2}}\,\big[\vhat{z}_{i}\vhat{z}_{j}-\onethird\delta_{ij}\big]
\,,
\nonumber\\
t_{ij}^{2,\pm 1}
&=&
\mp\frac{1}{\sqrt{2}}\, 
	\left[(\vhat{x}\pm i\vhat{y})_{i}\,\vhat{z}_{j}
		+  \vhat{z}_{i}(\vhat{x}\pm i\vhat{y})_{j}
	\right]
\,,
\nonumber\\
t_{ij}^{2,\pm 2}
&=&
\mp\frac{1}{2}(\vhat{x}\pm i\vhat{y})_{i}(\vhat{x}\pm i\vhat{y})_{j}
\,.
\eer
These spherical tensors satisfy the relations,
\be\label{dv}
\left(t^{J,M}_{ij}\right)^{*}=t^{J,-M}_{ij}
\,,
\ee 
are orthogonal,
\be\label{dw}
\sum_{ij}\,t^{J,M}_{ij}\,t_{ij}^{J',M'^{*}}
=
\delta_{JJ'}\,\delta_{MM'}
\,,
\ee 
and are related to the spherical harmonic functions for the direction $\vhat{a}$ by
\be\label{dx}
\cY_{JM}(\vhat{a}) 
= \left\{
\begin{matrix}
\sqrt{\frac{3}{4\pi}}\,t^{0,0}_{ij}\,\vhat{a}_{i}\vhat{a}_{j}\,;\quad J=0\,,\cr
\sqrt{\frac{3}{8\pi}}\,\epsilon_{ilm}\,t_{lm}^{1,M}\,\vhat{a}_{i}\,;\quad J=1\,,\cr
\sqrt{\frac{15}{8\pi}}\,t_{ij}^{2,M}\,\vhat{a}_{i}\vhat{a}_{j}\,;\quad J=2\,.
\end{matrix}
\right.
\ee 
The quantization axis $\vhat{z}$ defining the orientation of the order parameter modes is
determined by the propagation direction $\vhat{q}$ in the zero-field limit, as is clear from
Eqs. (\ref{dr}) and (\ref{ds}) in which $\vhat{q}$ is the only direction entering the equations
of motion for $\vec{d}^{\pm}$.

Here we are only concerned with the $J=2^{+},2^{-}$, and $0^{-}$ modes. Their dispersion
relations are obtained by projecting the amplitudes $D^{\pm}_{JM}$ out of Eqs. (\ref{dr}) and
(\ref{ds}). Note that the real ($\vec{d}^{+} $) and imaginary ($\vec{d}^{-} $) modes are
decoupled from one another, a consequence of particle-hole symmetry which is built into the
quasiclassical equations. In the limit $q\rightarrow 0$ the modes with different $(J,M)$ also
decouple. In particular, the $J=2^+$ modes satisfy the homogenous equation,
\be\label{dy}
\lambda(\omega)\left[\omega^{2}-\frac{8}{5}\Delta^{2}\right]\,
	D_{2,M}^{+}(\vec{q}=0,\omega) = 0
\,,
\ee 
for all $M =\{-2,...,2\}$. The five-fold degeneracy of the $J=2^+$ modes is
a consequence of rotational invariance of the B-phase, while the homogeneity
of Eqs. (\ref{dr}) and (\ref{dy}), in particular
the absence of any coupling to the density or longitudinal current
follows from particle-hole symmetry. The degeneracy is partially lifted
for modes with non-zero pair momentum $\vec{q}$;
\be
\omega_{M}^{2}=\frac{8}{5}\Delta^{2}+c_{M}^{2}q^{2}
\,,
\ee 
where the velocities $c_M$ are approximately given by 
$c_{\pm 2}^{2}\simeq\frac{1}{5}v_{f}^{2}$,
$c_{\pm 1}^{2}\simeq\frac{2}{5}v_{f}^{2}$, and 
$c_{0}^{2}\simeq\frac{7}{15}v_{f}^{2}$ 
for $q v_f\ll\Delta$ \cite{vdo63}.\footnote{These results were obtained by several
other authors \cite{nag75,wol77,bru80b}. The effects of quasiparticle
interactions, magnetic fields and textures on the mode velocities
have been considered by \cite{com82,vol84b,fis86,fis88b}.} 
The degeneracy of the $J=2^+$ modes is fully lifted by a magnetic field \cite{tew79a}. In the
field-dominated regime, $\Delta >\gamma H\gg(qv_{f})^{2}/\Delta$, the
modes exhibit a linear Zeeman splitting \cite{sch81,sau82},
\be\label{dz}
\omega_{M}=\sqrt{8/5}\,\Delta + M\,g_{2^{+}}\,\omega_{L}
\,,
\ee 
where 
\be
g_{2^{\pm}}=\frac{1}{12}\left[1\pm \frac{\left(1-y(T)\right)}{\lambda(\omega_{2^{\pm}})}\right] 
\,,
\ee  
is the Land\'e factor for the mode, 
\be\label{ea}
\omega_{L} = \frac{\gamma H}{1+ \onethird F_{0}^{a}\left[2 + y(T)\right]}
\,,
\ee 
is the effective Larmor frequency, and $y(T)$ is the Yosida function. The five-fold Zeeman
splitting was observed by Avenel, {\it et al.} \cite{ave80}, and provided clear identification
of the of the absorption peaks discovered by Giannetta {\it et al.} \cite{gia80} and Mast {\it
et al.} \cite{mas80} as the $J=2^{+}$ modes.

In summary, if the Zeeman splitting in a small magnetic field
($\gamma H\ll\Delta$), and the effects of dispersion and 
damping due to quasiparticle collisions are taken into account, then the factor 
$(\omega^{2}-\frac{8}{5}\Delta^{2})$ in Eq. (\ref{dy}) is replaced by 
$(\omega^{2} + 2i\omega\Gamma - \omega_{M}^{2} - c_{M}^{2}q^{2})$, 
where $1/\Gamma$  is the lifetime and $c_{M}$ determines the phase velocity 
of the $J=2^{+}$, $M$ mode. Complete expressions for the mode frequencies, 
Land\'e g-factors and mode velocities including Fermi-liquid and higher-order
pairing effects are given in Refs. \cite{sau81b,sau82,fis86}.

Although the $J=2^+$ mode is uncoupled to sound within the quasiclassical theory for the linear
response, there is nevertheless a weak coupling of the $J=2^{+}$ modes to zero sound arising
from the small, but nonvanishing particle-hole asymmetry of $^3$He. In particular, if we retain
the weak energy dependence of the density of states, 
$N(\xi_{\vec{p}})\simeq N(0) + N'(0)\xi_{\vec{p}} + \ldots$, 
then the right side of Eq. (\ref{dy}) is replaced by a term of the form,
\be
\int\frac{d\Omega_{\vec{p}}}{4\pi}\,\zeta(\omega,\eta)\,
	\left[\vec{\Delta}(\omega\varepsilon^{+}+\eta\varepsilon^{-})\right]_{i}\,\vhat{p}_{j}
\ee 
where $\zeta\sim N'(0) T_c/N(0)\lambda\sim T_c /E_f$ is a measure of the
particle-hole asymmetry of $^3$He \cite{koc81}.

We now derive dynamical equations describing the linear coupling of the 
$J=0^{-}$ and $J=2^{-}$ modes to density oscillations. These equations are later
used in the derivation of the nonlinear constitutive equations, Eqs. (\ref{au}) and (\ref{av}).
The $J=0^{-}$ mode describes oscillations in the phase of the equilibrium order 
parameter and is the Goldstone mode associated with broken gauge symmetry. 
If Eq. (\ref{ds}) is contracted with $t_{ij}^{0,0}$ the result is %

\be\label{eb}
\int\frac{d\Omega_{\vec{p}}}{4\pi}
	\lambda(\omega,\eta)
	\left[\Delta^{2}(\omega\varepsilon^{+}+\eta\varepsilon^{-}) -
    \onehalf(\omega^{2}-\eta^{2})\vec{\Delta}\cdot\vec{d}^{-}
	\right] = 0
\,,
\ee 
which to first order in
$1/s^{2}=(qv_{f}/\omega)^{2}\simeq(v_{f}/c_{1})^{2}\sim 1/F^{s}_{0}$ gives 
\be\label{ec}
\left(\omega^{2}-\onethird(qv_{f})^{2}\right)\,\frac{D^{-}_{0,0}}{\sqrt{3}}
=
2\Delta\left(\omega\varepsilon_{0}+\onethird qv_{f}\varepsilon_{1}\right) 
\,.
\ee 
For $q=0$ and $\omega\ll\Delta$ Eq. (\ref{ec}) reduces to the Josephson equation, 
$i\partial_{t}\varphi=-2\varepsilon_{0}$, where $\varphi$ is the phase of
the local equilibrium order parameter,
$\vec{\Delta}(\vhat{p},t)=\vec{\Delta}(\vhat{p})\,e^{i\varphi(t)}$ 
$\simeq\vec{\Delta}(\vhat{p})\left(1 + i\varphi\right)$, in which case 
$D_{0,0}^{-}/\sqrt{3} = 2\Delta\varphi$; and 
$\varepsilon_{0}$ is the scalar shift in the quasiparticle energy. For $q\ne 0$, 
Eq. (\ref{ec}) is identified as the Bogoliubov-Anderson
mode in the limit $\omega\ll\Delta$.

Similarly, the contraction of Eq. (\ref{ds}) with $t_{ij}^{2,M^{*}}$ gives the 
dispersion relation for the $J=2^-$ modes. In the long-wavelength limit,
\ber\label{ed}
\frac{1}{6}\lambda(\omega)
\left[\omega^{2}-\omega_{M}(q)^{2}\right]\,D_{2,M}^{-}(q,\omega) 
\negthickspace\negthickspace\negthickspace\negthickspace
&=&
\negthickspace\negthickspace\negthickspace\negthickspace
\int\frac{d\Omega_{\vec{p}}}{4\pi}\,
	t_{ij}^{2,M^{*}}\,\vhat{p}_{i}\vhat{p}_{j}\,
	\lambda(\eta,\omega)
\hspace*{1cm}
\\
\negthickspace\negthickspace\negthickspace\negthickspace
&\times&
\negthickspace\negthickspace\negthickspace\negthickspace
	\bigg[\Delta(\omega\varepsilon^{+}+\eta\varepsilon^{-}) 
	-\onehalf\left(\omega^{2}-\eta^{2}\right)\,\frac{D^{-}_{0,0}}{\sqrt{3}}
	\bigg] 
\,,
\nonumber
\eer
where
\be 
\omega_{M}^{2}=\frac{12}{5}\Delta^{2}+c_{M}^{2}q^{2}
\,,
\ee 
are the $J=2^{-},M$ mode frequencies. The mode velocities are to a good approximation the same
as those given above for the $J=2^{+}$ modes, and the right side determines the coupling of the
$J=2^{-}$ modes to the density fluctuations. To leading order in $(qv_f/\omega)$ we can neglect
the $q$-dependence of $\lambda(q,\omega)$. To further simplify matters we assume that $A^s_l =
0$ for $l\ge 2$. Only the $l=0$ and $l=1$ projections of the mean fields, $\varepsilon_0$ and
$\varepsilon_1$, which are proportional to the density and current fluctuations
\be\label{ee}
\varepsilon_{0}=\onehalf A_{0}^{s}\,\delta g_{0}\,,
\qquad
\varepsilon_{1}=\onehalf A_{1}^{s}\,\delta g_{1}
\,,
\ee 
contribute to Eq. (\ref{ed}). These expressions are combined
with the particle conservation law given in Eq. (\ref{ae}) to relate
$\varepsilon_1$ to $\delta g_{0}$. Then to leading order in $1/s^{2}$ 
the phase mode and the density fluctuation are related by
\be\label{ef}
\frac{D^{-}_{0,0}}{\sqrt{3}}\simeq\left(\frac{\Delta}{\omega}\right)\,\delta g_{0}(q,\omega) 
\,,
\ee 
and, thus the equations of motion for the $J=2^-$ modes become,
\be\label{eg}
\left(\omega^{2}-\omega_{M}(q)^{2}\right)\,D_{2,M}^{-}(q,\omega)  =
\frac{6}{5}\,G_{M}^{*}\,\frac{\Delta\,(c_{1}q)^{2}}{\omega(1+F_{0}^{s})}\,
\delta g_{0}(q,\omega)
\,,
\ee 
where $G_{M}=\sqrt{\frac{2}{3}}\,\delta_{M,0}$. In zero field the density
and current fluctuations excite only the $J=2^{-},M=0$ mode (with $\vhat{q}$ being the 
quantization axis for the excited pairs). However, in a magnetic field, 
$\Delta >\gamma H\gg(qv_{f})^{2}/\Delta$, 
the quantization axis is fixed by the magnetic field, 
$\vhat{z}=\ul{\vec{R}}(\vhat{n},\theta)\cdot\vhat{H}$, 
and the $J=2^-$ modes exhibit a Zeeman effect; 
$\omega_{M}=\sqrt{12/5}\,\Delta + M\,g_{2^{-}}\,\omega_{L}$. 
Equation (\ref{eg}) still holds, but with the coupling strengths given by
\be
G_{M}=\sqrt{8\pi/15}\,\cY_{2M}(\vhat{q})
\,.
\ee 
For propagation parallel to the quantization axis only the $M=0$ mode is excited;
however, for $\vhat{q}\ne\vhat{z}$, all five modes can couple
to the density wave. The Zeeman splitting of the $J=2^{-}$ modes has recently
been observed by Movshovich, {\it et al.} \cite{mov88}.

As discussed in Sec. (\ref{Conservation}),
the phase velocity and attenuation of zero sound are determined by
the longitudinal components of the stress tensor, $\delta\Pi\sim\delta g_{2}$. 
Taking $l=2$ moment of Eq. (\ref{dna}) gives 
\be\label{eh}
\delta g_{2}(q,\omega) = \frac{2}{5} 
\frac{\left(1-\lambda(\omega)\right)}{1+F_{0}^{s}}
	\left(\frac{c_{1}q}{\omega}\right)^{2}
	\delta g_{0}(q,\omega)+\frac{\omega\lambda(\omega)}{5\Delta}
	\sum_{M}G^{M}\,D_{2,M}^{-}(q,\omega) 
\,,
\ee 
to leading order in $1/s^2$. Combining Eq. (\ref{eh}) with the wave equation, Eqs. (\ref{aj})
and (\ref{stress}), we obtain the dispersion relation for collisionless sound,
\ber
\omega^{2}
&=&
c_{1}^{2}q^{2}\big[1 + 2\xi(q,\omega)\big]
\,,
\\
\label{dispersion}
\xi \equiv \frac{\delta\Pi}{\delta n}
&=&
\frac{2}{5}
\left(\frac{c_{1}q}{\omega}\right)^{2}
\left(\frac{1}{1+F^{s}_{0}}\right) 
\\
&\times&
\left[
\left(1-\lambda(\omega)\right) + \frac{3}{5}\lambda(\omega)
\sum_{M}\vert G_{M}\vert^{2}
\frac{\omega^{2}}{(\omega +i\Gamma)^{2} - \omega_{M}(q)^{2}}
\right]
\,.
\nonumber
\eer 
The first term proportional to $(1-\lambda(\omega))$ gives a shift in the phase
velocity for $\omega <2\Delta$ from non-condensate excitations. In the limit 
$\omega\rightarrow 0$, $T\rightarrow 0$ this shift vanishes, as does the shift in
phase velocity due to the off-resonant excitation of the $J=2^-$ modes represented
by the second set of terms in (\ref{dispersion}). At high frequencies, 
$\omega\sim\omega_{M}$, the $J=2^-$ modes contribute significantly to both the phase
velocity and damping of sound,
\ber 
\hspace*{-1cm}
\frac{\delta v(\omega)}{c_{1}}\bigg\vert_{2^{-}\,\mbox{\footnotesize modes}}
\hspace*{-10mm}
&=&
\hspace*{-3mm}
\mbox{Re}\,\xi_{2^{-}\,\mbox{\footnotesize modes}}
	\simeq\frac{6}{25}\,\frac{\lambda(\omega)}{1+F^{s}_{0}}\,
	\sum_{M}\vert G_{M}\vert^{2}
	\left(\frac{\omega^{2}}{\omega^{2}-\omega^{2}_{M}}\right) 
\,,
\\
\hspace*{-1cm}
\alpha_{2^{-}\,\mbox{\footnotesize modes}}
\hspace*{-3mm}
&=&
\hspace*{-3mm}
-q\,\mbox{Im}\,\xi_{2^{-}\,\mbox{\footnotesize modes}}
	\simeq\frac{6\pi}{25}\,\frac{\omega}{c_1}\frac{\lambda(\omega)}{1+F^{s}_{0}}\,
	\sum_{M}\vert G_{M}\vert^{2}
	\delta(\omega^{2}-\omega^{2}_{M}) 
\,.
\eer 
In addition to the resonant absorption and anomalous dispersion of zero sound for
$\omega\simeq\omega_{M}$, the $J=2^-$ modes also modify the pair-breaking threshold. 
For $\omega>2\Delta$ the pair-breaking attenuation arises from 
$\mbox{Im}\,\lambda\sim 1/\sqrt{\omega -2\Delta}$. However, the singularity at $\omega = 2\Delta $ is suppressed by the
off-resonant coupling to the $J=2^-$ mode, so that
\be
\alpha_{\mbox{\footnotesize pair-breaking}}\simeq 
	\frac{6}{25}\,\frac{\omega}{c_{1}}\, 
	\frac{\mbox{Im}\,\lambda(\omega)}{1+F^{s}_{0}}\,
	\left(\frac{\omega^{2}-4\Delta^{2}}{\omega^{2}-\frac{12}{5}\Delta^{2}}\right)
	\sim\sqrt{\omega -2\Delta}
\,,
\ee
in agreement with W\"olfle \cite{wol76} and Serene \cite{ser74}, and qualitative 
agreement with experimental measurements of the pair-breaking 
attenuation \cite{gia80,mei83a,dan83}. Finally, we remark that if we include the
particle-hole asymmetry corrections, then the dispersion function $\xi(q,\omega)$ 
also includes resonant coupling to the $J=2^+$ modes of the form \cite{koc81},
\ber\label{Coupling_to_RSQ}
&&
\hspace*{-3.3cm}
\xi_{2^{+}} =\frac{6}{25}
	\left(\frac{c_{1}q}{\omega}\right)^{2}
	\left(\frac{1}{1+F^{s}_{0}}\right)\,
	[\zeta(\omega)]^2\,
	\sum_{M}\vert G_{M}\vert^{2}
	\left(\frac{\omega^{2}}{(\omega+i\Gamma)^{2}-\omega_{2^{+}M}(q)^{2}}\right)
\,,
\nonumber\\
\mbox{with}\,\zeta\sim\frac{T_c}{E_f}\,\lambda
\,.
&&
\eer

\sec{Nonlinear response}{Nonlinear}

In Sec. (\ref{Conservation}) we argued that even without exact particle-hole symmetry a
nonlinear coupling between the $J=2^+$ modes and zero sound exists, and is of the form
\ber
\label{ej}
&&
\hspace*{-17mm}
\delta g_{2}(\omega) =
	\sum_{M}\frac{1}{\Delta^2}
	\int d\nu\,\Xi^{(1)}_{M}(\omega,\nu,\omega-\nu)\,
	\delta g_{0}(\nu)\,D^{+}_{2,M}(\omega-\nu)\,,
\\ 
\label{ek}
&&
\hspace*{-17mm}
\frac{\bar{\lambda}(\omega)}{6}				\negthickspace
	\left[\omega^{2}-\omega_{M}^{2}\right]	\negthickspace
	D^{+}_{2,M}(\omega)						\negthickspace
	=										\negthickspace
	\frac{1}{\Delta^2}\negthickspace\int \negthickspace d\nu
	\Xi^{(2)}_{M}\negthickspace(\omega,\negthickspace\nu,
	             \negthickspace\omega\negthickspace-\negthickspace\nu)
	\delta g_{0}(\nu)\delta g_{0}(\omega\negthickspace-\negthickspace\nu) 
\,.
\eer 
The wave vector dependence of the nonlinear terms is
suppressed for clarity. In fact the constitutive equations can be
written in the forms given in (\ref{au}) and (\ref{av}), as the kernels 
$\Xi^{\left( {1}\right)}_{M}\left( {\omega ,\nu ,\omega -\nu}\right) $ 
and $\Xi^{\left( {2}\right)}_{M}\left( {\omega ,\nu ,\omega -\nu}\right) $ 
are determined by the same coupling function,
\ber\label{ax}
A_{M}(\vec{q},\omega;\vec{s},\nu;\vec{q}-\vec{s},\omega-\nu) 
=
\frac{1}{5}\left(\frac{c_{1}s}{\nu}\right)^{2}\,Z_{M}(\vhat{q},\vhat{s})\,
\tilde{A}(\omega,\nu ,\omega-\nu)
\,,
\eer 
where the frequency and temperature dependence of the coupling strength is
contained in the factor $\tilde{A}(\omega,\nu,\omega-\nu)$,
which can be written as a linear combination of Tsuneto functions, 
$\lambda(x)$, evaluated at the frequencies 
$x=\{\omega,\nu,\omega-\nu\}$ with coefficients
that are algebraic functions of $\omega$, $\nu$, and $\omega-\nu$ and the energy
gap $\Delta(T)$. The dependence of $A_{M}$ on $\vhat{q}$ and 
$\vhat{s}$, the propagation directions of the two sound waves, and
the magnetic quantum $M$ is contained in the factor
\be\label{az}
Z_{M}(\vhat{q},\vhat{s}) 
	= (\vhat{q}\cdot\vhat{s})\,\vhat{s}_{i}\,t^{2,M}_{ij}\vhat{q}_{j}
	- \onethird\,\vhat{q}_{i}\,t^{2,M}_{ij}\,\vhat{q}_{j}
	- \onethird\,\vhat{s}_{i}\,t^{2,M}_{ij}\,\vhat{s}_{j}
\,.
\ee 
The explicit expression for $\tilde {A} $ is derived below. 

The right sides of Eqs. (\ref{ej}) and (\ref{ek}) come from contributions to the quasiclassical
propagator which are second order in the nonequilibrium mean fields, and are derived from Eq.
(\ref{cq}). The explicit calculation of $\delta\hat{g}^{(2)}$ is lengthy, particularly at finite
wavevector; however, gauge and Galilean invariance and the generalization of the Onsager-type
relations, Eqs. (\ref{cy}) and (\ref{cz}), for the nonlinear response, reduce the number of
independent terms. Furthermore, it is necessary to calculate $\Xi^{(1)} $ and $\Xi^{(2)}$ only
to leading order in $1/s^{2}$ since the kernels of Eqs. (\ref{ej}) and (\ref{ek}) need only be
evaluated at $q=0$.

Since we are interested here in weak nonlinearities arising from the coupling of sound to the
$J=2^{+}$ modes, the density and current induce oscillations of the same frequency in the mean
fields, $\varepsilon^+$ and $\varepsilon^-$, in the phase of the order parameter (the $J=0^{-}$
mode), and the (off-resonant) $J=2^{-}$ mode. To derive Eqs. (\ref{ej}) and (\ref{ek}) it is
simplest to eliminate the phase of the order parameter by a gauge transformation. We then write
the nonequilibrium self energy in the form
\be
\label{el}
\delta\hat{\sigma} 
	= \varepsilon^{+}\hat{1}
	+ \varepsilon^{-}\hat{\tau}_{3}
	+ \vec{d}^{-}\cdot\hat{\tau}_{3}\,\hat{\vec{\delta}}
	+ \vec{d}^{+}\cdot\hat{\vec{\delta}}
\,, 
\ee 
where $\varepsilon^+$ and $\varepsilon^-$ are related to the density oscillation 
by $\varepsilon^{+}=\onehalf A^{s}_{0}\,\delta g_{0}$ 
and 
$\varepsilon^{-} = \onehalf A^{s}_{1}\,\omega\,\left(\vhat{p}\cdot\vhat{q}\right)$ 
$\delta g_{0}/(1+F^{s}_{0})$.
The oscillations in the imaginary part of the order parameter, $\vec{d}^{-}$, contain only a
contribution from the off-resonant excitation of the $J=2^{-}$ mode (from Eq. (\ref{eg})), and
is related to the density oscillation by
\be
\label{en}
\vec{d}^{-}(\vec{q},\omega) = 
	\frac{\mu(\omega)}{1+F_{0}^{s}}\,
	\left(\frac{c_{1}q}{\omega}\right)^{2}
	\vec{R}(\vhat{q})\,\delta g_{0}(q,\omega)
\,,
\ee 
where 
$\mu(\omega)={{6}\over{5}}\omega\Delta/\left(\omega^{2}-{{12}\over{5}}\Delta^{2}\right)$
and $\vec{R}(\vhat{q})=\vhat{q}(\vhat{p}\cdot\vhat{q})-\onethird\vhat{p}$.

These linear relations are used below to obtain the response of $^3$He-B arising from the
nonlinear coupling of sound to the $J=2^+$ modes. To obtain these nonlinear couplings we expand
the $4\times 4$ matrix propagator $\delta\check{g}$ and the self energy $\delta\hat{\sigma}$ in
terms of the matrices $\{\hat{\gamma}_{a}\}$ as in Eq. (\ref{cd}), then the expansion
coefficients for the second-order contribution to the propagator obtained from Eq. (\ref{cq})
are given by
\ber
\label{ep}
\delta g_{a}(\vec{q},\omega)  
&=& 
	\sum_{b,c=1}^{16}
	\int\frac{d\nu}{2\pi}
	\int\frac{d^{3}s}{(2\pi)^{3}}\, 
\\
&\times&		
		e_{a}\,X_{abc}(\omega,\nu,\omega-\nu)\,
		\delta\sigma_{b}(\vec{s},\nu)\,
		\delta\sigma_{c}(\vec{q}-\vec{s},\omega-\nu)
\,, 
\nonumber
\eer 
where 
\ber
\hspace*{-5mm}
X_{abc}(\omega,\nu,\omega-\nu) 
&=&
\bar{X}_{abc}(\omega,\nu,\omega-\nu) 
+
\bar{X}_{acb}(\omega,\nu,\omega-\nu)
\,,
\label{eq}
\nonumber\\
\hspace*{-5mm}
\bar{X}_{abc}(\omega,\nu,\omega-\nu) 
&=&
\mu^{RRK}_{abc} + \mu^{RKA}_{abc} + \mu^{KAA}_{abc}
\,,
\eer
and $\mu^{QST}_{abc}$ involves a product of the three equilibrium propagators $\hat{G}^{Q}$,
$\hat{G}^{S}$ and $\hat{G}^{T}$ [$Q,S,T=\{R,K,A\}$],
\ber\label{er}
\hspace*{-5mm}
\mu^{QST}_{abc}(\omega,\nu,\omega-\nu) =
	\int\frac{d\epsilon}{2\pi i}\int d\xi_{\vec{p}}\,
	\mbox{Tr}\bigg[
   	&\hspace*{-3mm}\hat{\gamma}_{a}\hspace*{-3mm}&
	\hat{G}_{0}^{Q}(\xi_{\vec{p}}+\onehalf\eta,\epsilon+\onehalf\omega)
\nonumber\\
   &\hspace*{-3mm}\hat{\gamma}_{b}\hspace*{-3mm}&
	\hat{G}_{0}^{S}(\xi_{\vec{p}}-\onehalf\eta+\eta',\epsilon-\onehalf\omega+\nu)
\nonumber\\
   &\hspace*{-3mm}\hat{\gamma}_{c}\hspace*{-3mm}&
	\hat{G}_{0}^{T}(\xi_{\vec{p}}-\onehalf\eta,\epsilon-\onehalf\omega)
	\bigg]
\,,
\eer
with $\eta=v_{f}\,\vhat{p}\cdot\vec{q}$ and $\eta'=v_{f}\,\vhat{p}\cdot\vec{s}$. The expression
for $X_{abc}$ contains six terms corresponding to the possible time orderings of the
propagators. It follows from its definition that
\be\label{es}
X_{abc}(\omega,\nu,\omega-\nu) = X_{acb}(\omega,\omega-\nu,\nu)
\,.
\ee 
In addition the identities $\mbox{Tr}[\hat{A}\hat{B}]=\mbox{Tr}[\hat{B}\hat{A}]$ and
$h^{R}=h^{A^{*}}$ imply
\ber
\label{et} 
\mbox{Re}\,X_{abc}(\omega,\nu,\omega-\nu) 
&=&
\mbox{Re}\,X_{bac}(-\nu,-\omega,\omega-\nu) 
\nonumber\\
&=&
\mbox{Re}\,X_{cba}(-\omega+\nu,\nu-\omega)
\,,
\eer 
which are the nonlinear generalizations of the Onsager-like
relations given in (\ref{cy}). As is the case for
the linear response functions further relations follow from gauge,
Galilean and rotational invariance. It is straight forward to show
that Galilean invariance implies 
\ber
&&
\hspace*{-5mm}
\nu\,X_{a2b}(\omega,\nu,\omega-\nu) 
=
\eta'\,X_{a1b}(\omega,\nu,\omega-\nu) 
\,,
\label{eu}
\\
&&
\hspace*{-5mm}
(\omega-\nu)\,X_{a2b}(\omega,\nu,\omega-\nu) 
=
(\eta -\eta')\,X_{a1b}(\omega,\nu,\omega-\nu) 
\,,
\label{ev}
\eer 
and thus, the mean fields, $\varepsilon^+$ and $\varepsilon^-$, appear only in the 
combinations 
$\nu\varepsilon^{+}(\vec{s},\nu)+\eta'\varepsilon^{-}(\vec{s},\nu)$ and 
$(\omega-\nu)\varepsilon^{+}(\vec{q}-\vec{s},\omega-\nu)+(\eta-\eta')
\varepsilon^{-}(\vec{q}-\vec{s},\omega-\nu)$. 
Consequently, it is convenient to introduce the quantity
\be\label{ex}
\omega\theta(\vec{q},\omega) \equiv
	\omega\,\varepsilon^{+}(\vec{q},\omega)+\eta\,\varepsilon^{-}(\vec{q},\omega)
\,.
\ee 
We now write down the general form for the distribution function and pair amplitudes, 
$\delta g^{+}$, $\delta\vec{f}^{+}$ and $\delta\vec{f}^{-}$, which are second order in the
non-equilibrium mean fields. Particle-hole symmetry and the relations (\ref{eu}) and (\ref{ev})
imply that there are only three independent coupling tensors $X_{11j}$,$X_{1lj}$ and $X_{lmj}$
(where $j=10,11,12$ and $l,m=14,15,16$) which appear in the expressions for $\delta g^{+}$,
$\delta\vec{f}^{-}$ and $\delta\vec{f}^{+}$. Gauge and rotational invariance further imply that
these coupling tensors can be written in the form
\ber\label{ey}
X_{11j}
&=&
A\,\Delta_{j} X_{1lj} = B\,\Delta_{l} \Delta_{j} + C\,\delta_{lj}
\nonumber\\
X_{lmj}
&=&
E\,\Delta_{l}\delta_{mj} + F\,\Delta_{m}\delta_{jl} + 
G\,\Delta_{j}\delta_{lm} + H\,\Delta_{l}\Delta_{j}\Delta_{m}
\,.
\eer
The identity $X_{lmj}(\omega,\nu,\omega -\nu) = X_{mlj}(-\nu,-\omega,\omega -\nu)$ 
further implies
\be\label{ez}
E(\omega,\nu,\omega-\nu) = -F(-\nu,-\omega,\omega-\nu) 
\,.
\ee
The second-order terms in $\delta g^{+}$, $\delta\vec{f}^{-}$ and $\delta\vec{f}^{+}$ are
then given in terms of these seven coefficients,
\ber\label{fa}
\hspace*{-20mm}
\delta g^{+(2)}(\omega) \hspace*{-1mm} = \hspace*{-1mm} \int d\nu\,
\Big[
&\,& \hspace*{-6mm}
A(\omega,\nu,\omega-\nu)\,\theta(\nu)\vec{\Delta}\cdot\vec{d}^{+}(\omega-\nu) 
\nonumber\\
&\hspace*{-10mm}+&\hspace*{-6mm}
B(\omega,\nu,\omega-\nu)\,\vec{\Delta}\cdot\vec{d}^{-}(\nu)\,
                           \vec{\Delta}\cdot\vec{d}^{+}(\omega-\nu)
\nonumber\\
&\hspace*{-8mm}+&\hspace*{-6mm}
C(\omega,\nu,\omega-\nu)\,\vec{d}^{-}(\nu)\cdot\vec{d}^{+}(\omega-\nu)
\Big]
\,,
\eer
\ber\label{fb}
\hspace*{-10mm}
\delta\vec{f}^{-(2)}(\omega) \hspace*{-1mm} = \hspace*{-1mm} \int d\nu
\Big[ 
&\hspace*{-5mm}\,&\hspace*{-3mm}
B(-\nu,-\omega,\omega -\nu)\,
  \theta(\nu)\,\vec{\Delta}\cdot\vec{d}^{+}(\omega-\nu)\,\vec{\Delta}
\nonumber\\
&\hspace*{-5mm}+&\hspace*{-3mm}
C(-\nu,-\omega,\omega -\nu)\,\theta(\nu)\,\vec{d}^{+}(\omega -\nu) 
\nonumber\\
&\hspace*{-5mm}+&\hspace*{-3mm}
E(\omega,\nu,\omega-\nu)\,\vec{d}^{-}(\nu)\cdot\vec{d}^{+}(\omega-\nu)\,\vec{\Delta}
\nonumber\\
&\hspace*{-5mm}+&\hspace*{-3mm}
F(\omega,\nu,\omega-\nu)\vec{\Delta}\cdot\vec{d}^{-}(\nu)\,\vec{d}^{+}(\omega-\nu)
\nonumber\\
&\hspace*{-5mm}+&\hspace*{-3mm}
G(\omega,\nu,\omega-\nu)\,\vec{\Delta}\cdot\vec{d}^{+}(\omega-\nu)\,\vec{d}^{-}(\nu) 
\nonumber\\
&\hspace*{-5mm}+&\hspace*{-3mm}
H(\omega,\nu,\omega-\nu)\,\vec{\Delta}\cdot\vec{d}^{-}(\nu)\,
                          \vec{\Delta}\cdot\vec{d}^{+}(\omega-\nu)\,\vec{\Delta}
\Big]
\,,
\eer
\ber\label{fc}
\hspace*{-10mm}
\delta\vec{f}^{+(2)}(\omega) =  \int d\nu
\Big[ 
&\hspace*{-5mm}\,&\hspace*{-3mm}
A(\nu-\omega,\nu,-\omega)\,\theta(\nu)\,\theta(\omega-\nu)\,\vec{\Delta}
\nonumber\\
&\hspace*{-5mm}+&\hspace*{-3mm}
B(\nu-\omega,\nu,-\omega)\,\vec{\Delta}\cdot\vec{d}^{-}(\nu)\theta(\omega-\nu)\,\vec{\Delta}
\nonumber\\
&\hspace*{-5mm}+&\hspace*{-3mm}
B(-\nu,\omega-\nu,-\omega)\,\theta(\nu)\,\vec{\Delta}\cdot\vec{d}^{-}(\omega-\nu)\,\vec{\Delta}
\nonumber\\
&\hspace*{-5mm}+&\hspace*{-3mm}
C(\nu-\omega,\nu,-\omega)\theta(\omega-\nu)\,\vec{d}^{-}(\nu)
\nonumber\\
&\hspace*{-5mm}+&\hspace*{-3mm}
C(-\nu,\omega-\nu,-\omega)\,\theta(\nu)\,\vec {d}^{-}\left( {\omega -\nu}\right) 
\\
&\hspace*{-5mm}+&\hspace*{-3mm}
E(\nu-\omega,\nu,-\omega)\,\vec{\Delta}\cdot\vec{d}^{-}(\omega-\nu)\,\vec{d}^{-}(\nu) 
\nonumber\\
&\hspace*{-5mm}+&\hspace*{-3mm}
F(\nu-\omega,\nu,-\omega)\,\vec{\Delta}\cdot\vec{d}^{-}(\nu)\,\vec{d}^{-}(\omega-\nu) 
\nonumber\\
&\hspace*{-5mm}+&\hspace*{-3mm}
G(\nu-\omega,\nu,-\omega)\,\vec{d}^{-}(\nu)\cdot\vec{d}^{-}(\omega-\nu)\,\vec{\Delta}
\nonumber\\
&\hspace*{-5mm}+&\hspace*{-3mm}
H(\nu-\omega,\nu,-\omega)\,
	\vec{\Delta}\cdot\vec{d}^{-}(\nu)\,\vec{\Delta}\cdot\vec{d}^{-}(\omega-\nu)\,\vec{\Delta}
\Big]
\,.
\nonumber
\eer
We have used Eqs. (\ref{eq}) and (\ref{er}) for $X_{abc}(\omega,\nu,\omega-\nu)$ to calculate 
the seven coupling functions $A,B,C,E,F,G$ and $H$ for $q=0$. Each of these functions
can be written in terms of the Tsuneto function, $\lambda(\omega)$,
\ber\label{fd}
A(\omega,\nu,\omega-\nu)	
&\hspace*{-2mm}=&\hspace*{-3mm} 
							  -4\Delta^{2}\,k_{1}(\omega,\nu,\omega-\nu) 
							+ (\omega-\nu)\,k_{2}(\omega,\nu,\omega-\nu) 
\nonumber\\
B(\omega,\nu,\omega-\nu)	
&\hspace*{-2mm}=&\hspace*{-3mm}
							  2\omega\,k_{1}(\omega,\nu,\omega-\nu) 
\nonumber\\
C(\omega,\nu,\omega-\nu)	
&\hspace*{-2mm}=&\hspace*{-3mm}
 							  \frac{-1}{\omega\Delta}\,
							  \left[\nu^{2}\bar{\lambda}(\nu)
							-		(\omega-\nu)^{2}\bar\lambda(\omega-\nu)\right]
\nonumber\\							
&\hspace*{+4mm}-&\hspace*{-3mm}					
							2(\omega-\nu)\Delta^{2}\,k_{1}(\omega,\nu,\omega-\nu)
\nonumber\\
E(\omega,\nu,\omega-\nu)	
&\hspace*{-2mm}=&\hspace*{-3mm}
 							  \frac{-2}{\nu\Delta}
							  \left[\omega\bar{\lambda}(\omega)
							-       (\omega-\nu)\bar {\lambda}(\omega-\nu)
							  \right] 
\nonumber\\
F(\omega,\nu,\omega-\nu)
&\hspace*{-2mm}=&\hspace*{-3mm}
							\frac{-2}{\omega\Delta}
							\left[\nu\bar{\lambda}(\nu)+(\omega-\nu)
							      \bar{\lambda}(\omega-\nu)
							\right] 
\nonumber\\
G(\omega,\nu,\omega-\nu) 
&\hspace*{-2mm}=&\hspace*{-3mm}
							- 4\Delta^{2}\,k_{1}(\omega,\nu,\omega-\nu)
							+ k_{3}(\omega,\nu,\omega-\nu) 
\nonumber\\
H(\omega,\nu,\omega-\nu) 
&\hspace*{-2mm}=& 
							  4\,k_{1}(\omega,\nu,\omega-\nu)
\,,
\eer 
where 
\ber
\label{fe} 
\hspace*{-1cm}
k_{n}(\omega,\nu,\omega-\nu)\Delta
&\hspace*{-3mm}=&\hspace*{-3mm} 
k_{n}(-\nu,-\omega,\omega-\nu)\,\Delta
= 
\frac{2}{\omega\nu(\omega-\nu)} 
\nonumber\\
&\hspace*{-3mm}\times&\hspace*{-3mm}
\left[
\omega^{n}\bar{\lambda}(\omega) 
+
(-\nu)^{n}\,\bar{\lambda}(\nu) - (\omega-\nu)^{n}\,\bar{\lambda}(\omega-\nu) 
\right] 
\,.
\eer
The detailed derivation of these functions can be found in Ref. \cite{mck88}.

From the nonlinear propagators given in Eqs. (\ref{fa}), (\ref{fb}), and (\ref{fc}) we obtain
the constitutive relations given in Eqs. (\ref{ej}) and (\ref{ek}). The nonlinear contribution
to the longitudinal stress tensor is given in part by the $l=2$ projection of Eq. (\ref{fa}),
\ber
&&
\hspace*{-2cm}
\int\frac{d\Omega_{\vec{p}}}{4\pi}\,P_{2}(\vhat{p}\cdot\vhat{q})\delta g^{+(2)}(\omega) 
	=
\label{fg} 
\nonumber\\
&&
\sum_{M}\,Z_{M}(\vhat{q},\vhat{s})
	\int d\nu\,L(\omega,\nu,\omega-\nu)\,\delta g_{0}(\nu)\,D^{+}_{2,M}(\omega-\nu)
\,, 
\eer
where
\be
\label{fh} 
\hspace*{-2mm}
L(\omega,\nu,\omega-\nu)
   = {{6}\over{35(1+F_{0}^{s})}}\,\left(\frac{c_{1}s}{\nu}\right)^{2}
	 \Bigg[\Delta\,A + \mu(\nu)\big({{2}\over{3}}\,B\,\Delta^{2} + {{7}\over{6}}\,C\big)\Bigg] 
\,,
\ee
and $Z_{M}$ is given by Eq. (\ref{az}). There is also a contribution to the stress tensor from
the $J=2^{-}$, $M=0$ mode [see Eq. (\ref{eh})] which, as a result of the nonlinear term in Eq.
(\ref{fb}), couples to the $J=2^{+}$ modes. This coupling is given by 
\ber\label{fj}
{{1}\over{6}}\bar{\lambda}(\omega)
\left(\omega^{2}-{{12}\over{5}}\Delta^{2}\right)\,D^{-}_{2,0}(\omega)
&\hspace*{-2mm}=&\hspace*{-3mm} 
	-\sqrt{{{3}\over{2}}}\int\frac{d\Omega_{\vec{p}}}{4\pi}\,
	\vec{R}(\vhat{q})\cdot\delta\vec{f}^{-(2)}(\omega)
\nonumber\\
&\hspace*{-35mm}=&\hspace*{-18mm} 
	-\onefifth\,{{\sqrt{3/2}}\over{(1+F_{0}^{s})}}
\sum_{M}\,Z_{M}(\vhat{q},\vhat{s})
\\
&\hspace*{-35mm}\times&\hspace*{-20mm}
\int d\nu\,\left(\frac{c_{1}s}{\nu}\right)^{2}\,K(\omega,\nu,\omega-\nu)\delta g_{0}(\nu) 
		   D^{+}_{2,M}(\omega-\nu)
\,,
\nonumber
\eer 
where
\ber\label{fk}
\hspace*{-7mm}
K(\omega,\nu,\omega-\nu) 
&=&
\bigg[\frac{4}{7}\Delta^{2}B + C\big](-\nu,-\omega,\omega-\nu) 
\nonumber\\
&+& 
{{2}\over{3}}\mu(\nu)\;\big[E+F+G+{{4}\over{7}}\Delta^{2}H\bigg](\omega,\nu,\omega-\nu) 
\,.
\eer 
In deriving (\ref{fg}) and (\ref{fj}) we made use of the identities
\ber\label{fl}
\onefifth Z_{M}(\vhat{q},\vhat{s}) 
&=&
\int\frac{d\Omega_{\vec{p}}}{4\pi}\,P_{2}(\vhat{p}\cdot\vhat{q})\,
	\vhat{p}_{i}\,t_{ij}^{2,M}\,R_{j}(\vhat{s})
\nonumber\\
&=&
\int\frac{d\Omega_{\vec{p}}}{4\pi}\,P_{2}(\vhat{p}\cdot\vhat{s})\,
	\vhat{p}_{i}\,t_{ij}^{2,M}\,R_{j}(\vhat{q})
\nonumber\\
&=&
\threehalves
\int\frac{d\Omega_{\vec{p}}}{4\pi}\,\vec{R}(\vhat{q})\cdot\vec{R}(\vhat{s})\,
	\vhat{p}_{i}\,t_{ij}^{2,M}\,\vhat{p}_{j}
\nonumber\\
&=&
\sevensixths
\int\frac{d\Omega_{\vec{p}}}{4\pi}\,P_{2}(\vhat{p}\cdot\vhat{q})\,P_{2}(\vhat{p}\cdot\vhat{s})\,
	\vhat{p}_{i}\,t_{ij}^{2,M}\,\vhat{p}_{j}
\,.
\eer
The resulting nonlinear stress is 
\be\label{fm}
\delta g_{2}(\omega)
=
	\onefifth\omega\bar{\lambda}(\omega)\Delta\,\sqrt{\threehalves}\,D^{-}_{2,0}(\omega) 
+
	\int\frac{d\Omega_{\vec{p}}}{4\pi}\,P_{2}(\vhat{q}\cdot\vhat{s})\,
	\delta g^{+(2)}(\omega)
\,,
\ee 
which upon combining Eqs. (\ref{fg}), (\ref{fj}) and (\ref{fm}) gives Eq. (\ref{ej}) with
\ber
\label{fn}
\Xi_{M}^{(1)}(\omega,\nu,\omega-\nu) 
	= 
		\frac{1}{5(1+F_{0}^{s})}
		\left(\frac{c_{1}s}{\nu}\right)^{2}\,
		Z_{M}(\vhat{q},\vhat{s})\,
		\tilde{A}(\omega,\nu,\omega-\nu) 
\,,
\eer 
where
%
\ber\label{fp}
\tilde{A}(\omega,\nu,\omega-\nu) 
&=&
\sixsevenths\Delta^{3}\,A(\omega,\nu,\omega-\nu) 
\\
&+&
\foursevenths\Delta^{4}\,
	\big[\mu(\nu)\,B(\omega,\nu,\omega-\nu)-\mu(\omega)\,B(-\nu,-\omega,\omega-\nu)\big]
\nonumber\\
&+&
\Delta^{2}\,
	\big[\mu(\nu)\,C(\omega,\nu,\omega-\nu)-\mu(\omega)\,C(-\nu,-\omega,\omega-\nu)\big] 
\nonumber\\
&-&
\twothirds\Delta^{3}\,\mu(\omega)\mu(\nu)
	\big[E+F+G+\foursevenths{H}\,\Delta^{2}\big](\omega,\nu,\omega-\nu) 
\,.
\nonumber
\eer
Thus, the constitutive equation can be written as (\ref{au}),
\be
\label{nonlinear1}
\hspace*{-3mm}
\delta\Pi(\omega) = \frac{1}{(1+F^{s}_{0})\Delta^{2}}
	\sum_{M}\int d\nu\,A_{M}(\omega,\nu,\omega-\nu)\,\delta n(\nu)\,D^{+}_{2,M}(\omega-\nu) 
\,, 
\ee 
with $A_M$ given by Eqs. (\ref{ax}) and (\ref{fp}). It follows directly from Eqs. (\ref{fd}) and
(\ref{fp}) that the coupling function $A_{M}$ satisfies
\be\label{bb}
A_{M}(\omega,\nu,\omega-\nu) = A_{M}(-\nu,-\omega,\omega-\nu) 
\,.
\ee

In a similar way the coupling functions $\Xi^{(2)}_{M}$ in Eq. (\ref{ek}) 
are found by substituting Eq. (\ref{fc}) into
\be
\label{fq}
\hspace*{-5mm}
\onesixth\bar{\lambda}(\omega)
	\left(\omega^{2}+2i\omega\Gamma-\omega_{M}^{2}-c_{M}^{2}q^{2}\right)D^{+}_{2,M}
=
-	\int\frac{d\Omega_{\vec{p}}}{4\pi}
		t_{ij}^{2,M^{*}}\,\vhat{p}_{i}\delta\vec{f}_{j}^{+(2)}(\omega) 
\,.
\ee 
Then to leading order in $1/s^{2}$
\be
\label{fr}
\Xi^{(2)}_{M}(\omega,\nu,\omega-\nu) = \frac{1}{1+F^{s}_{0}}
	\left\vert\frac{c_{1}(\vec{q}-\vec{s})}{\omega-\nu}\right\vert^{2}\,
	A_{M}(\nu,-\omega,\nu-\omega)^{*}
\,,
\ee 
so that the second constitutive equation becomes,
%
\ber
\label{nonlinear2}
&&
\hspace*{-8mm}
\lambda(\omega)\,
	\left[\omega^{2}+2i\omega\Gamma-\omega_{2^{+},M}(q)^{2}\right] D^{+}_{2,M} 
\\
&&
=
\frac{6}{N(E_{f})^{2}}
\int d\nu\,\left\vert\frac{c_{1}(\vec{q}-\vec{s})}{\omega-\nu}\right\vert^{2}\, 
	A_{M}(\nu-\omega,\nu,-\omega)^{*}\,
	\delta n(\nu)\,\delta n(\omega-\nu) 
\,.
\nonumber
\eer
Equations (\ref{nonlinear1}) and (\ref{nonlinear2}), together with the wave equation, Eq.
(\ref{aj}), describe the weak nonlinear interaction of the $J=2^{+}$ modes with the density
fluctuations of zero sound.

Before we discuss the consequences of these nonlinear constitutive equations we give a physical
explanation of why the same coupling function $A_{M}$ appears in both Eqs. (\ref{nonlinear1})
and (\ref{nonlinear2}), and also satisfies the identity given in Eq. (\ref{bb}). Consider three
wave packets with carrier frequencies ($\omega_1$, $\omega_2$, $\omega_3$) all propagating along
the $z$-axis. The density and order parameter fluctuations can be written
\ber\label{bc}
\delta{n}(z,t) &=& N_{1}(z,t)\,e^{i(\omega_{1}t-q_{1}z)} 
				+ N_{2}(z,t)\,e^{i(\omega_{2}t-q_{2}z)}
				+ \mbox{c.c.}
\,,
\nonumber\\
D_{2,M}^{+}(z,t) &=& E_{M}(z,t)\,e^{i(\omega_{3}t-q_{3}z)}
				 + \mbox{c.c.}
\,.
\eer
We assume that the sound pulses are sufficiently short that the quasiparticle damping can be
neglected ({\it i.e.} $\frac{1}{N_{j}}\,\pder{N_{j}}{t}\gg\Gamma\,,j=1,2$), that only one of the
five $J=2^+$ modes is excited by the two sound waves, and that the three-wave resonance
conditions for this mode hold: $\omega_{M}=\omega_{3}=\omega_{1}+\omega_{2}$,
$\vec{q}_{3}=\vec{q}_{1}+\vec{q}_{2}$.
Substituting Eq. (\ref{bc}) into Eqs. (\ref{nonlinear1}), (\ref{nonlinear2}), and (\ref{aj});
and making a slowly varying envelope approximation, in which the amplitudes $N_{1}$, $N_{2}$ and
$E_{M}$ are assumed to vary on time and length scales that are large compared to
$1/\omega_{j}\sim 1/\Delta$ and $1/q_{j}\sim c_{1}/\Delta$, we obtain the following
equations
%
\ber
\label{bd}
i\left(\pder{}{t}-v_{1}\pder{}{z}\right)\,N_{1}
=
\frac{\omega_{1}A_{M}(\omega_{1},-\omega_{2},\omega_{3})}
	 {\Delta(1+F_{0}^{s})}\,
	E_{M}\,N_{2}^{*}
\,,
\nonumber\\
i\left(\pder{}{t}-v_{2}\pder{}{z}\right)\,N_{2}
=
\frac{\omega_{2}A_{M}(\omega_{2},-\omega_{1},\omega_{3})}
	 {\Delta(1+F_{0}^{s})}\,
	E_{M}\,N_{1}^{*}
\,,
\nonumber\\
\frac{i}{6}\bar\lambda(\omega_{3})
\left(\pder{}{t}-v_{M}\pder{}{z}\right)\,E_{M}
=
\frac{A_{M}(-\omega_{1},-\omega_{2},-\omega_{3})^{*}}
	 {2\Delta\omega_{3}N(E_f)^{2}}\,
	N_{1}\,N_{2}
\,,
\label{bd3}
\eer
where $v_{1}$ $v_{2}$ and $v_{M}$ are the group velocities of the sound waves and collective
mode, respectively.\footnote{Under certain conditions these equations admit {\it soliton}
solutions. However, these solitons are quite different from the zero-sound solitons in $^3$He-B
considered by Sauls \cite{sau81d}.}

In a three-wave resonance the total number of quanta should be conserved, {\it i.e.} for each
phonon of frequency $\omega_{1}$ that is destroyed a second phonon of frequency $\omega_{2}$ is
also destroyed and a real squashon is created. Thus, we expect a continuity equation of the form
\be
\label{be}
 \left(\pder{}{t}-v_{1}\pder{}{z}\right)\frac{U_{1}}{\omega_{1}} = 
 \left(\pder{}{t}-v_{2}\pder{}{z}\right)\frac{U_{2}}{\omega_{2}} =
-\left(\pder{}{t}-v_{M}\pder{}{z}\right)\frac{U_{M}}{\omega_{3}}
\,,
\ee
to hold, where $U_{j}/\omega_{j}$ is the number density of quanta of the mode
with frequency $\omega_j$. It is straight forward to show (see {\it e.g.} 
Ref. \cite{bay78}) that the energy density of zero-sound phonons is
\be
\label{bf}
U_{j}=\frac{N(E_{f})\vert N_{j}\vert^{2}}{1+F_{0}^{s}} 
\,, 
\ee 
while that of the real squashons is
\be
\label{bg}
U_{M}=\twothirds N(E_{f})\bar{\lambda}(\omega_{M})\,\omega_{M}^{2}\,\vert D^{+}_{M}\vert^{2}
\,.
\ee
The conservation equations for the quanta follow directly from the three equations for the mode
amplitudes and the identity given in (\ref{bb}). Equations (\ref{be}) also occur in nonlinear
optics \cite{shen84} where they are called the Manley-Rowe relations after similar equations
derived for nonlinear microwave circuits \cite{man56}.
\vspace*{3mm}
\sec{Stimulated Raman scattering and two-phonon absorption by the 
         J=2$^+$ modes}{Stimulated}
   
We now consider the three-wave resonance equations for two density waves interacting with the
$J=2^{+}$ modes in more detail. The dynamics of the three-wave resonance is described by the
wave equation, Eq. (\ref{aj}), which each of the two sound waves satisfies, with the
constitutive equations given by Eq. (\ref{nonlinear1}) and the equation of motion for the the
$J=2^{+}$ order parameter modes, Eq. (\ref{nonlinear2}). The density fluctuation resulting from
two sound waves is
\be
\label{fs}
\delta{n}(\vec{x},t) = \tilde{N}_{1}(\vec{x},t) + \tilde {N}_{2}(\vec{x},t) + \mbox{c.c.}
\,,
\ee
\be
\label{ft}
\tilde{N}_{j}(\vec{x},t) = N_{j}(\vec{x},t)\,e^{i(\omega_{j}t-\vec{q}_{j}\cdot\vec{x})}\,;
\quad j=1,2
\,.
\ee 
If the wave amplitudes vary slowly on the time scale of the collective mode lifetime,
$1/\Gamma$, then Eq. (\ref{nonlinear2}) can be solved for $D_{2,M}^{+}(\vec{x},t)$. The
collective mode amplitude then contains terms oscillating with frequencies
$\omega_{1}+\omega_{2}$, $\omega_{1}-\omega_{2}$, $2\omega_{1}$, $2\omega_{2}$ and $\omega=0$.
One of these terms will dominate if its frequency 
$\omega_{a}$ and wavevector $\vec{q}_{a}$ satisfy the resonance condition, 
\be
\label{fu}
\omega_{a}^{2} = \omega_{M}^{2} + c^{2}_{M}\,q_{a}^{2}
\,.
\ee 
First consider the case where the sum or difference of the frequencies of the two sound waves
is approximately equal to one of the collective mode frequencies. The collective mode
amplitudes are then given by
%
\be
\label{fv}
D_{2,M}^{+}(\vec{x},t)
=
\frac{A_{M}(\omega_{1},-\omega_{2},\omega_{1}+\omega_{2})^{*}\,
	  \tilde{N}_{1}(\vec{x},t)\,\tilde{N}_{2}(\vec{x},t)}
	 {N(E_f)^2\,\Delta\,\,\phi^{M}(\vec{q}_{1}+\vec{q}_{2},\omega_{1}+\omega_{2})}
+ \mbox{c.c.}
\,,
\ee 
for $\omega_{1}+\omega_{2}\simeq\omega_{M}$, or
\be
\label{fw}
D_{2,M}^{+}(\vec{x},t)
=
\frac{A_{M}(\omega_{1},\omega_{2},\omega_{1}-\omega_{2})^{*}\,
	  \tilde{N}_{1}(\vec{x},t)\,\tilde{N}_{2}(\vec{x},t)}
	 {N(E_f)^2\,\Delta\,\,\phi^{M}(\vec{q}_{1}-\vec{q}_{2},\omega_{1}-\omega_{2})}
+ \mbox{c.c.}
\,,
\ee 
if $\omega_{1}-\omega_{2}\simeq\omega_{M}$. The denominator is defined by
\be
\label{fx}
\phi^{M}(q,\omega)
=
\onesixth\bar{\lambda}(\omega)
	\left[\omega^{2} + 2i\Gamma\omega - \omega_{M}^{2} - c_{M}^{2}q^{2}\right] 
\,.
\ee 
If these solutions for the collective mode amplitudes and the density fluctuation given in Eq.
(\ref{fs}) are substituted into the expression (\ref{nonlinear1}) for the stress, then the
latter contains terms oscillating with frequency $\omega_{1}$, $\omega_{2}$,
$\omega_{1}+2\omega_{2}$, and $2\omega_{1}+\omega_{2}$ when
$\omega_{1}+\omega_{2}\simeq\omega_{M}$; and with frequency $\omega_{1}$, $\omega_{2}$,
$2\omega_{1}-\omega_{2}$ and $2\omega_{2}-\omega_{1}$ when
$\omega_{1}-\omega_{2}\simeq\omega_{M}$.
It is shown below that the terms with frequency $\omega_{1}$ and $\omega_{2}$ dominate all
others. The total stress can be written
\be
\label{fy}
\delta\Pi(\vec{x},t) 
= \delta\Pi_{L}(\vec{x},t) 
+ \delta\Pi_{1}(\vec{x},t) 
+ \delta\Pi_{2}(\vec{x},t) 
\,,
\ee 
where the first term contains the contribution from the linear coupling to the collective modes
and the second and third terms are nonlinear contributions with frequencies $\omega_{1}$ and
$\omega_{2}$, respectively. The latter are given by
\ber
\label{fz}
\hspace*{-11mm}
\frac{\delta\Pi_{1}}{\tilde{N}_{1}}	&=& 
	\left[\chi^{(3)}(\omega_{1},-\omega_{2},\omega_{1}+\omega_{2}) 
		+
		  \chi^{(3)}(\omega_{1},\omega_{2},\omega_{1}-\omega_{2})
	\right]
\vert N_{2}\vert^{2}
\,,
\nonumber\\
\hspace*{-11mm}
\frac{\delta\Pi_{2}}{\tilde{N}_{2}}	&=& 
	\left[\chi^{(3)}(\omega_{2},-\omega_{1},\omega_{1}+\omega_{2}) 
		+
		  \chi^{(3)}(\omega_{2},\omega_{1},\omega_{2}-\omega_{1})
	\right]
\vert N_{1}\vert^{2}
\,,
\eer
where the nonlinear susceptibility $\chi^{(3)}$ is given by
\be
\label{ga}
\hspace*{-3mm}
\chi^{(3)}(\omega,\nu,\omega-\nu)
\negthickspace = \negthickspace
\frac{1}{N(E_f)^2\Delta^{2}(1+F_{0}^{s})}			\negthickspace
	\left\vert\frac{c_{1}q}{\omega}\right\vert^{2}	\negthickspace
	\sum_{M}\negthickspace\frac{\vert A^{M}(\omega,\nu,\omega-\nu)\vert^{2}}
				 			   {\phi^{M}(\vec{q}-\vec{s},\omega-\nu)}.
\ee
We temporarily neglect the mode dispersion in Eq. (\ref{ga}) and substitute Eq. (\ref{fy}) with
Eq. (\ref{fz}) into the wave equation Eq. (\ref{aj}). Then if the wave amplitudes $N_{1}$ and
$N_{2}$ are nearly static and vary on a length scale that is long compared to the wavelength of
sound,
\ber
\label{gb}
\left(\pder{}{z} + \alpha_{1}\right)\,N_{1}(z) &=& iq_{1}\,\chi{(3)}_{\pm}\,|N_{2}|^2\,N_{1}
\nonumber\\
\left(\pder{}{z} + \alpha_{2}\right)\,N_{2}(z) &=& iq_{2}\,\chi{(3)}_{\pm}\,|N_{1}|^2\,N_{2}
\,,
\eer
for parallel waves propagating in the $z$ direction, where $\alpha_{1}$ ($\alpha_{2}$) is the
linear attenuation coefficient of sound with frequency $\omega_{1}$ ($\omega_{2}$).

The number density of zero-sound phonons with frequency $\omega_{j}$ is given by
$n_{j}=U_{j}/\omega_{j}\,(j=1,2)$, where $U_{j}$ is the sound energy density given by Eq.
(\ref{bf}). This latter expression can be combined with Eq. (\ref{gb}) for the sound wave
envelopes to give
\ber\label{gc}
\left(\pder{}{z} + 2\alpha_{1}\right)\,n_{1} = -  K_{\pm}\,n_{1}\,n_{2}
\,,
\nonumber\\
\left(\pder{}{z} + 2\alpha_{2}\right)\,n_{2} = \mp K_{\pm}\,n_{1}\,n_{2}
\,,
\eer
where
\ber
\label{gd} 
\hspace*{-5mm}
K_{\pm}
&=&
	\left(\frac{1}{1+F_{0}^{s}}\right)^2
	\frac{\omega_{1}\omega_{2}}{c_{1}U_{c}}\,
	\mbox{Im}\,\sum_{M}
		\left\{\frac{\vert A^{\pm}_{M}\vert^{2}}{\phi_{M}(\vec{0},\omega_{1}\pm\omega_{2})}
		\right\}
\,,
\nonumber\\
\hspace*{-5mm}
\phi_{M}
&=&
	\onesixth\bar{\lambda}(\omega)
			 \left[\omega^{2} + 2i\omega\Gamma - \omega_{M}^{2} \right] 
\,,
\nonumber\\
\hspace*{-5mm}
A_{M}^{\pm}
&=&
	A_{M}(\omega_{1},\mp\omega_{2},\omega_{1}\pm\omega_{2}) 
\,,
\eer 
and $U_{c}=\onehalf N(E_{f})\Delta^{2}$ is proportional to the superfluid condensation energy
density. Equations similar to Eq. (\ref{gc}) occur in nonlinear optics; the plus and minus
signs corresponding to {\it two-photon absorption} and {\it stimulated Raman scattering},
respectively (see {\it e.g.} Ref. \cite{shen84}, pp. 148 and 203). In the limit of negligible
linear attenuation $(\alpha_{1} = \alpha_{2} = 0)$ Eqs. (\ref{gc}) imply that $n_{1}(z)\pm
n_{2}(z)$ is constant. These are the Manley-Rowe relations discussed in Sec. (\ref{Nonlinear}).
The full analytic solution of Eqs. (\ref{gc}) in this limit can be found in Ref. \cite{shen84}.

If one of the sound waves is much more intense than the other then Eqs. (\ref{gc}) can be also
solved analytically.$\negmedspace$ This limit is known as the {\it undepleted pump
approximation} because the number of phonons in the pump wave can be taken to be constant. If
the high-frequency wave is the high-intensity pump wave ({\it i.e.} $\omega_{1} >
\omega_{2}\,,n_{1}\gg n_{2}$), then the solution to Eq. (\ref{gc}) is
\be
\label{gf}
n_{2}(z) = n_{2}(0)\,e^{(\mp{g}_{\pm}-2\alpha_{2})z}
\,, 
\ee 
where $g_{\pm}=K_{\pm}{n}_{0}$ and $n_{1}(z)=n_{0}$. In the case of stimulated Raman scattering
$(\omega_{1}-\omega_{2}\simeq\omega_{M})$, the wave with lower frequency, the {\it Stokes
wave}, can be amplified if the pump wave is of sufficient intensity so that $g_{-} >
2\alpha_{2}$. In the opposite limit, where the lower frequency wave is the high-intensity pump
wave $(n_{2}=n_{0}\gg{n}_{1})$, the solution of Eq. (\ref{gc}) gives
\be
\label{gg}
n_{1}(z) = n_{1}(0)\,e^{(-g_{\pm} - 2\alpha_{1})z}
\,.
\ee 
The attenuation of the high-frequency wave for the case
$(\omega_{1}-\omega_{2}\simeq\omega_{M})$ is known in optics as the {\it inverse Raman effect}.

In the above discussion, the mode dispersion in the third-order susceptibility, given in Eq.
(\ref{ga}), was neglected in order to point out similarities with nonlinear optics. We now show
that, as first pointed out by Koch and W\"olfle \cite{koc81} for the linear acoustic response,
the mode dispersion effects the size and width of the peaks in the sound attenuation due to the
collective modes.

The part of the stress with frequency $\omega_{1}$ due to nonlinear effects can be written as
\ber
\label{gh}
\hspace*{-10mm}
\frac{\delta\Pi(\omega_{1})}{N_{1}(\omega_{1})}
	&=& 
	\left(\frac{c_{1}q_{1}}{\omega_{1}}\right)^{2} 
	\left(\frac{c_{1}q_{2}}{\omega_{2}}\right)^{4}
\nonumber\\
	&\times&
	\sum_{M}
	\frac{\Upsilon^{+}_{M}}{\phi_{M}(\vec{q}_{1}+\vec{q}_{2},\omega_{1}+\omega_{2})}
	+
	\frac{\Upsilon^{-}_{M}}{\phi_{M}(\vec{q}_{1}-\vec{q}_{2},\omega_{1}-\omega_{2})}
\,,
\eer
where
\be
\label{gk}
\Upsilon^{\pm}_{M}=	\frac{1}{50(1+F_{0}^{s})^{2}}
					\vert\tilde{A}(\omega_{1},\mp\omega_{2},\omega_{1}\pm\omega_{2})
					Z_{M}(\vhat{q}_{1},\vhat{q}_{2})\vert^{2}\,
					\frac{U_{2}}{U_{c}}
\,.
\ee
If we neglect the dispersion of the pump wave, with frequency $\omega_2$ and (nearly constant)
energy density $U_2$, {\it i.e.} $\omega_2 = c_1 q_2$, then Eq. (\ref{gh}) has a similar form
to that for the linear coupling of sound to the collective modes except that here the coupling
is proportional to energy density of the pump wave. Combining Eq. (\ref{gh}) with the wave
equation, Eq. (\ref{aj}), we set $q_{1}=k_{1}+i\alpha_{1}$ and solve for $\delta v_{1}$, the
shift in the phase velocity $v_{1}=\omega_{1}/k_{1}$ and $\delta\alpha_{1}$, the nonlinear
contribution to the attenuation. The change in phase velocity due to the nonlinear coupling to
collective modes is to leading order in $1/s^{2}$
\be\label{gl}
\frac{\delta v_{1}}{c_{1}}
=
\sum_{M}\,
		\Upsilon^{+}_{M}\,\cF_{M}(\omega_{1}+\omega_{2}) 
+ 		\Upsilon^{-}_{M}\,\cF_{M}(\omega_{1}-\omega_{2}) 
\,,
\ee 
where
\be
\label{gm}
\cF_{M}(\omega) = \mbox{Re}\,\phi_{M}(\omega)^{-1}
=
\frac{6(\omega^{2}-\omega^{2}_{M})}
	 {\bar\lambda(\omega)\left[(\omega^{2}-\omega_{M}^{2})^{2}+(2\omega\Gamma)^{2}\right]}
\,.
\ee 
Solving for $\delta\alpha_{1}$ to leading order in $1/s^{2}$ we obtain
\be
\label{gn}
\frac{\delta\alpha_{1}}{k_{1}} =
\sum_{M}\,\cG_{M}(\Upsilon^{+}_{M},\omega_{1}+\omega_{2}) 
+ 		\,\cG_{M}(\Upsilon^{-}_{M},\omega_{1}-\omega_{2}) 
\,,
\ee
where
\be
\label{gp}
\cG_{M}(X,\omega)
\negthickspace = \negthickspace
\frac{\omega\Gamma{X}} 
	 {\onetwelveth\bar\lambda(\omega)
	 \left[(\omega^{2}-\omega_{M}^{2})^{2} + (2\omega\Gamma)^{2}\right] 
     \negthickspace + \negthickspace 
     \big(\frac{c_{M}}{c_{1}}\big)^{2}X[\omega_{1}\omega 
	 +(\omega^{2}-\omega_{M}^{2})]}
\,.
\ee 
It follows from Eqs. (\ref{gl}) and (\ref{gn}) that the phase velocity and attenuation have
well defined signatures when $\omega_{1}+\omega_{2}=\omega_{M}$ and
$\omega_{1}-\omega_{2}=\omega_{M}$, corresponding to two-phonon absorption and Raman
scattering, respectively. The size of the peaks in the sound attenuation are given by
\be
\label{pq}
\left(\frac{\alpha_{1}}{k_{1}}\right)_{\mbox{\footnotesize{peak}}}
=
\frac{\Gamma\Upsilon^{\pm}_{M}}
     {\onethird\bar{\lambda}(\omega_{1}\pm\omega_{2})(\omega_{1}\pm\omega_{2})\,\Gamma^{2}
     + (c_{M}/c_{1})^{2}\omega_{1})\Upsilon^{\pm}_{M}}
\,,
\ee 
and the widths $W$ of the peaks are given by 
\be
\label{gr}
W^2 = 	\bigg\vert \Gamma^2 
	+ 
		\frac{3\omega_{1}}{(\omega_{1}\pm\omega_{2})\,
		\bar{\lambda}(\omega_{1}\pm\omega_{2})}
		\left(\frac{c_{M}}{c_{1}}\right)^{2}\,
		\Upsilon^{\pm}_{M}
		\bigg\vert
\,,
\ee 
where $\Gamma$ is the width due to quasiparticle damping and $W$ depends on the collective mode
velocity, $c_M$, and the strength of the nonlinear coupling to sound.

\begin{figure}[!]
\includegraphics[width=0.95\textwidth]{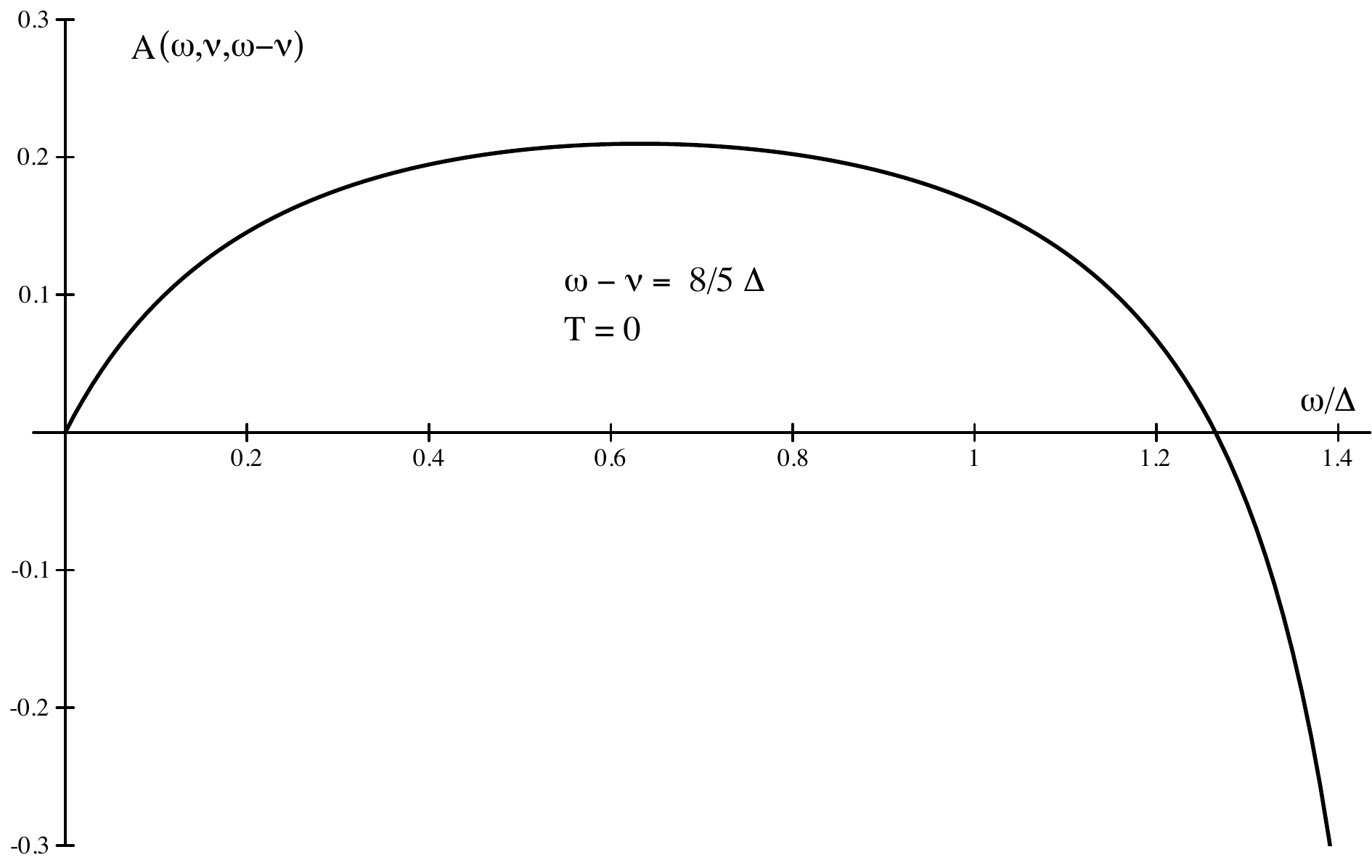}
\caption{
The dependence of the nonlinear coupling constant $\tilde{A}(\omega,\nu,\omega-\nu)$ on the
frequency $\omega$ for $\omega-\nu=\sqrt{8/5}\Delta$ and zero temperature.
}
\label{Figure:Nonlinear-Coupling}
\end{figure}

\begin{figure}[!]
\includegraphics[width=0.95\textwidth]{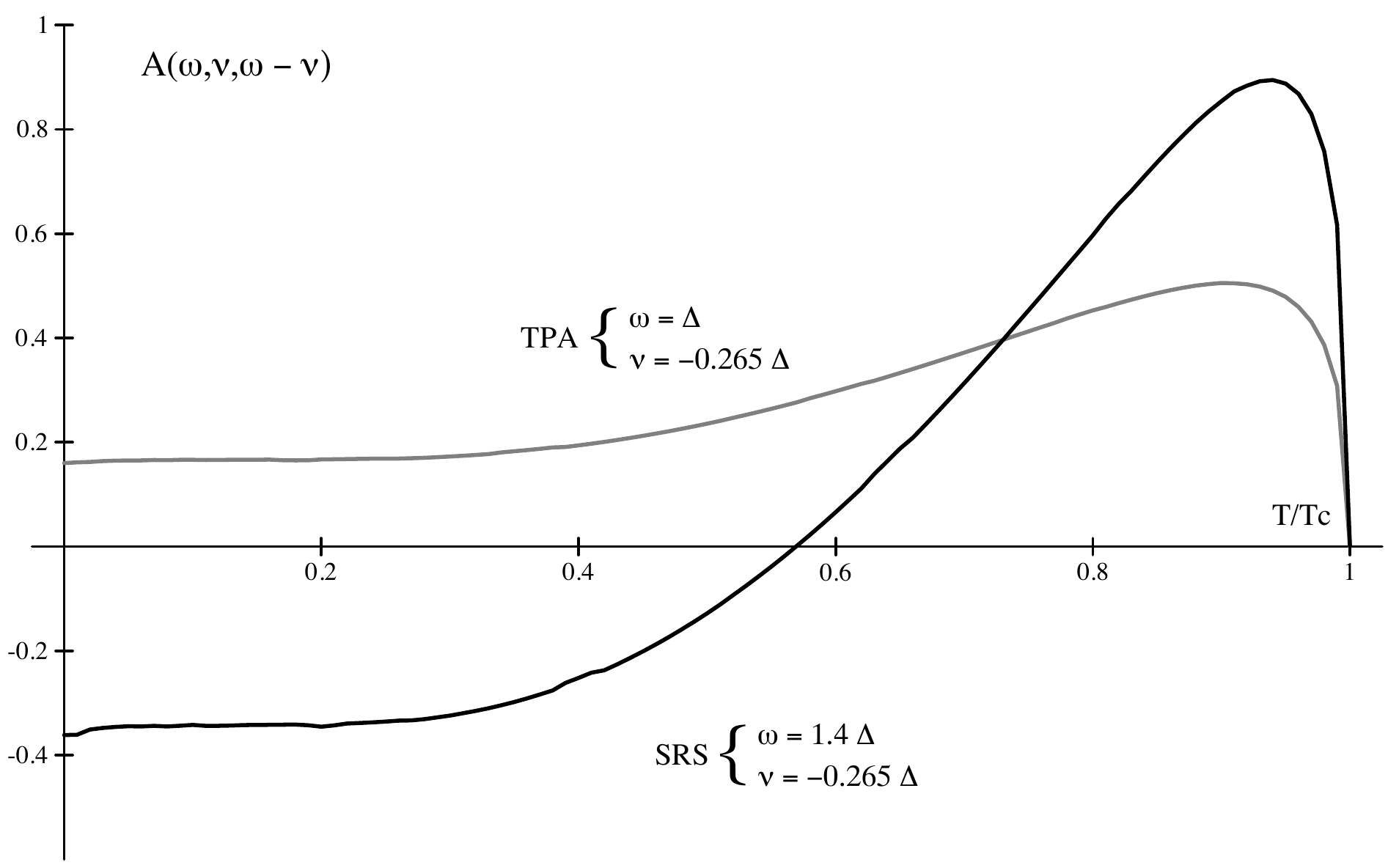}
\caption{
The temperature dependence of the nonlinear coupling constant
$\tilde{A}(\omega,\nu,\omega-\nu)$ for 
$\omega-\nu=\sqrt{8/5}\Delta$; a) $\omega=\Delta$ (two-phonon absorption), 
b) $\omega=1.4\Delta$ (stimulated Raman scattering).
}
\label{Figure:Nonlinear-Coupling_vs_T}
\end{figure}

We now consider the dependence of the nonlinear coupling function $A_{M}$, given by Eq.
(\ref{ax}), on various parameters. The temperature and frequency dependences of $A_{M}$, and
therefore $\Upsilon_{M}^{\pm}$, are contained in the factor
$\tilde{A}(\omega_{1},\mp\omega_{2},\omega_{1}\pm\omega_{2})$. The dependence of the
dimensionless coupling function $\tilde{A}(\omega,\nu,\omega-\nu)$ on the frequency $\omega$ at
zero temperature, when $\omega-\nu=\sqrt{8/5}\Delta$, is shown in 
Fig. \ref{Figure:Nonlinear-Coupling}.
Note that $\tilde{A}$ is zero at $\omega=0$ and $\sqrt{8/5}\Delta$. The temperature dependence
of $\tilde{A}(\omega,\nu,\omega-\nu)$ with $\omega-\nu=\sqrt{8/5}\Delta$ for two-phonon
absorption, with $\omega=\Delta$, is shown in Fig. \ref{Figure:Nonlinear-Coupling_vs_T}(a); and
for Raman scattering, with $\omega=1.4\Delta$, in Fig. \ref{Figure:Nonlinear-Coupling_vs_T}(b).
Note that the coupling function for Raman scattering vanishes near $T=0.6\,T_{c}$. Since both
coupling functions are approximately constant at low temperatures $T\leq\,0.4\,T_{c}$ it
follows from Fig. \ref{Figure:Nonlinear-Coupling} that in this temperature range nonlinear
effects will be relatively small if either $\omega_{1}$ or
$\omega_{2}\leq\,0.05\Delta\simeq\,0.1\,k_{\text{B}}T_{c}$. Outside this range of frequencies,
and away from $ \omega_{1} $ or $\omega_{2}\simeq\sqrt{12/5}\Delta$ (where the pump and signal
waves are resonant with the $J=2^{-}$ mode), the function $\tilde{A}(\omega,\nu,\omega-\nu)$
depends weakly on frequency.

The angular dependence of $\Upsilon^{\pm}_{M}$ is determined by
$Z_{M}(\vhat{q}_{1},\vhat{q}_{2})$ given in Eq. (\ref{az}), which in turn depends on
$\vhat{z}$, the quantization axis of the modes. This direction is determined by the relative
size of the Zeeman energy, $\gamma H$, and the collective mode dispersion energy $Q_{\pm}\equiv
v_{f}^{2}(\vec{q}_{1}\pm\vec{q}_{2})^{2}/\Delta$. If $Q_{\pm}\gg\gamma{H}$ then $\vhat{z}$ is
parallel to $\vec{q}_{1}\pm\vec{q}_{2}$, while if $Q_{\pm}\gg\gamma{H}$, then $\vhat{z}$ is
parallel to the ``rotated'' magnetic field $\ul{\vec{R}}(\vhat{n},\theta)\cdot\vec{H}$. The
angular dependence reduces to a simple form when the two wavevectors are either parallel or
antiparallel,
\be
\label{gs}
\vert Z_{M}(\vhat{q},\pm\vhat{q})\vert^{2}
=
\frac{8\pi}{135}
\vert{\cY_{2M}(\vhat{q})}\vert^{2}
\,,
\ee 
where $\cY_{2M}(\vhat{q})$ is a spherical harmonic. 
For $\vhat{q}=\pm\vhat{z}\,\vert\,Z_{M}\,\vert^{2}=\twotwentysevenths\,\delta_{M,0}$.

The sizes of the anomalies in the sound velocity and attenuation depend on the magnitudes of
$\Upsilon^{\pm}_{M}$ and $\Gamma$. The height of the attenuation peaks increases linearly with
$\Upsilon^{\pm}_{M}/\Gamma$ provided that $\Gamma\ll\Gamma_{d}$. Clearly
$\Upsilon^{\pm}_{M}/\Gamma$ will be largest for large pump wave energy densities, low pressures
(where $F_{0}^{s}$ and $\Gamma$ are smallest) and low temperatures, where $\Gamma$ is small.
However, when the ratio $\Upsilon^{\pm}_{M}/\Gamma$ becomes sufficiently large that $\Gamma$ is
comparable to the dispersion width $\Gamma$ the heights of the attenuation peaks tend to
limiting values and their widths increase as $\Upsilon^{\pm}_{M}$ increases.
   
The damping of the order parameter collective modes due to quasiparticle collisions. $\Gamma$ is
an increasing function of pressure and temperature. The expression for $\Gamma$ derived by
W\"olfle \cite{wol76} depends in a complicated way on the quasiparticle scattering amplitudes,
which are not well known. However, the lifetime of the modes, $1/\Gamma$, is roughly determined
by the quasiparticle lifetime, $\tau$, which at very low temperatures $(T\ll T_{c})$ can be
written
\be
\label{gu}
\frac{1}{\tau} = A(P)\,\left(\frac{T_{c}}{T}\right)^{3/2}\,e^{-\Delta /T}
\,,
\ee 
where $A$ is a function of pressure \cite{ein84}. Although this expression is not valid at
higher temperatures it is sufficiently accurate for the semi-quantitative results required here.
We assume that the pressure dependence of $A(P)$ is that of $1/\tau_{N}$, where $\tau_{N}$ is
the quasiparticle lifetime in the normal state,
\ber
\frac{1}{\tau_{N}}=\frac{(T/\mbox{mK})^{2}}{0.6-0.01\,P}
\,,
\nonumber
\eer
where $P$ is the pressure in bars \cite{ein84}. Thus, we assume 
\be
\label{gx}
A(P)=\frac{A_{0}\,(T_{c}/\mbox{mK})^{2}}{0.6-0.01\,P}
\,,
\ee
where $A_{0}$ is determined by fitting Eq. (\ref{gu}) to the data given by Halperin \cite{hal82}
for the lifetime of the $J=2^{-}$ mode at 13 bar. Expressions (\ref{gu}) and (\ref{gx}) are used
for the mode lifetime $\Gamma$ in the calculations discussed below. In general, as the pressure
increases the nonlinear features in the temperature become smaller and broader.

\begin{figure}[!]
\includegraphics[width=0.95\textwidth]{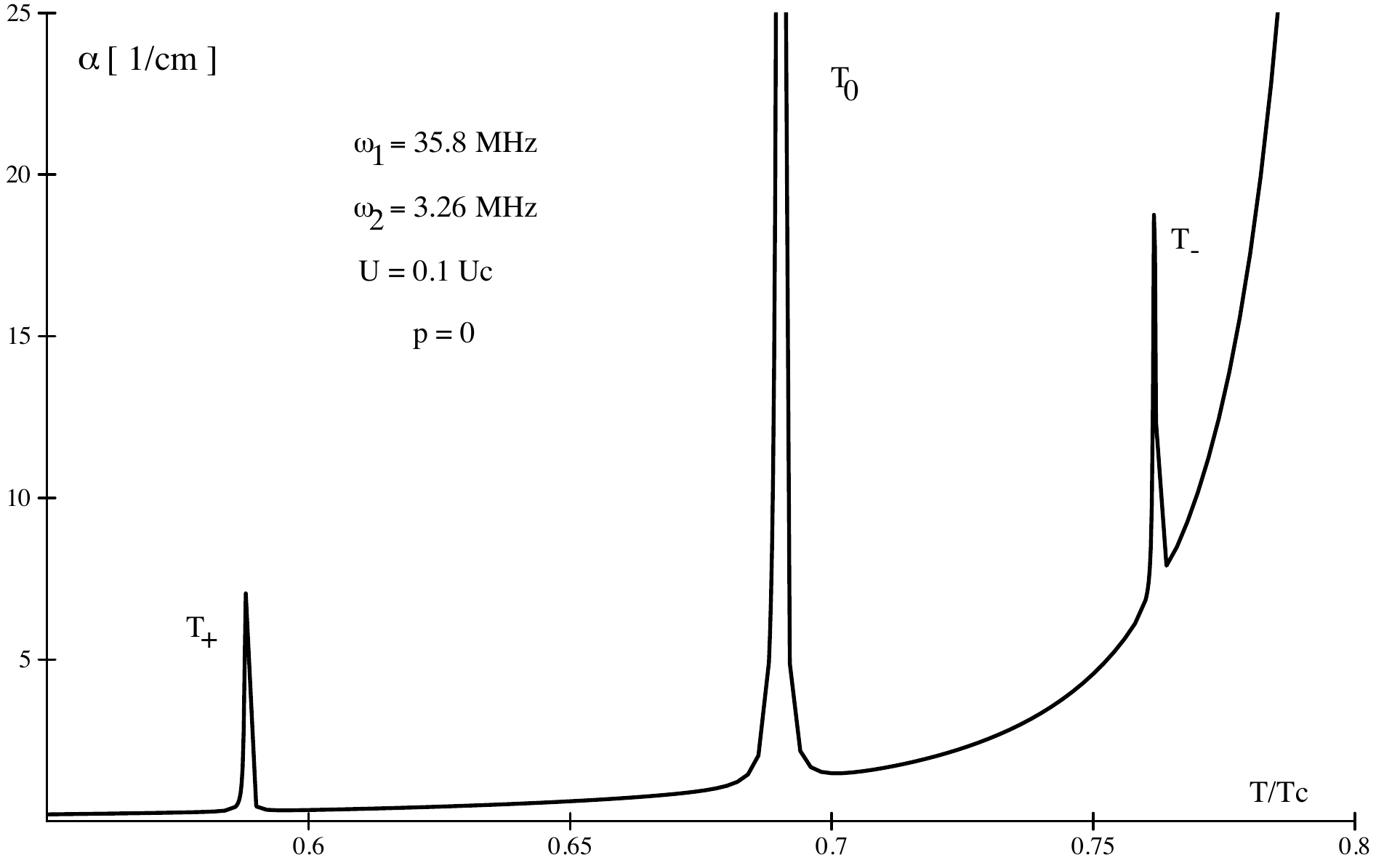}
\includegraphics[width=0.95\textwidth]{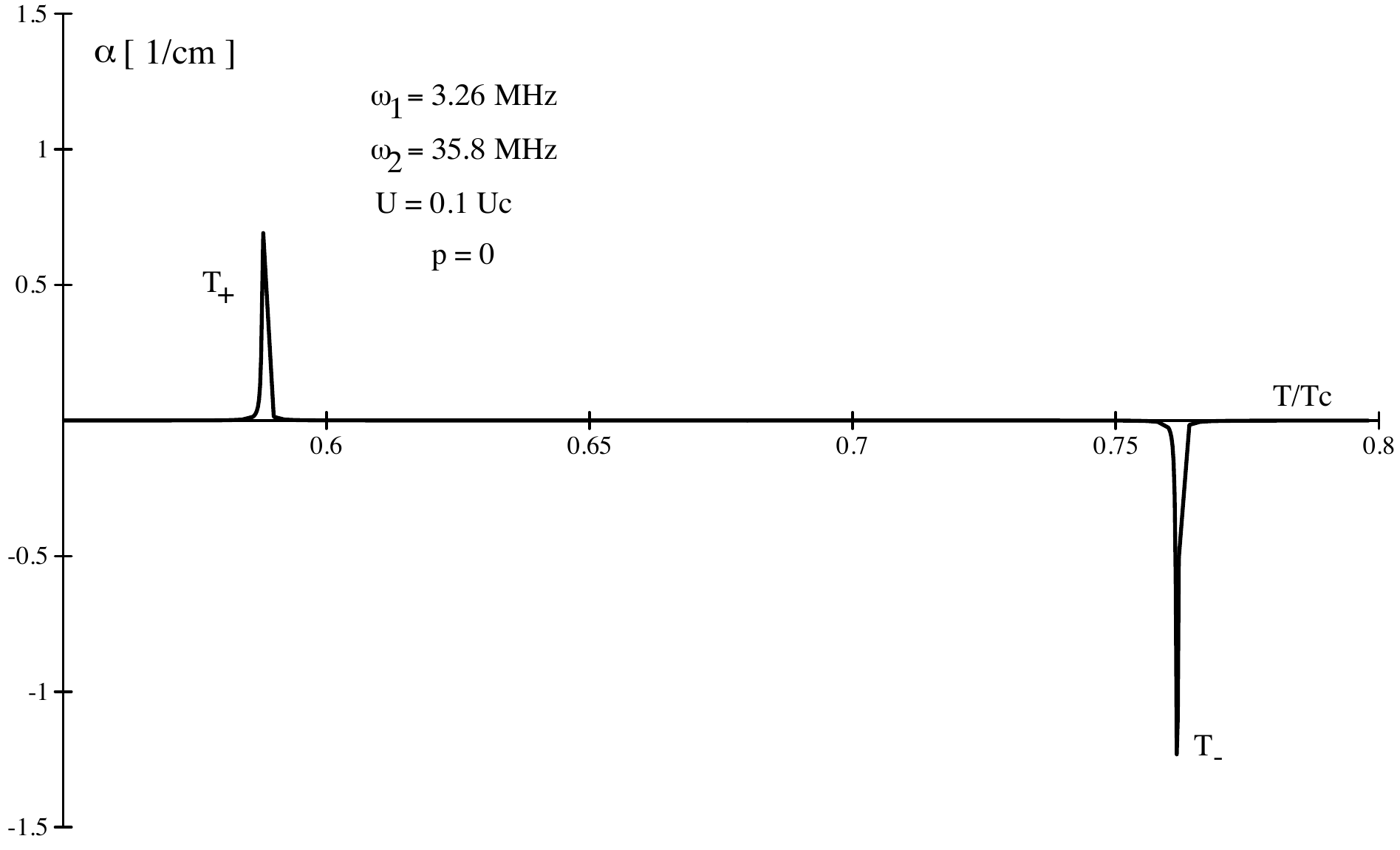}
\caption{
The predicted temperature dependences at zero pressure of the attenuation/amplification (in
units of $cm^{-1}$) of a signal zero-sound wave with frequency $\omega_{1}$ in the presence of a
parallel pump wave of frequency $\omega_{2}$ and energy density $U=0.1\,U_{c}$;
a) $\frac{\omega_{1}}{2\pi}=35.8\,\mbox{MHz}$, $\frac{\omega_{2}}{2\pi}=3.26\,\mbox{MHz}$,
b) $\frac{\omega_{1}}{2\pi}=3.26\,\mbox{MHz}$, $\frac{\omega_{2}}{2\pi}=35.8\,\mbox{MHz}$. 
The peaks at $T/T_{c}\sim 0.58$ and $0.76$ in both cases are due to nonlinear resonances. The
central peak at $T/T_{c}=0.68$ is the linear absorption from the $J=2^{+}$ mode.
}
\label{Figure:Nonlinear-Attenuation_vs_T}
\end{figure}

The attenuation as a function of temperature, at zero pressure,
for a sound wave of frequency $35.8\,\mbox{MHz}$ 
$(\hbar\omega_{1}/k_{\text{B}}T_{c}=1.87)$ in the presence of a 
parallel pump wave with frequency $3.26\,\mbox{MHz}$ 
$(\hbar\omega_{2}/k_{\text{B}}T_{c}=0.17)$ and energy density 
$U/U_{c}=0.1$ is shown in Fig. \ref{Figure:Nonlinear-Attenuation_vs_T}(a).
The large central peak is due to the linear coupling of the $J=2^{+}$ mode to the higher
frequency wave and occurs at a temperature $T_{0}$ such that
$\sqrt{8/5}\,\Delta(T_{0})=\hbar\omega_{1}$. The two peaks to the left and right of the central
peak are due to two-phonon absorption and the inverse Raman effect, respectively, and occur at
temperatures $T_{+}\simeq 0.58T_{c}$ and $T_{-}\simeq 0.76T_{c}$, determined by
$\sqrt{8/5}\,\Delta(T_{\pm})=\hbar(\omega_{1}\pm\omega_{2})$. The background attenuation is due
to the linear coupling of sound to the $J=2^{-}$ mode, off-resonance.

The temperature dependence at zero pressure of the attenuation (amplification) of a sound wave
with frequency $\frac{\omega_{1}}{2\pi}=3.26\,\mbox{MHz}$ in the presence of a pump wave with
frequency $\frac{\omega_{2}}{2\pi}=35.8\,\mbox{MHz}$ and energy density $U/U_{c}=0.1$ is shown
in Fig. \ref{Figure:Nonlinear-Attenuation_vs_T}(b). The peak at $T_{+}$ is again due to
two-phonon absorption. However, \emph{amplification} occurs for temperatures near $T_{-}$
because of stimulated Raman scattering: phonons of frequency $\omega_{2}$ decay into phonons
with lower frequency $\omega_{1}$ and real squashons of frequency $\omega_{M}$. No absorption
peak from the linear coupling of the $J=2^{+}$ mode to the low frequency $\omega_{1}$ wave
appears in Fig. \ref{Figure:Nonlinear-Attenuation_vs_T}(b) because it occurs at a temperature
$T_{2}$ (such that $\sqrt{8/5}\,\Delta(T_{2})=\hbar\omega_{2}$), which is larger than 
$0.8T_{c}$.

\begin{figure}[!]
\includegraphics[width=0.95\textwidth]{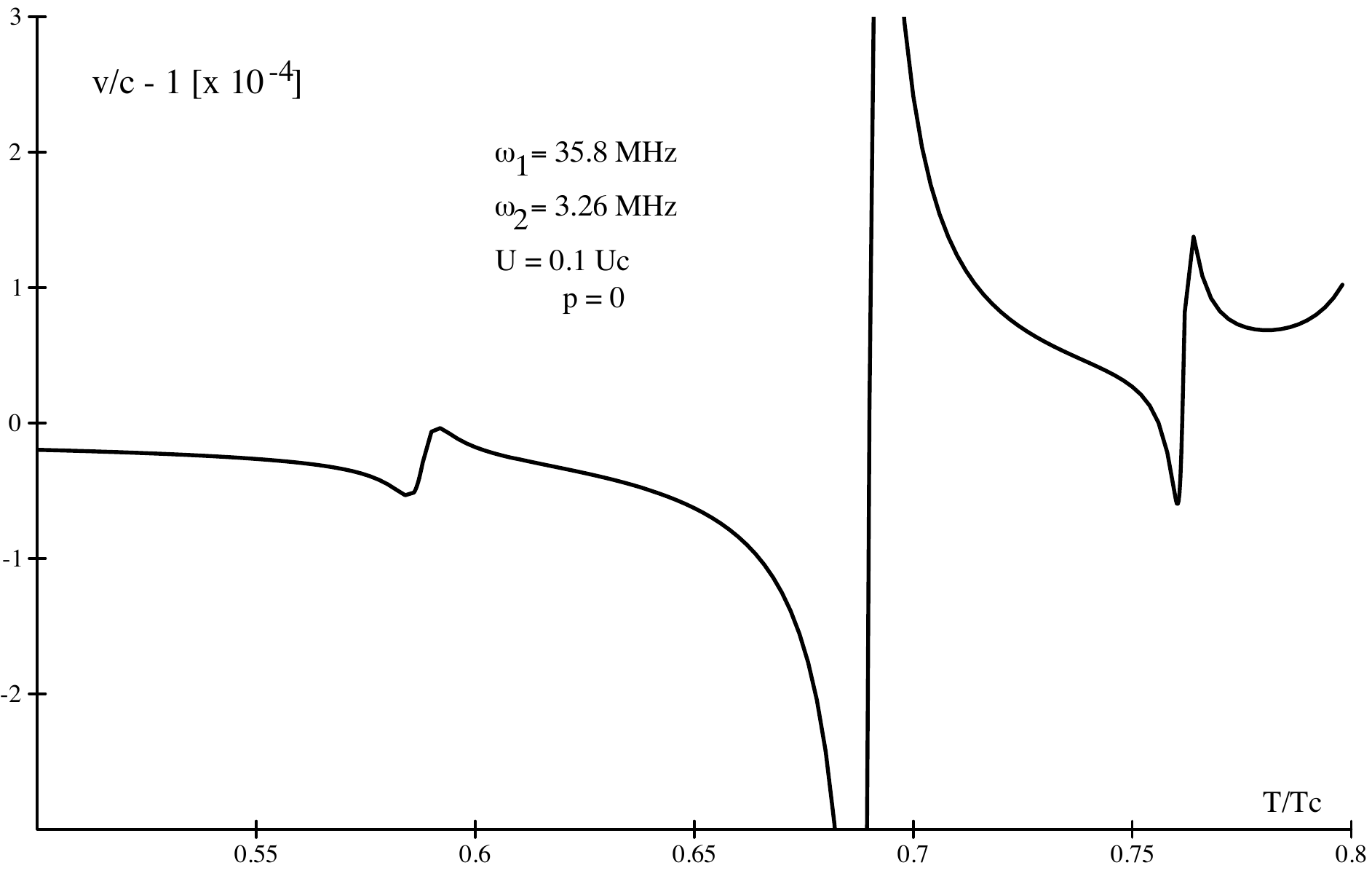}
\includegraphics[width=0.95\textwidth]{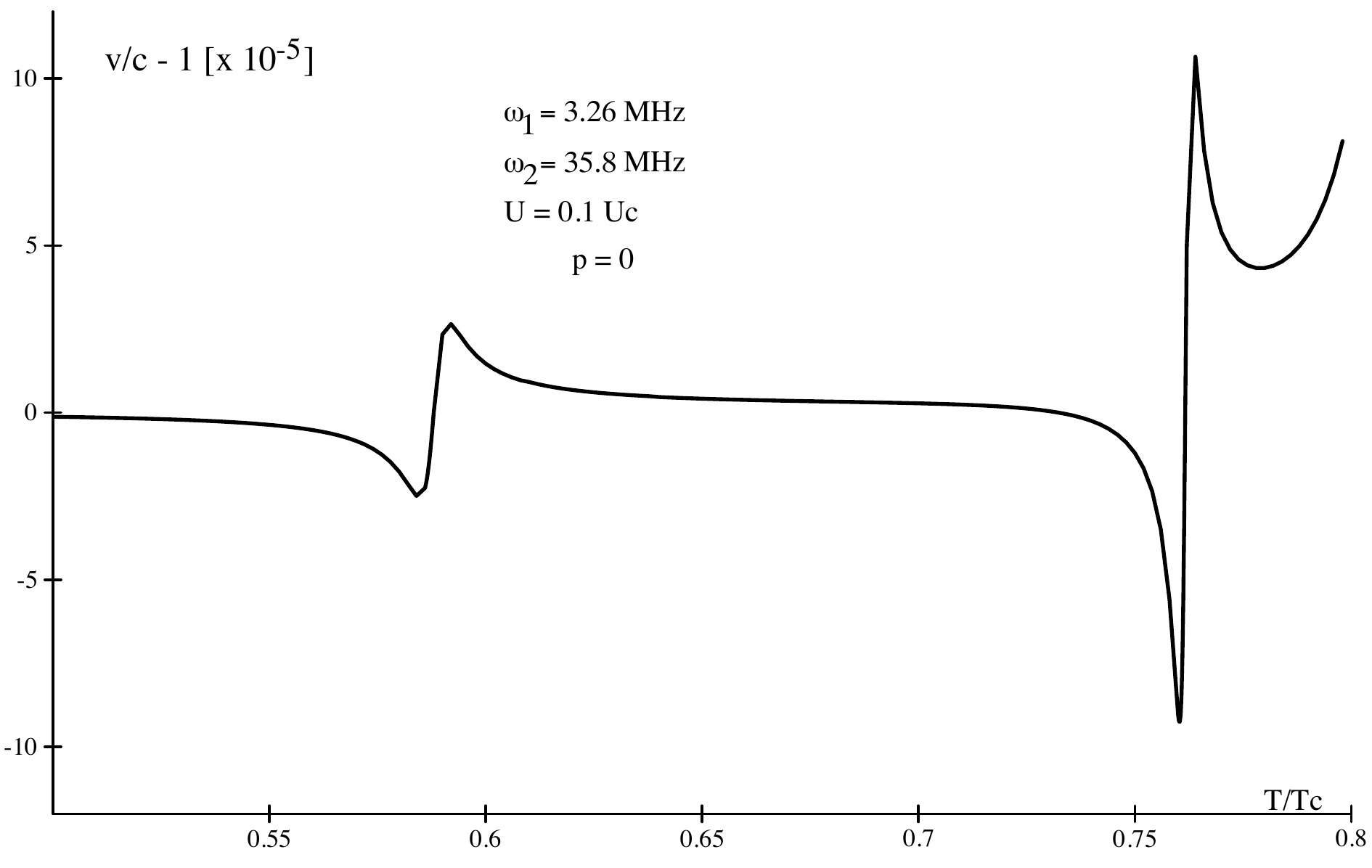}
\caption{
The predicted temperature dependence of $v(\omega,T)/c_{1}$,
the change in phase velocity of a zero-sound wave due to linear and nonlinear 
coupling to the $J=2^{+}$ mode. All parameter values are the same as in Figs.
\ref{Figure:Nonlinear-Attenuation_vs_T}(a) and (b).
}
\label{Figure:Nonlinear-Velocity_vs_T}
\end{figure}

Figures \ref{Figure:Nonlinear-Velocity_vs_T}(a) and \ref{Figure:Nonlinear-Velocity_vs_T}(b) show
the changes in phase velocity of a zero sound wave of frequency $\omega_{1}$ due to its linear
and nonlinear interaction with the $J=2^{+}$ modes in the presence of a parallel wave of high
intensity $(U\simeq 0.1U_{c})$ and frequency $\omega_{2}$. In
\ref{Figure:Nonlinear-Velocity_vs_T}(a),
$\omega_{1}/2\pi=35.8\,\mbox{MHz}\,(\hbar\omega_{1}/k_{\text{B}}T_{c}=1.87)$ and
$\omega_{2}/2\pi=3.26\,\mbox{MHz}\,(\hbar\omega_{2}/k_{\text{B}}T_{c}=0.17)$, while these two
frequencies are reversed in Fig. \ref{Figure:Nonlinear-Velocity_vs_T}(b). The features at
temperature $T_{+}\simeq 0.58T_{c}$ are due to two-phonon absorption by the $J=2^{+}$ mode. The
features at $T_{-}\simeq 0.76T_{c}$ in Fig. \ref{Figure:Nonlinear-Velocity_vs_T}(a) and (b) are
due to the inverse Raman effect and stimulated Raman scattering, respectively. Note that in both
Figs. \ref{Figure:Nonlinear-Attenuation_vs_T} and \ref{Figure:Nonlinear-Velocity_vs_T} the
sound frequencies have been chosen so that both nonlinear resonances appear in a reasonable
temperature range and do not overlap with the large anomalies due to linear coupling to the
$J=2^{-}$ mode or pair-breaking.

Although the sizes of the nonlinear absorption and velocity anomalies increase with the energy
density of the pump wave, there are obvious constraints on the amount of sound energy that is
desirable in an experiment. Just as the superfluid state is destroyed by a superflow with
velocity larger than the depairing critical velocity, it is also destroyed by sound waves with
amplitude larger than some critical value. We expect that this occurs when $U\approx U_{c}$. In
addition the pump wave causes heating of the superfluid at a maximum rate
\be
\label{gy}
C_{p}\frac{dT}{dt}=2\,\alpha\,c\,U 
\,,
\ee 
where $C_p$ and $dT/dt$ are the heat capacity and rate of temperature increase of the $^3$He,
and $\alpha$, $c$, and $U$ are the attenuation, phase velocity and energy density of the pump
wave. In an experiment the sound energy density must be sufficiently small that the cryostat can
keep the superfluid at a fixed temperature during the period of time in which the sound waves
propagate through the cell. In the large amplitude sound experiments in $^3$He-B of Polturak
{\it et al}. \cite{pol81a}, they used the measured heating rate and estimated the sound energy
density to be about one percent of the superfluid condensation energy density, {\it i.e.}
$U/U_{c}\simeq 0.01$. This may be a more realistic estimate than the value of $U/U_{c}=0.1$ used
in Figs. \ref{Figure:Nonlinear-Attenuation_vs_T} and \ref{Figure:Nonlinear-Velocity_vs_T}, shown 
for illustrative purposes.

\begin{figure}[!]
\includegraphics[width=0.95\textwidth]{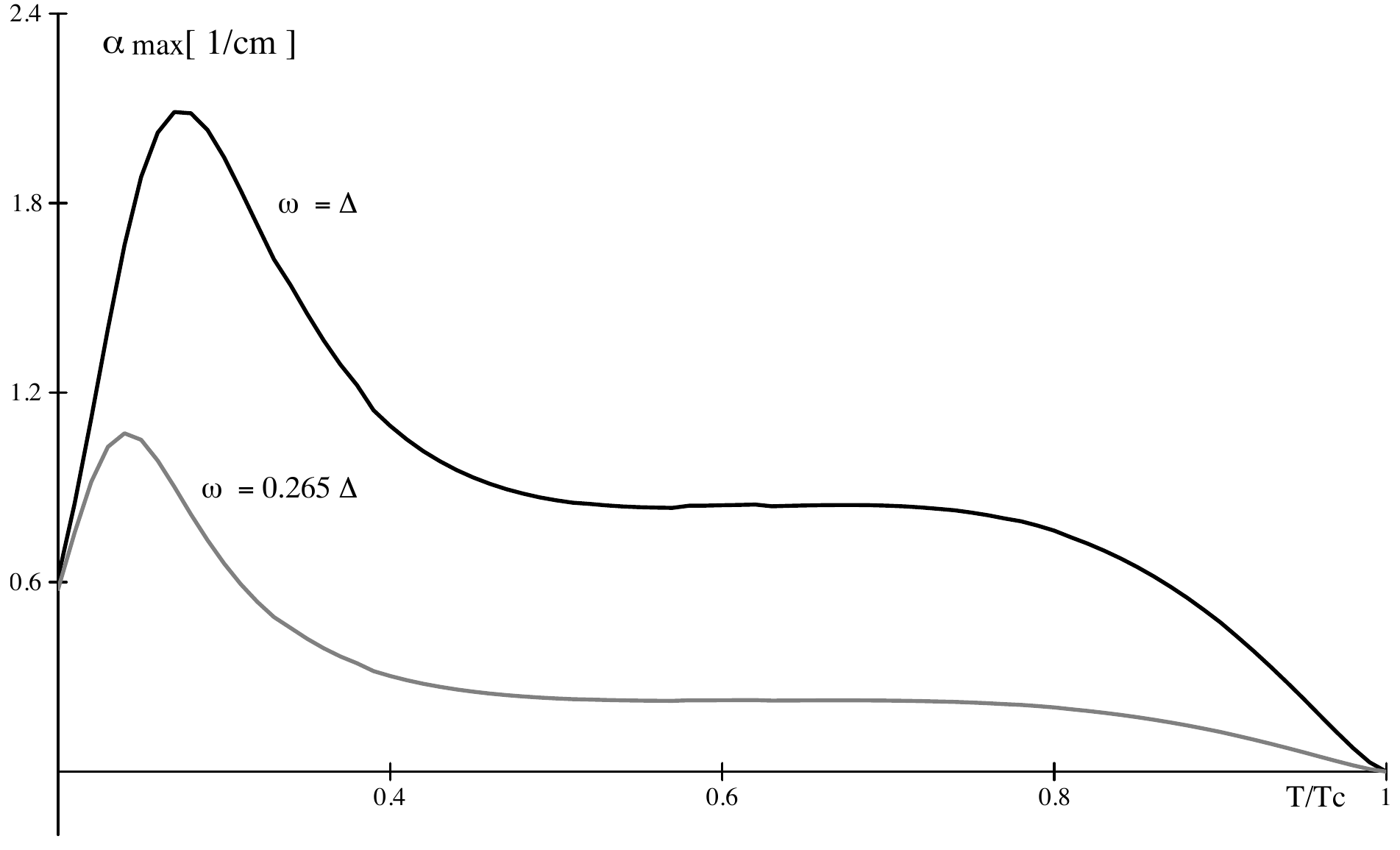}
\caption{
The temperature dependence at zero pressure of the peak
in the signal wave attenuation due to two-phonon absorption by the
$J=2^{+}$ mode for a pump wave of energy density $U=0.01\,U_{c}$ 
and a signal wave of frequency 
a) $\omega=\Delta$, and 
b) $\omega=0.265\Delta$.
}
\label{Figure:Two-Phonon_Absorption_vs_T}
\end{figure}

\begin{figure}[!]
\includegraphics[width=0.95\textwidth]{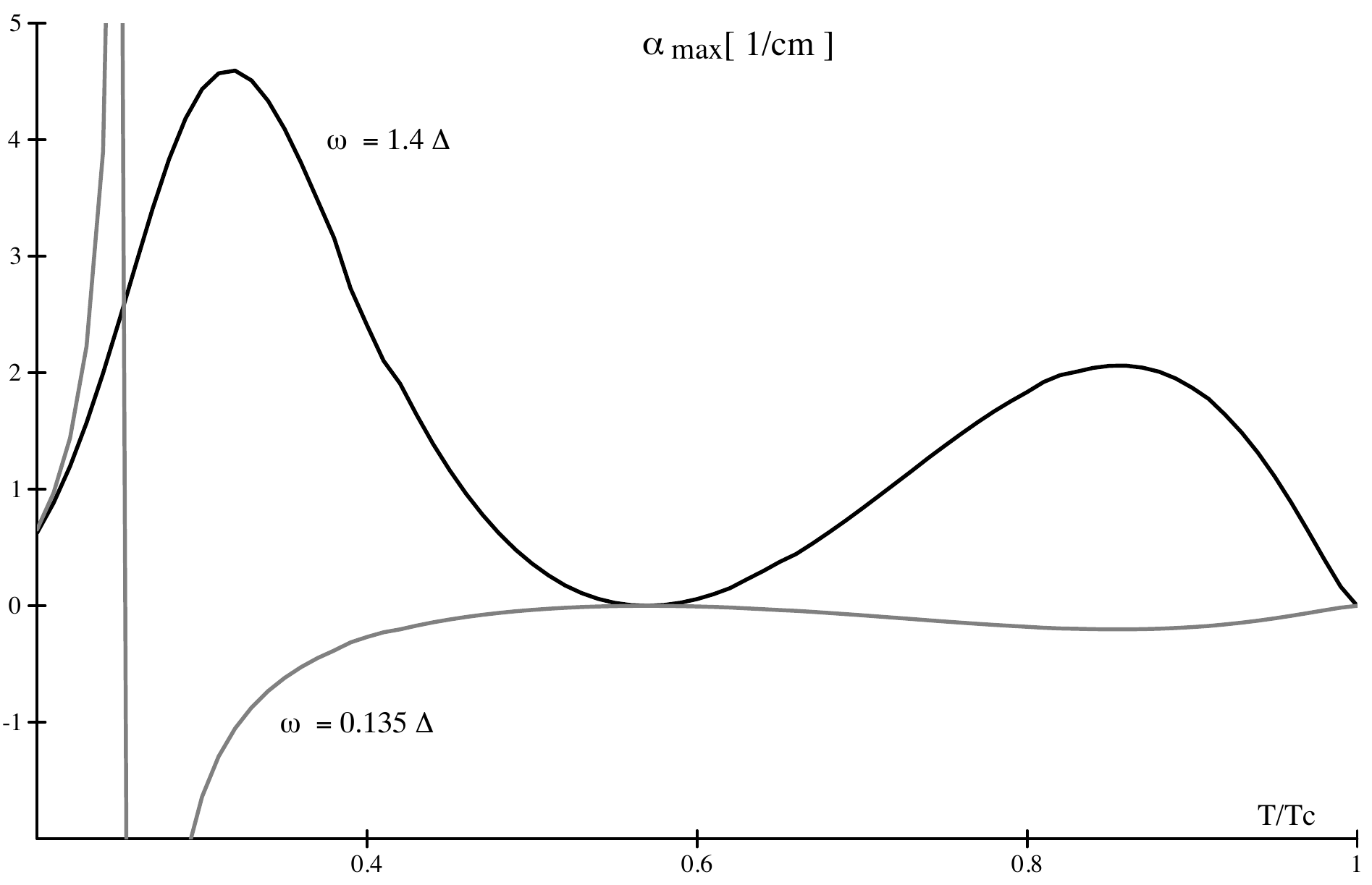}
\caption{
The predicted temperature dependence at zero pressure of a) the attenuation peak due to the
inverse Raman effect for $\omega=1.4\Delta$, and
b) the amplification peak due to stimulated Raman scattering for $\omega=0.135\Delta$. The pump
wave energy density is $U=0.01\,U_{c}$.
}
\label{Figure:Raman_Absorption_vs_T}
\end{figure}

Figures \ref{Figure:Two-Phonon_Absorption_vs_T}(a) and (b) show the temperature dependence of
the height of the attenuation peaks associated with two-phonon absorption for a pump wave of
energy density of $0.01\,U_c$. Corresponding calculations for the Raman peaks are shown in Figs.
\ref{Figure:Raman_Absorption_vs_T}(a) and (b).
The Raman peaks are expected to be small for temperatures close to $0.6 T_c$ because, as shown
in Fig. \ref{Figure:Nonlinear-Coupling_vs_T}(a), the coupling function vanishes near this
temperature. The maximum attenuation due to the nonlinear coupling of sound to the $J=2^{+}$
modes is of order $1\,\rm{cm}^{-1}$ over a wide temperature range. Since changes in the
attenuation of order $0.1\,\rm{cm}^{-1}$, and changes in the phase velocity of order one part in
$10^6$ are detectable, these nonlinear anomalies should be observable for accessible pump-wave
energy densities.


\vspace*{3mm}
\sec{Generation of third-harmonic and anti-Stokes waves}{Generation}

When we considered the three-wave resonance equations in the quasi-steady-state approximation,
we noted that in addition to the terms in the nonlinear stress tensor with the frequency of the
pump wave, $\omega_1$, and the Stokes wave, $\omega_2$, there are also oscillations with
frequencies $2\omega_{1}-\omega_{2}$ and $2\omega_{2}-\omega_{1}$ which generate sound waves
with the same frequencies. The analogous waves in nonlinear optics are known as anti-Stokes
waves. These waves have much lower intensity than the pump wave and Stokes wave, which is why we
neglected them earlier. Nevertheless, the anti-Stokes waves may be observable if the sound path
is sufficiently short to reduce the destructive interference caused by dispersion. Our
discussion is similar to Yariv's treatment (Ref. \cite{yariv75}, p. 421) of the effect of
dispersion on optical second-harmonic generation.

The term in the nonlinear stress tensor with frequency $\omega_{a}=2\omega_{1}-\omega_{2}$ and
wave vector $2\vec{q}_{1}-\vec{q}_{2}$ is
\be
\label{gz}
\delta\Pi_{a}(\vec{x},t)
=
\chi_{a}\left[\tilde{N}_{1}(\vec{x},t)\right]^{2}\,\tilde{N}_{2}(\vec{x},t)^{*}
\,,
\ee 
where
%
\ber
\label{ha}
\hspace*{-8mm}
\chi_{a} 
&=& 
	\frac{1}{N(E_f)^{2}(1+F_{0}^{s})\Delta^{2}}\left(\frac{c_{1}q_{1}}{\omega_{1}}\right)^{2}
\nonumber\\
\hspace*{-8mm}
&\times&
	\sum_{M}\,\frac{A_{M}(2\omega_{1}-\omega_{2},\omega_{1},\omega_{1}-\omega_{2})\, 
		            A_{M}(\omega_{1},\omega_{2},\omega_{1}-\omega_{2})^{*}}
			  {\phi_{M}(\vec{q}_{1}-\vec{q}_{2},\omega_{1}-\omega_{2})}
\,.
\eer
The density oscillation generated by this driving term,
\be
\label{hb}
\delta{n}(\vec{x},t) = N_{a}(\vec{x},t)\,e^{-i(\omega_{a}t-\vec{q}_{a}\cdot\vec{x})}
\,,
\ee 
satisfies the wave equation
\be
\label{hc}
\left(\pder{^{2}}{t^2} - c_{1}^{2}\nabla^{2}\right)\delta{n}
 = 
2\,c_{1}^{2}\nabla^{2}\,\big[\delta\Pi_{a}+\delta\Pi_{L}\big] 
\,,
\ee 
where $\delta\Pi_{L}$ is the contribution to the stress tensor from the linear coupling of the
anti-Stokes wave to the collective modes. In the slowly varying envelope approximation the
anti-Stokes wave obeys
\be
\label{hd}
\pder{N_{a}}{z} 
= iq_{a}\,\chi_{a}\,\left[\tilde{N}_{1}\right]^{2}\,\tilde{N}_{2}^{*}\,e^{i\Delta{q}z} 
\,,
\ee 
where $\Delta q\equiv q_{a}-2q_{1}+q_{2}$ and $\omega_{a}=v(\omega)\,q_{a}$ where $v(\omega)$ is
the frequency-dependent phase velocity for sound in the linear response limit. For frequencies
away from the $J=2^+$ collective mode resonances the dispersion is dominated by the
contributions from the excitations and off-resonant $J=2^{-}$ collective mode, and is given by
\be
\label{he}
\frac{v(\omega)}{c_{1}}-1 
	= \mbox{Re}\left(\frac{\delta \Pi_{L}}{\delta{n}}\right) 
	= \frac{6\lambda(\omega)}{25(1+F_{0}^{s})}\,
	  \frac{(\omega^{2}-4\Delta^{2})}{(\omega^{2}-12\Delta^{2}/5)}
\,,
\ee 
from Eq. (\ref{dispersion}). The wave-vector mismatch
\be
\label{hf}
\Delta{q} 
	= \omega_{2}\left(\frac{1}{v(\omega_{2})} - \frac{1}{v(\omega_{a})}\right)
	-2\omega_{1}\left(\frac{1}{v(\omega_{1})} - \frac{1}{v(\omega_{a})}\right)
\,,
\ee 
is then nonzero because of dispersion. In order to integrate Eq. (\ref{hd}) we assume that the
sound path is sufficiently short, that the attenuation of the pump and Stokes waves is
sufficiently small so that their amplitudes $\tilde{N}_{1}$ and $\tilde{N}_{2}$ are nearly
constant, and that the boundary condition for the anti-Stokes wave is $N_{a}(0)=0$. The solution
is
\be
\label{hg}
N_{a}(z) = q_{a}\chi_{a}\,\tilde{N}_{1}(0)^2\,\tilde{N}_{2}(0)\, 
		   \frac{\left(e^{i\Delta{q}z}-1\right)}{\Delta{q}}
\,.
\ee 
and the relative intensity of the anti-Stokes wave to the Stokes wave is 
\be
\label{hh}
\frac{I_{a}(z)}{I_{2}} = \vert{Gz}\vert^{2}\,J_{0}\bigg(\onehalf\Delta{q}z\bigg)^{2} 
\,,
\ee 
where $J_{0}(x) = \sin(x)/x$ is the zeroth-order spherical Bessel function, and
$G=q_{a}\chi_{a}\tilde{N}_{1}(0)^{2}$. Thus, the anti-Stokes wave is destroyed by interference
if the wave-vector mismatch or path length $L$ are sufficiently large, {\it i.e.}
$\Delta{q}L\gg 1$.
For frequencies $\omega_{1},\omega_{2}\sim\Delta$, the velocity mismatch is
$[v(\omega_{1})-v(\omega_{2})]/c_{1}\sim 1/F_{0}^{s}$, so for low pressures, where $F_{0}^{s}$
is smallest, the wave-vector mismatch is relatively large, $\Delta{q}\sim 10\,\mbox{cm}^{-1}$.
Thus, in order for anti-Stokes wave {\it not} to be destroyed by interference, the length of the
sound path should be less than $1/\Delta{q}\simeq 1\,\mbox{mm}$.\footnote{It may be possible to
choose the frequencies, $\omega_{1},\omega_{2}$, the temperature and the pressure judiciously so
that $\Delta q\approx 0$.}

The amplification of the Stokes wave was shown to be about $1\,\mbox{cm}^{-1}$ for a pump wave
with energy density $0.1\,U_{c}$. Using this energy we estimate the relative intensities of the
anti-Stokes and Stokes waves to be $I_{a}/I_{2}\simeq 10^{-3}$ for path lengths the order of a
millimeter. Thus, although the Stokes wave is much larger than the anti-Stokes wave it may still
be observable, provided it is not overdamped. If $\omega_{1}>\omega_{2}$ then the condition for
stimulated Raman scattering, $\omega_{1}-\omega_{2}\simeq\sqrt{8/5}\Delta$, implies that
$\omega_{a}>2\Delta$, and consequently the anti-Stokes wave would be strongly attenuated by pair
breaking. On the other hand, if $\omega_{2}>\omega_{1}$ and
$\omega_{2}-\omega_{1}\simeq\sqrt{8/5}\Delta$, then the anti-Stokes wave is not damped by pair
breaking.

The generation of {\it third harmonics} is
similarly limited by dispersion. Serene \cite{ser84} has noted
that the $J=2^{+}$ modes can give rise to a
significant third-harmonic of the density. If there is a high-intensity
sound wave, with frequency $\omega_1$ and wave vector $\vec {q}_1$,
\be
\label{hj}
\delta{n}(\vec{x},t) = \tilde {N}_{1}(\vec{x},t)\,e^{-i(\omega_{1}t-\vec{q}_{1}\cdot\vec{x})}
\,,
\ee 
whose second harmonic is resonant with the $J=2^{+}$ modes ({\it i.e.}
$2\omega\simeq\sqrt{8/5}\Delta$), then in the quasi-steady-state approximation the amplitudes of
these modes are given by
%
\be 
D^{+}_{2,M}(\vec{x},t) 
	= 
	\frac{A_{M}(\omega_{1},-\omega_{1},2\omega_{1})}
		 {N(E_f)^2\,\Delta\,\phi^{M}(2\vec{q}_{1},2\omega_{1})}
	\left|\frac{c_{1}q_{1}}{\omega_{1}}\right|^{2}
	\Big[\tilde{N}_{1}(\vec{x},t)\Big]^{2} + \mbox{c.c.}
\,.
\ee 
If this amplitude is inserted into the nonlinear stress tensor given by 
Eq. (\ref{nonlinear1}), together with Eq. (\ref{hj}), then a third-harmonic 
(frequency $3\omega_{1}$ and wavevector $3\vec{q}_{1}$) is obtained,
\be
\label{hl}
\delta\Pi(\vec{x},t) = \chi_{3}\Big[\tilde{N}_{1}(\vec{x},t)\Big]^{3}
\,,
\ee 
where
%
\be
\label{hm}
\chi_{3} 
	= 															\negthickspace\negmedspace
		\sum_{M}\frac{1}{N(E_f)^2\,(1+F_{0}^{s})\,\Delta^{2}} 	\negthickspace\negmedspace
 		\left|\frac{c_{1}q_{1}}{\omega_{1}}\right|^{2}			\negthickspace\negmedspace
		\left[
		\frac{A_{M}(3\omega_{1},\omega_{1},2\omega_{1}) 
			  A_{M}(\omega_{1},-\omega_{1},2\omega_{1})^{*}}
			 {\phi_{M}(2\vec{q}_{1},2\omega_{1})}
		\right].
\ee
Repeating the analysis of the of the relative intensity of the anti-Stokes waves for the
third-harmonic, it follows that the ratio of the intensity of the third harmonic to that of the
first harmonic is given by
\be
\label{hn}
\frac{I_{3}(z)}{I_{1}}
	= 	\left|G_{3}z\right|^{2}\,J_{0}\bigg(\onehalf\Delta{q}z\bigg)
\,,
\ee 
where the wave-vector mismatch is $\Delta{q}=q_{3}-3q_{1}$, and
$G_{3}=q_{3}\chi_{3}\,N_{1}(0)^{2}$. We estimate that for the third harmonic not to be destroyed
by the interference, the sound path must be shorter than $1/\Delta{q}\simeq 1\,\mbox{mm}$, as in
the case of the anti-Stokes wave. Similarly, for reasonable values of parameters the intensity
of the third harmonic is estimated to be $I_{3}/I_{1}\simeq 10^{-4}$. Although the neglect of
the third harmonic in the previous section was justified it may also be of sufficient intensity
to be observable. However, in practice it may be difficult to distinguish the third harmonic
signal generated by the nonlinear response of the superfluid from the signals generated by the
nonlinearities in the experimental apparatus \cite{ser84}.

The results presented here for third-harmonic generation differ from those of Serene in two
respects; (i) here we consider the detrimental effect of dispersion on the third-harmonic
signal, and (ii) Serene's estimates for the nonlinear coupling constants are several orders of
magnitude larger than those presented here. This discrepancy has been resolved. Serene
\cite{ser88} did not evaluate the complicated integral he obtained for the coupling to the third
harmonic, rather he assumed the properly scaled integral to be of order unity. However, Kopp
\cite{kop87b} has evaluated the integral in Serene's paper and shown that it vanishes to leading
order in $1/s^{2}\sim 1/F_{0}^{s}$. The method used here to derive the collective mode equations
shows that this cancellation to leading-order in $1/s^{2}$ is a consequence of gauge and
Galilean invariance.

\medskip
\sec{Conclusion}{Conclusion}

We have attempted to give a reasonably complete discussion
of the relationship of collisionless sound propagation and attenuation
to the order parameter collective modes of superfluid $^3$He-B.
These modes, which reflect the symmetries of the
normal state as well as the broken gauge and relative rotational symmetries
of the condensate, play a central role in the high-frequency response
of the superfluid. The acoustic spectroscopy based on the linear coupling
of the $J=2^{\pm}$ modes is well developed.
The Zeeman effect for both sets of $J=2$ modes
has been observed, as well as the effects of gap distortion and textures
on the attenuation and velocity spectra of sound. Measurements of
the temperature and field dependences of the $J=2^{\pm}$
modes have been used to obtain quantitative results
for the gap as well as the quasiparticle interactions that renormalize
the collective mode frequencies. In short, the linear response is
fairly well understood.

Much less is known about the nonlinear acoustic response of
superfluid $^3$He. We have presented the underlying
theory, based on the quasiclassical equations of superfluid 
$^3$He, of the weak nonlinear response of superfluid
$^3$He-B for the case of a three-wave resonance
between two sound waves of different frequencies and the 
$J=2^{\pm}$ modes, as well as the generation
of higher harmonic sound waves. The $J=2^{+}$ modes couple nonlinearly to
two zero-sound waves even
though exact particle-hole symmetry of the normal Fermi liquid is
built into the quasiclassical equations. The predictions for the nonlinear
absorption and velocity anomalies resulting from the $J=2^{+}$ modes depend
on relatively well understood material properties of $^3$He.

\medskip
\subsection*{\large\sl Acknowledgements}
We thank our colleagues Bill Halperin, Thilo Kopp, John Ketterson,
Dierk Rainer and Joe Serene for their many insights and helpful comments
on sound propagation, collective modes and the quasiclassical theory
of superfluid $^3$He.

\vspace*{2cm}
\sec{Corrections to the Published Manuscript}{Corrections}

\begin{enumerate}
	\item The labeling of the Nambu matrices in Eq. \ref{ca} is corrected 
	      from $\mu=\{1,2,3,4\}$ to $\mu=\{0,1,2,3\}$.
	\item Sign in Eq. \ref{cl} is corrected: $+\eta\rightarrow -\eta$.
	\item Added subscript to Eq. \ref{equlibrium-normalizations}.
	\item In the line above Eq. \ref{dl}, the corrected sentence reads:
	      ``... for $\sigma_{a}\ne\varepsilon^{+}$ ({\it i.e.} $a\ne 1$) ...''.
	\item  In the sentence above Eq. \ref{QC-transformed}, the corrected sentence reads,
		   ``... local spin rotations generated by 
			$\hat{\Lambda}=\vec{\theta}\cdot\hat{\vec{\Sigma}}$ ...''.
	\item Corrected coupling strength in Eq. \ref{Coupling_to_RSQ}:
			 $\zeta(\omega)\rightarrow[\zeta(\omega)]^2$.
	\item Corrected Eq. \ref{fd} line 1: $A(\omega,\nu,\omega-\nu)=-4\Delta^2\ldots$.
	\item Corrected Eq. \ref{fd} line 5: $E(\omega,\nu,\omega-\nu)=\frac{-2}{\nu\Delta}\ldots$.
	\item In the 2nd line of Eq. \ref{fl}, 
		  $P_{2}(\vhat{p},\vhat{s})\rightarrow P_{2}(\vhat{p}\cdot\vhat{s})$.
	\item Corrected Eq. \ref{nonlinear1} to agree with Eq. \ref{au}, 
	      which is the correct form.
	\item Corrected the RHS of the 2nd line of Eq. \ref{bd}:
	 		$A_{M}(\omega_{2},-\omega_{1},\omega_{2})$ $\rightarrow$
	 		$A_{M}(\omega_{2},-\omega_{1},\omega_{3})$.
	\item In Eq. \ref{hj}:
		  $e^{-(i\omega_{1}-\vec{q}_{1}\cdot\vec{x})}$
		  $\rightarrow$ $e^{-i(\omega_{1}t-\vec{q}_{1}\cdot\vec{x})}$.
\end{enumerate}
\end{document}